**Review on Generalized Dynamics of Soft-Matter Quasicrystals and Its Applications**

Tian-You Fan[1,*], Wenge Yang[2,#] and Xiao-Hong Sun[3,**]

[1] School of Physics, Beijing Institute of Technology, Beijing 100081, China

[2] Center for High Pressure Science and Technology Advanced Research, Shanghai 201203, China

[3] School of InformationEngineering, Zhengzhou University, Zhengzhou,450006, China

email:
*Tian-You Fan (tyfan2013@163.com),corresponding author
#Wenge Yang (yangwg@hpstar.ac.cn)
**Xiao-Hong Sun (iexhsun@zzu.edu.cn)

**Abstract**

This article provides a detailed review on the generalized dynamics of soft-matter quasicrystals developed recent years. Comparing to solid quasicrystals consisted mainly with metallic alloys, soft-matter quasicrystals have been observed in liquid crystals, polymers, colloids, nanoparticles and surfactants, which indicates quite different formation mechanism. Based onLandau-Anderson theory and group representation theory, we have studied the symmetry, symmetry breaking and elementary excitations for the observed and possible soft-matter quasicrystals. We further proposed one more elementary excitation – fluid phonon additional to the phonon and phason for solid quasicrystals, to quantitatively describe the dynamics of soft-matter quasicrystals. The general governing equations and the solutions of the dynamics evolution on the distribution, deformation and motion of the new phase are studied, which reveal quite distinguishing dynamic behaviour with those of conventional fluids and solid quasicrystals. Someapplications are introduced, which show this is an important field of new materials and materials science.

## 1. Introduction

Solid quasicrystal (or metallic alloy quasicrystals) was first reported by Shechtman et al in a metallic Al-Mn alloyin 1984 [1] with a forbidden crystallographic (icosahedral)symmetry but a long rangequasiperiodic order. For a long time, scientistsbelieved that atoms have to be packed periodically to form a crystal, and only crystallographic symmetry (230 space groups) should be present. Shechtman's discovery uncovered a new type of condensed matter structural configuration, which was eventually recognized by the 2011 Nobel Prize in Chemistry [2].Following this discovery, many other kinds of quasicrystalshave been synthesized [3-7] and even natural quasicrystalswerefound [8]. It is clear thatthe traditional crystallography and associated mathematical representation theory need to be modified to describe the structures and properties of this matter.Immediately after the discovery,Bak [9, 10] developed the Landau-Anderson symmetry breaking theory to describe the elasticity behaviour of quasicrystals, whereas a new elementary excitation---phasonwas introduced apart from phonon degree of freedom. From the group representation theory point of view, the quasicrystals can be described by higher-dimensional crystallographyfollowing Janssen's N-dimensional group theory [11], where the phonon degree of freedom is described as $\mathbf{u}$ and the phason as $\mathbf{w}$. Then the density of ordered phase according to Landau-Anderson theory [12,13] is taken to be of the form

$$\rho(\boldsymbol{r}) = \sum \rho_G e^{i\boldsymbol{G}\cdot\boldsymbol{r}} \quad (1)$$

where

$$\rho_G = |\rho_G| e^{i\left(\boldsymbol{G}^{\parallel}\cdot\boldsymbol{u}+\boldsymbol{G}^{\perp}\cdot\boldsymbol{w}\right)} \quad (2)$$

$\boldsymbol{G}^{\parallel}$is the reciprocalvector in the physical space and $\boldsymbol{G}^{\perp}$ is the conjugate vectorin the perpendicular space. Under Janssen's group theory framework, Ding et al [14], and Yang et al [15] determined all of independent nonzero elastic constants for two- and three- dimensional quasicrystalsobserved by that time, Wang et al [16] completed the similar work for one-dimensional quasicrystals with group representation theory. Applications of the linear elasticity theory on defects have been applied to symmetry pattern simulations and comparison to experimental observations at atomic scale [17-19].With these works, the generalized theory of elasticity of solid quasicrystals has been set up, the review of Hu et al [20] gave a complete summary about this, in which some analytic solutions of dislocations of the matter are obtained. Based on these developments,Fan and his group [21] further extended the mathematical theory including the comprehensive solving system of elasticity of solid quasicrystals, their solutions ranging from different initial and boundary conditions to various configurations of one-, two- and three-dimensionalquasicrystalsare given, in which some new mathematical physics equations and methods are created.

Almost at the same time toBak's work, Lubensky et al [22] studied the viscosity of solid quasicrystals. For that they developed the hydrodynamics theory for quasicrystals in which the authors revealed that the phonons represent wave propagation, while the phasons represent diffusion. They also developed the Poisson bracket method in condensed matter physics. Their contributions are one of basis of the present discussion like those listed in Refs [9-11, 14-21]. However this theory given by Lubensky et al is much more complicated, and it is quite difficult to solve the initial-boundary value problems of the equations of Lubensky et al.So far the report from Cheng and Fan et al [23] is the only application case for penta-/deca-gonal system in strict sense of hydrodynamics solution of Lubensky theory, in which some significant results are given for the first time. Before the strict solving to the equations of Lubensky et al, Rochal and Lorman [24] and Fan et al [25-27] suggested the simplified form of the equations of Lubensky et al, i.e., the linearized form of the equations, or so-called phonon-phason dynamics, Rochal and Lorman gave some discussions in equality while Fan et al presented systematical quantitative solutions. Although these studies revealed the wave propagation behaviour of phonons and that of diffusion of phasons of solid quasicrystals, these are not the strict hydrodynamics analysis.

Most metallic alloy quasicrystals are very brittle at low and intermediate temperature, but they dramatically deform plastically at high temperature. As reported by Messerschmidt [28], the study onthe plasticity of solid quasicrystals,like dislocations and their motion at high temperature is very important for understanding the deformation mechanism of this matter,but the theory of the plasticity has not been advanced so far [29]due to the complexity of the problem up to now and lack of enough experimental results.

While the study on solid quasicrystals and their mechanical behaviour has been developed and met great difficulties, the soft-matter quasicrystalswere observed extensively in liquid crystals, polymers, colloids, nanoparticles and surfactants and so on [30-41]. This is an important event in 21th century chemistry**.**The soft-matter quasicrystals are formed through self-assembly of spherical building blocks by supramolecules, compounds and block copolymers and so on, which is associated with chemical process and is quite different from that of solid quasicrystals. Thenewstructure presents both natures of soft matter andquasicrystals.Soft matter is an intermediate phase between ideal solid and simple fluid, which exhibits fluidity as well as complexity as pointed out by de Gennes [42], while quasicrystals are highly ordered phase.We can say the soft-matter quasicrystals are complex fluid with quasiperiodic symmetry. Refs [43-51] reviewed soft-matter quasicrystals from different angles on their formation mechanism, structure stability, thermodynamics and the correlation between Frank-Kasper phase and quasicrystals etc.To describe the dynamics of soft-matter quasicrystals, how can one generalize the knowledge from regular softmatter, and what is the difference from those for simple fluids and solid quasicrystals?

To describe the fluidity, a new elementary excitation--- fluid phonon is introduced [52-55] besides the phonon and phason as described above for solid quasicrystal study, in which the concept of fluid phonon was originated from Landau school [56]. As the introducing of the fluid phonon, the equation of state is discussed in the present research, which is also a difficult topic in soft matter study. Similar like solid quasicrystals, we start with Landau-Anderson theory for soft-matter quasicrystals study, and utilizegroup theory and group representation theory to configure the energy terms.

For all two-dimensional solid quasicrystals discovered so far (5-, 8-, 10-, and 12-fold symmetry), only two dimensional perpendicular space is needed to describe the phason term in equation (2) (we name this category as first type ofsoft-matter quasicrystals in this paper). For describing possible 2d quasicrystals with 7-, 9, 14- and 18-fold symmetry, Hu et al [18,57] and Yang et al [58] proposed to use two two-dimensional perpendicular spaces to describe two kinds of phasonelementary excitations from group representation theory point of view (we name this category as second type of soft-matter quasicrystals), although no such a solid quasicrystal has been reported yet.It was exciting to see the report of a 18-fold symmetry colloids by Fisher et al [39] after 17 years of the prediction of Hu et al [18,57] and Yang et al [58] firstly, this shows the power of the group theory. By group representation theory, Ding et al [12], Hu et al [18,57] and Yang et al [58] have determinedall nonzero independent elastic constants of two-dimensional quasicrystals, which help to construct the Hamiltonians of elementary excitations of soft-matterquasicrystals.

The elementary excitations ---phonons, phasons and fluid phonon,their interactions and the equation of state constitute the basis of the generalized dynamics of the matter. The governing equations (nonlinear partial differential equation set) of the dynamics for various soft-matter quasicrystalssystems observed so far and possible in the future will be generated. Applications on describing the motion of the matter, the prices of the solutions---ranging from transient response, flow past obstacle to topological defect and metric defect will be presented. Through these studies, the results provide a good way to examine the existing theory, predication of the distribution, deformation and motion of the matter and present the giant differences in dynamic responses between soft-matter quasicrystals and solid quasicrystals. The dynamics based on the Landau phenomenological theory and group representation theory may help soft-matter quasicrystals study to be become another quantitative branch in soft matter science after that of liquid crystalsand provide a tool for stress analysis of some fundamental samples in needs of experiments and applications. An application of the matter to photon band-gap is also introduced in the Subsection 6.6 in addition.

## 2. Symmetry, symmetry breaking, elementary excitations of soft-matter quasicrystals

The formation mechanism and matter constitution of soft-matter quasicrystals are quite different from those of solid quasicrystals, but they share the similar non-crystalline symmetries as the latter. So far, the soft-matter quasicrystals have the highest symmetry in soft matter. In the structure and property studies,the symmetry breaking and elementary excitations play a central role in understanding the fundamental mechanism of static and dynamic procedure.

The observed soft-matter quasicrystals so far are all two-dimensional quasicrystals. Fig. 1 shows a 12-fold symmetry diffraction pattern whichwas frequently observed in many kinds of soft matter[30-41]. From the group theory point of view, the similar symmetry patterns with 5-, 8- and 10-fold are possible candidates for soft-matter quasicrystals as well. Their symmetry properties have been studied by Janssen [11], Ding et al [14], Yang et al [15,17-19], Hu et al [20] in detail, the point groups are listed in Table 1.

For describing a planar symmetrical lattice with crystalline allowed 1-, 2-, 3-, 4- and 6-fold symmetry, two linear independent unit lattices are enough, for which we consider a 2d crystallographic space $E^2$ is sufficient to describe all lattice in the real (physical) space, i.e., $E^2 = E_P^2$. But for the aforementioned 5-, 8-, 10- and 12- fold symmetry, four linear independent unit lattices are needed to describe the structure, i.e., a 4d space $E^4$ is required to describe the 2d quasiperiodic structure in real space, we label one additional 2d space as perpendicular space $E_{\perp}^2$, i.e., $E^4 = E_P^2 \oplus E_{\perp}^2$. We call this class as the first kind of quasicrystals. So far, all solid quasicrystals belong to the first kind. The later discovered 18-fold symmetrical soft-matter quasicrystals, it requests 6 linear independent unit lattices to describe the two-dimensional real space lattice, for which we need to introduce two additional 2d perpendicular spaces, we label them as $E_{\perp 1}^2$ and $E_{\perp 2}^2$, i.e.,

$$E^6 = E_P^2 \oplus E_{\perp 1}^2 \oplus E_{\perp 2}^2. \qquad (3)$$

We call this class as the second kind of quasicrystals [57, 58].

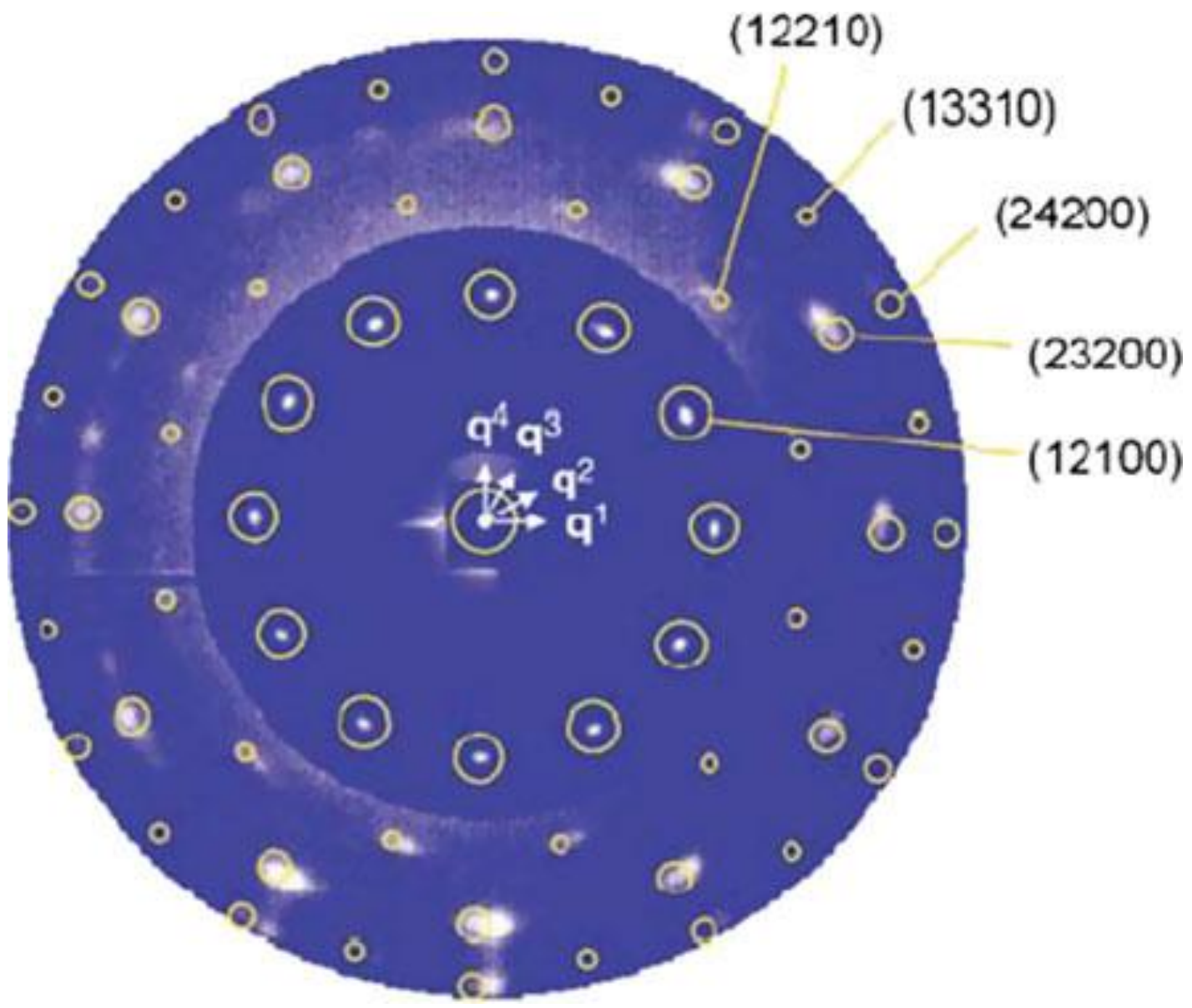

Fig. 1 Diffraction pattern of 12-fold symmetry quasicrystals in soft matter [39].

Table 1 Symmetry systems and their possible point groups for first kind of two-dimensional quasicrystals

| Systems | Names(for solid) | Point groups |
|---|---|---|
| 5-fold symmetry | Pentagonal quasicystals | $5, \bar{5}$ |
| | | $5m, 52, \bar{5}m$ |
| 10-fold symmetry | Decagonal quasicrystals | $10, \overline{10}, 10/m$ |
| | | $10mm, 1022, \overline{10}m2, 10/mmm$ |
| 8-fold symmetry | Octagonal quasicrystals | $8, \bar{8}, 8/m$ |
| | | $8mm, 822, \bar{8}m2, 8/mmm$ |
| 12-fold symmetry | Dodecagonal quasicrystals | $12, \overline{12}, 12/m$ |
| | | $12mm, 1222, \overline{12}m2, 12/mmm$ |

The diffraction pattern of observed soft-matter quasicrystals with 18-fold symmetry is shown in Fig.2.The mathematic Penrose tiling description on this type of quasicrystals is depicted in Fig. 3.The symmetry group representation methods of second kind of two-dimensional quasicrystals have been studied by Hu et al [56], Yang et al [57] and Tang and Fan refer to [55]. Based on the Schoenflies method, the point groups of 7-, 14-, 9- and 18-fold symmetry quasicrystals are listed in Table 2.

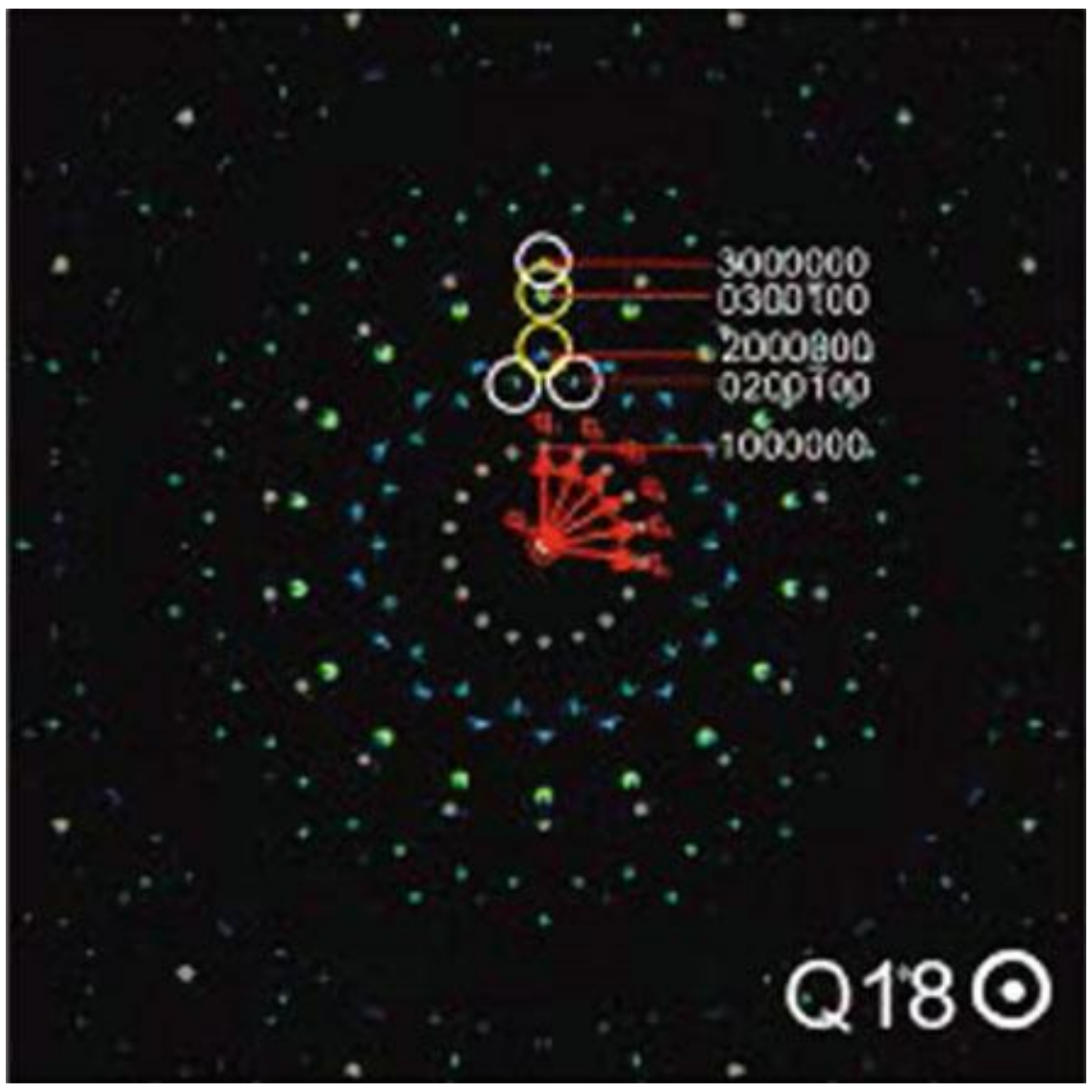

Fig.2 Diffraction pattern of soft-matter quasicrystals with 18-fold symmetry. [39]

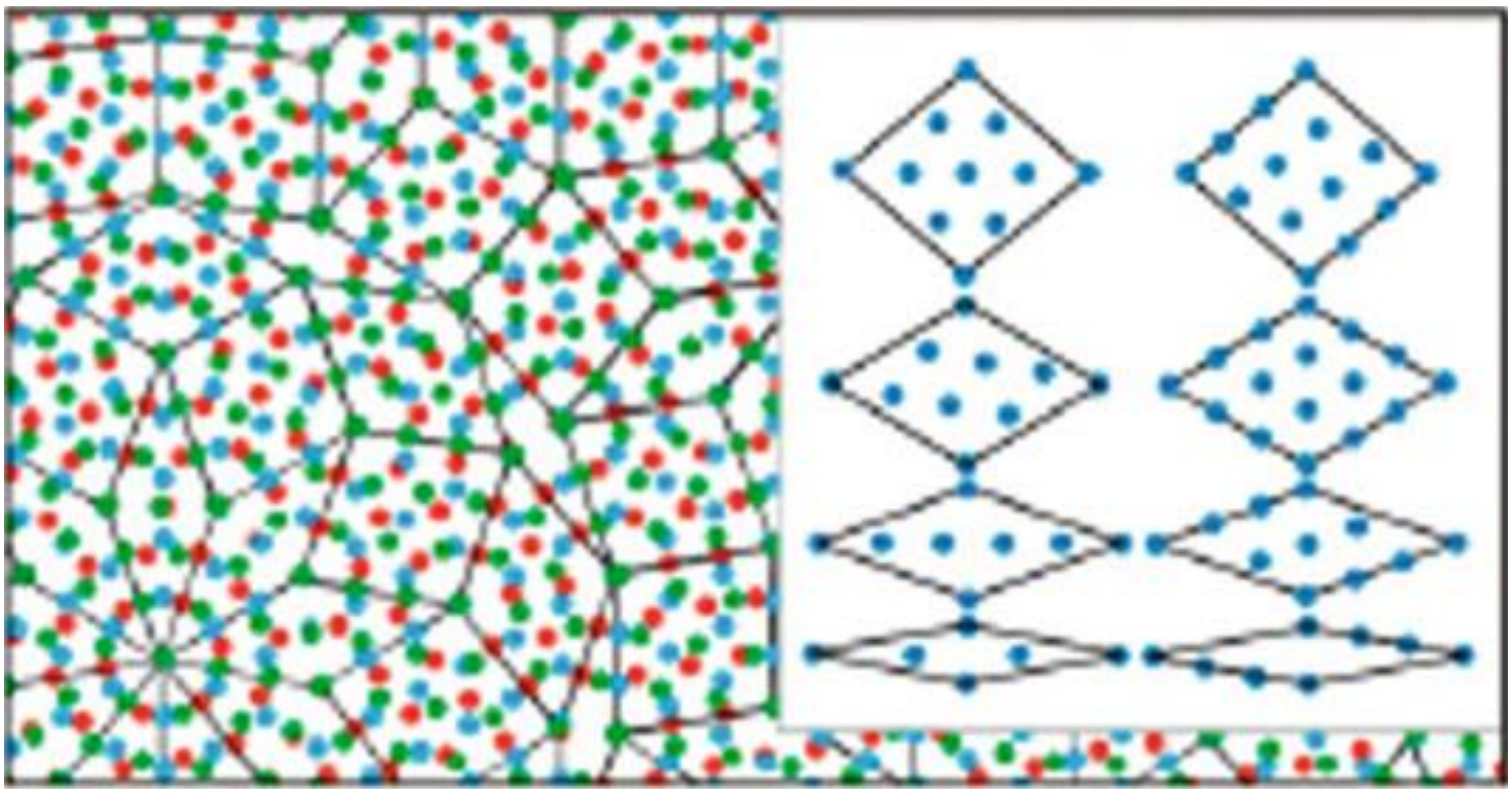

Fig. 3 The mathematic description of 18-fold planar symmetry with Penrose tiling. [39]

Table 2. Point groups of allsecond kind of two-dimensional quasicrystals

| Axis | Plus $m_h$ | Plus $m_v$ | Plus $2_h$ | Plus multi-operators |
|---|---|---|---|---|
| 7 | $\overline{14}$ | $7m$ | 72 | $\overline{14}m2$ |
| $\bar{7}$ | | $\bar{7}m$ | $\bar{7}m$ | |
| 14 | $14/m$ | $14mm$ | 1422 | $14/mmm$ |
| $\overline{14}$ | | $\overline{14}m2$ | $\overline{14}m2$ | |
| 9 | $\overline{18}$ | $9m$ | 92 | $\overline{18}m2$ |
| $\bar{9}$ | | $\bar{9}m$ | $\bar{9}m$ | |
| 18 | $18/m$ | $18mm$ | 1822 | $18/mmm$ |
| $\overline{18}$ | | $\overline{18}m2$ | $\overline{18}m2$ | |

Group representation theory including the character tables of the second kind of two-dimensional quasicrystals is quite complicated, and the introduction about the derivation needs a very large volume of space, we will not discuss in detail here. The key results concerning this are the determination of quadratic invariants of strain tensors of phonons, first and second phasons and their couplings (i.e., the determination of all independent nonzero components of physical modulus tensors of the material) and the constitutive equations, which will be presented late in Sections 3 and 4, respectively.

The symmetry breaking of first kind of two-dimensional quasicrystals leads to two elementary excitations – phonon field $\mathbf{u}$ in physical space $E_P^2$ and phason field $\mathbf{w}$ in perpendicular space $E_\perp^2$, as described in equation (2), in which $\mathbf{u}$ represents wave propagation, and $\mathbf{w}$ represents diffusion.The symmetry breaking of second kind of two-dimensional quasicrystals leads to elementary excitations: phonon $\mathbf{u}$ in $E_P^2$ and two kinds of phasons $\mathbf{v}$ in $E_{\perp 1}^2$ and $\mathbf{w}$ in $E_{\perp 2}^2$, as mentioned in equation 3. Here and after, we call $E_{\perp 1}^2$ as the first perpendicular space, $E_{\perp 2}^2$ the second one. Based on the concept, the Landau-Anderson expansion (1) may be extended as

$$\rho(\mathbf{r}) = \sum_{\mathbf{G}\in L_R} \rho_{\mathbf{G}} \exp\{i\mathbf{G}\bullet\mathbf{r}\} = \sum_{\mathbf{G}\in L_R} |\rho_{\mathbf{G}}| \exp\{-i\Phi_{\mathbf{G}} + i\mathbf{G}\bullet\mathbf{r}\} \quad (4)$$

with the extended phase angular

$$\Phi_n = \mathbf{G}_n^{\|}\bullet\mathbf{u} + \mathbf{G}_n^{\perp 1}\bullet\mathbf{v} + \mathbf{G}_n^{\perp 2}\bullet\mathbf{w} \quad (5)$$

in which $\mathbf{G}_n^{\parallel}$ represents reciprocal lattice vector in parallel space $E_{\parallel}^2$, and $\mathbf{G}_n^{\perp 1}$ and $\mathbf{G}_n^{\perp 2}$ the reciprocal lattice vectors in the first and second perpendicular space $E_{\perp 1}^2$ and $E_{\perp 2}^2$, $\mathbf{u}$ the phonon displacement field in parallel space, and $\mathbf{v}$ and $\mathbf{w}$ the first and second phason displacement fields in the two perpendicular spaces, respectively.

Additionalto phonon and phason, soft-matter quasicrystalhasanother elementary excitation - fluid phonon, whose field is velocity $\mathbf{V}$, which is originated from Landau school [56], this is the substantive difference between soft-matter quasicrystals and solid quasicrystals from elementary excitations point of view.

## 3 The first kind of soft-matter quasicrystals

The two-dimensional solid quasicrystals discovered so far are pentagonal, octagonal, decagonal and dodecagonal quasicrystals. They present following common features:

1) One needs a set of four rationally independent reciprocal basis vectors to index the diffraction pattern with integers;

2) The basis vectors can be considered as projections from a 4-dimensional embedding space ($V$) upon to 2-dimensional physical space ($V_E$);

3) The space $V$ is the direct sum of $V_E$ and $V_I$, where $V_I$ is the orthogonal complementary space;

4) Four hydrodynamic degrees of freedom in phases can be parameterized by two2-dimensional vector fields. One of them is the phonon field (denoted by $\mathbf{u}$), and the other is the phason field (denoted by $\mathbf{w}$). This can be described by Landau-Anderson equation (2).

The experimentally observed 12-fold symmetry soft-matter quasicrystal presents also above characters.Although not reported yet, possible 5-, 8- and 10-fold symmetry soft-matter quasicrystals should have the similar characters too. These structures can be classified as the first kind of soft-matter quasicrystals. They have the elementary excitations---phonon, phason and fluid phononwith displacement $\mathbf{u}$ and $\mathbf{w}$ and velocity $\mathbf{V}$, respectively.

In the two-dimensional quasicrystals of $n$-fold symmetry($n = 5, 8, 10, 12$), atoms along the $n$-fold axis are arranged periodically, while atoms are arranged quasiperiodically in the plane perpendicular to this axis. If we take $z$-axis as the $n$-fold axis, then we have the field variables $u_x, u_y, u_z; \mathrm{w}_x, \mathrm{w}_y, \mathrm{w}_z = 0; \mathrm{V}_x, \mathrm{V}_y, \mathrm{V}_z$.

To describe distribution, deformation and motion of soft-matter quasicrystals, we need the following geometry quantities or kinetic quantities

$$\varepsilon_{ij}=\frac{1}{2}\left(\frac{\partial u_i}{\partial x_j}+\frac{\partial u_j}{\partial x_i}\right), w_{ij}=\frac{\partial w_i}{\partial x_j}, \dot{\xi}_{ij}=\frac{1}{2}\left(\frac{\partial V_i}{\partial x_j}+\frac{\partial V_j}{\partial x_i}\right) \quad (6)$$

in (6) the first term is called phonon strain tensor, the second phason strain tensor and the third the deformation rate tensor of fluid phonon, respectively.

We have the following constitutive relations

$$\left.\begin{aligned} \sigma_{ij} &= C_{ijkl}\varepsilon_{ik} + \mathrm{R}_{ijkl}\, w_{kl}, \\ H_{ij} &= K_{ijkl}w_{ij} + \mathrm{R}_{klij}\, \varepsilon_{kl}, \\ p_{ij} &= -p\delta_{ij} + \sigma'_{ij} = -p\delta_{ij} + \eta_{ijkl}\dot{\xi}_{kl}, \end{aligned}\right\} \quad (7)$$

in which $\sigma_{ij}$ called the phonon stress tensor, $C_{ijkl}$ the phonon elastic tensor, and $H_{ij}$ the phason stress tensor, $K_{ijkl}$ the phason elastic tensor , $\mathrm{R}_{ijkl}$ and $\mathrm{R}_{klij}$ the phonon-phason coupling elastic tensor, $p_{ij}$ the fluid stress tensor, $p$ the pressure, $\sigma'_{ij}$ the viscosity fluid stress tensor, $\eta_{ijkl}$ the viscosity coefficient tensor of fluid, respectively. The constants $C_{ijkl}, K_{ijkl}$ and $\mathrm{R}_{ijkl}$ and $\mathrm{R}_{klij}$ are determined by group representation theory, refer to Hu et el [20]. According to the group representation theory, for 12-fold symmetry quasicystals, phonon-phason coupling elastic tensor $\mathrm{R}_{ijkl}=0, \mathrm{R}_{klij}=0$, this shows there is no coupling between phonon and phason for the quasicrystal system.

With the above information, we obtain the energy of the individual quasicrystal system so the Hamiltonians of the relevant system, which help us to derive the dynamics equations by the Poisson bracket method of condensed matter physics, those will be discussed in Section 5 and Appendix.

## 4 Second kind of soft-matter quasicrystals

A 18-fold symmetry quasicrystal was first reported in soft-matter [39]. It belongs to second kind of quasicrystals which requires a six-dimensional embedding space to describe its lattice. The possible 7-, 14- and 9-fold symmetry quasicrystals, not yet reported by experiments, belong to the same class as well. The six-dimensional embedding space consists of parallel space $E_{\parallel}^2$ and two perpendicular spaces $E_{\perp 1}^2$ and $E_{\perp 2}^2$, i.e., we recall because of the importance(and readers are not so familiar)

$E^6 = E_{\|}^2 \oplus E_{\perp 1}^2 \oplus E_{\perp 2}^2$ （8）

We call $E_{\perp 1}^2$ as the first perpendicular space, $E_{\perp 2}^2$ the second one. Based on this concept, the Landau-Anderson expansion may be extended as

$\rho(\mathbf{r}) = \sum_{\mathbf{G}\in L_R} \rho_{\mathbf{G}} \exp\{i\mathbf{G}\bullet\mathbf{r}\} = \sum_{\mathbf{G}\in L_R} |\rho_{\mathbf{G}}| \exp\{-i\Phi_{\mathbf{G}} + i\mathbf{G}\bullet\mathbf{r}\}$ （9）

with the extended phase angle term (although it is repeat with (5), we recall again due to the importance)

$$\Phi_n = \mathbf{G}_n^{\|}\bullet\mathbf{u} + \mathbf{G}_n^{\perp 1}\bullet\mathbf{v} + \mathbf{G}_n^{\perp 2}\bullet\mathbf{w} \qquad (10)$$

here $\mathbf{G}_n^{\|}$ represents reciprocal lattice vector in parallel space $E_{\|}^2$, and $\mathbf{G}_n^{\perp 1}$ and $\mathbf{G}_n^{\perp 2}$ the reciprocal lattice vectors in the first $E_{\perp 1}^2$ and second $E_{\perp 2}^2$ perpendicular space , $\mathbf{u}$ the phonon displacement field in parallel space, and $\mathbf{v}$ and $\mathbf{w}$ the first and second phason displacement fields in the two perpendicular spaces, respectively.

In the casefor second kind of quasicrystals, the Janssen theory developed in 1992 [11] cannot be used to describe the structure from symmetry point of view. Later Hu et al in 1994 [57] and Yang et al in 1995 [58] suggested to adopt the six-dimensional embedding space hypothesis and extended the Landau-Anderson theory. All 7-, 9-, 14-, and 18-fold symmetry second kind of quasicrystals can be described with the equations (9) and (10). To describe the observed 18-fold symmetry and possible 7-, 9-, 14-fold symmetry soft-matter quasicrystals, the fluid phonon field $\mathbf{V}$ was suggested by Fan recently [52].

Only considering the x-y quasiperiodic plan withthe convention from Hu et al [57] and Yang et el [58], the displacements are defined as $\mathrm{u} = (u_x, u_y)$ , $\mathrm{v} = (v_x, v_y)$ , $\mathrm{w} = (w_x, w_y)$ and the corresponding strain fields can be derived as

$$\varepsilon_{ij} = \frac{1}{2}\left(\frac{\partial u_i}{\partial x_j} + \frac{\partial u_j}{\partial x_i}\right), \quad v_{ij} = \frac{\partial v_i}{\partial x_j}, w_{ij} = \frac{\partial w_i}{\partial x_j}, \qquad (11)$$

and the fluid velocity field $\mathrm{V} = (V_x, V_y)$ and the fluid deformation rate tensor

$$\dot{\xi}_{ij} = \frac{1}{2}\left(\frac{\partial V_i}{\partial x_j} + \frac{\partial V_j}{\partial x_i}\right) \quad (12).$$

Then the corresponding constitutive law is

$$
\left.\begin{aligned}
\sigma_{ij} &= C_{ijkl}\varepsilon_{kl} + r_{ijkl}v_{kl} + R_{ijkl}w_{kl} \\
\tau_{ij} &= T_{ijkl}v_{kl} + r_{klij}\varepsilon_{kl} + G_{ijkl}w_{kl} \\
H_{ij} &= K_{ijkl}w_{kl} + R_{klij}\varepsilon_{kl} + G_{klij}v_{kl} \\
p_{ij} &= -p\delta_{ij} + \sigma_{ij}^{'} = -p\delta_{ij} + \eta_{ijkl}\dot{\xi}_{kl}
\end{aligned}\right\} \quad (13)
$$

the meanings of $\sigma_{ij}$ , $C_{ijkl}$ , $p_{ij}$ and $\eta_{ijkl}$ are as those defined in previous section, $r_{ijkl}$ the phonon-first phason coupling (i.e., the $\mathrm{u}-\mathrm{v}$ coupling ) elastic constant tensor, $R_{ijkl}$ the phonon-second phason coupling (i.e., the $\mathrm{u}-\mathrm{w}$ coupling ) elastic constant tensor, $\tau_{ij}$ the stress tensor associated with phason strain tensor $v_{ij}$ , $T_{ijkl}$ the phason elastic constants corresponding to $\tau_{ij}-v_{ij}$ , the meanings of $H_{ij}$ and $K_{ijkl}$ are the same discussed beforehand, corresponding to field $\mathrm{w}$ , but it is second phason field now, and $G_{ijkl}$ the elastic constants of coupling (i.e., the $\mathrm{v}-\mathrm{w}$ coupling) between first-second phason fields. According to group representation theory [57,58], the independent nonzero elastic constants of phonons, phasons, coupling elastic constants between phonon-phason, first phason-second phason are determined.

With the above information, we obtain the energy of the individual quasicrystal system so the Hamiltonians of the relevant system, which help us to derive the dynamics equations by the Poisson bracket method of condensed matter physics [59-63], those will be discussed in Section 5 and Appendix.

## 5 Generalized dynamics of soft-matter quasicrystals

Soft matter is easy to deform, and responses readily to a weak fluctuation of externalfield. Soft-matter quasicrystals behaviour also like this. Studying the dynamics of the matter is significant. We focus on the dynamic analysis quantitatively on the distribution, deformation and motion of the new phase.

### 5.1 The general form of generalized dynamics of soft-matter quasicrystals

Based on the Poisson bracket method of condensed matter physics [59-63] we have the mass conservation equation or continuum equation

$$
\frac{\partial \rho}{\partial t} + \nabla_k \left(\rho V_k\right) = 0 \quad (14)
$$

the momentum conservation equations or generalized Navier-Stokes equations

$$\frac{\partial g_i(\mathrm{r},t)}{\partial t} = -\nabla_k(\mathrm{r})(V_k g_i) + \nabla_j(\mathrm{r})\left(-p\delta_{ij} + \eta_{ijkl}\nabla_k(\mathrm{r})g_l\right) - \left(\delta_{ij} - \nabla_i u_j\right)\frac{\delta H}{\delta u_j(\mathrm{r},t)} - \left(\nabla_i w_j\right)\frac{\delta H}{\delta w_j(\mathrm{r},t)} - \rho\nabla_i(\mathrm{r})\frac{\delta H}{\delta\rho(\mathrm{r},t)}, g_j = \rho V_j \quad (15)$$

the equations of motion of phonons due to the symmetry breaking

$$\frac{\partial u_i(\mathrm{r},t)}{\partial t} = -V_j\nabla_j(\mathrm{r})u_i - \Gamma_{\mathrm{u}}\frac{\delta H}{\delta u_i(\mathrm{r},t)} + V_i, \quad (16)$$

in which $\Gamma_{\mathrm{u}}$ represents phonon dissipation coefficient, and the equations of motion of phasons due to the symmetry breaking

$$\frac{\partial w_i(\mathrm{r},t)}{\partial t} = -V_j\nabla_j(\mathrm{r})w_i - \Gamma_{\mathrm{w}}\frac{\delta H}{\delta w_i(\mathrm{r},t)}, \quad (17)$$

in which $\Gamma_w$ represents phason dissipation coefficient. However the equation set up to now is not closed yet, because the number of field variables is greater than that of field equations. We must supplement one additional equation, the equation of state, i.e., the relation between fluid pressure and mass density:

$$p = f(\rho)$$

which is a difficult topic in the study of soft matter.

Wensink[64] studied the equation of state on columnar liquid crystals in one-dimensional case, but in our computation there are some difficulties by using it, then we [51] take following modifications as

$$p = f(\rho) = 3\frac{k_B T}{l^3\rho_0^3}\left(\rho_0^2\rho + \rho_0\rho^2 + \rho^3\right) \quad (18)$$

where $\rho_0$ is the initial density, or the rest mass density, $k_B$ the Boltzmann constant, $T$ the absolute temperature, $l$ the thickness of hard disks of the columnar liquid crystals in the original paper of Wensink [64], we here take it as characteristic size of soft-matter quasicrystals, in general this is a meso-characteristic size, $l = 1 \sim 100nm$, and our computation shows if take $l = 8 \sim 9nm$ the computational results are in the best accuracy. The equation of state with power function was also used by Landau [65] but for superfluid.

In equations (15-17), the $H$ denotes the Hamiltonian defined by

$$H = H[\Psi(\mathrm{r},t)] = \int\frac{\mathrm{g}^2}{2\rho}d^d\mathrm{r} + \int\left[\frac{1}{2}A\left(\frac{\delta\rho}{\rho_0}\right)^2 + B\left(\frac{\delta\rho}{\rho_0}\right)\nabla\bullet\mathrm{u}\right]d^d\mathrm{r} + F_{el}$$

$$= H_{kin} + H_{density} + F_{el} \quad (19)$$

$$F_{el} = F_{\mathrm{u}} + F_{\mathrm{w}} + F_{\mathrm{uw}}, \mathrm{g} = \rho\mathrm{V}$$

and $\mathrm{V}$ represents the fluid velocity field, $A, B$ the constants describing density variation. The last term of (19) represents elastic energies, which consists of phonons, phasons and phonon-phason coupling parts. For the first kind of soft-matter quasicrystals, $F_{el}$ can be obtained by summering following three terms

$$
\begin{aligned}
F_{\mathrm{u}} &= \int \frac{1}{2} C_{ijkl} \varepsilon_{ij} \varepsilon_{kl} d^{d}\mathrm{r} \\
F_{\mathrm{w}} &= \int \frac{1}{2} K_{ijkl} w_{ij} w_{kl} d^{d}\mathrm{r} \qquad (20) \\
F_{\mathrm{uw}} &= \int \left( R_{ijkl} \varepsilon_{ij} w_{kl} + R_{klij} w_{ij} \varepsilon_{kl} \right) d^{d}\mathrm{r}
\end{aligned}
$$

here the superscript $^{d}$ means the dimension of the object we are calculating , $C_{ijkl}$ the phonon elastic constant tensor, $K_{ijkl}$ phason elastic constant tensor, and $R_{ijkl}, R_{klij}$ the phonon-phason coupling elastic constant tensor, and the strain tensors $\varepsilon_{ij}, w_{ij}$ are defined by (6), and constitutive law is given by (7). The definition of (19) is a heritage of the work of Lubensky et al [22], and we did some developments in the applications in the soft-matter quasicrystals.

For the second kind of soft-matter quasicrystals, the elastic energy in equation (19) consists of following six terms

$$F_{el} = F_u + F_v + F_w + F_{uv} + F_{uw} + F_{vw}$$

$$
\begin{cases}
H = H[\Psi(\mathrm{r},t)] = \int \frac{\mathrm{g}^2}{2\rho} d^{d}\mathrm{r} + \int \left[ \frac{1}{2} A \left( \frac{\delta\rho}{\rho_0} \right)^2 + B \left( \frac{\delta\rho}{\rho_0} \right) \nabla \bullet \mathrm{u} \right] d^{d}\mathrm{r} + F_{el} \\
= H_{kin} + H_{density} + F_{el} \qquad (19') \\
F_{el} = F_u + F_v + F_w + F_{uv} + F_{uw} + F_{vw}, \mathrm{g} = \rho \mathrm{V}
\end{cases}
$$

and

$$F_u = \int \frac{1}{2} C_{ijkl} \varepsilon_{ij} \varepsilon_{kl} d^d \mathrm{r}$$

$$F_v = \int \frac{1}{2} T_{ijkl} v_{ij} v_{kl} d^d \mathrm{r}$$

$$F_w = \int \frac{1}{2} K_{ijkl} w_{ij} w_{kl} d^d \mathrm{r}$$

$$F_{uv} = \int \left( r_{ijkl} \varepsilon_{ij} v_{kl} + r_{klij} v_{ij} \varepsilon_{kl} \right) d^d \mathrm{r} \qquad (21)$$

$$F_{uw} = \int \left( R_{ijkl} \varepsilon_{ij} w_{kl} + R_{klij} w_{ij} \varepsilon_{kl} \right) d^d \mathrm{r}$$

$$F_{vw} = \int \left( G_{ijkl} v_{ij} w_{kl} + G_{klij} w_{ij} v_{kl} \right) d^d \mathrm{r}$$

The strain tensors are defined by (11) and the constitutive law is given by (13).

The dynamic equations (14)-(18) constitute a basis of generalized dynamics of soft-matter quasicrystals, which can quantitatively describe the distribution, deformation and motion of soft-matter quasicrystals. At the same time we developed corresponding computationtoolswith various initial-boundary conditions, and computation shows the equation set are self consistent and solvable mathematically. In this sense we confirm the equations and solving system provide an analytic tool for quantitatively describing the dynamic motion of the matter.This also shows in constructing the equations, the Landau-Anderson symmetry breaking, elementary excitation principle, group theory and group representation play a central role, of course, which are a heritage and development of Lubensky hydrodynamics of solid quasicrystals. This shows, the experience, methodology,and in particular the contributions from pioneer researchers, e.g. Bak, Lubensky et al, accumulated in the development of solid quasicrystals study summarized in Section 1 provide a solid base for the further development on the study of soft-matter quasicrystals.

### 5.2 Dynamics equations of soft-matter quasicrystals with 12-fold symmetry

Due to the frequent observation of soft-matter quasicrystals with 12-fold symmetry, it becomes the most important system to be investigated. Thethree-dimensional constitutive laws on phonon, phason and fluid phonon, can be constructed as follows [54]

$$\left.\begin{aligned}
&\sigma_{xx}=C_{11}\varepsilon_{xx}+C_{12}\varepsilon_{yy}+C_{13}\varepsilon_{zz}\\
&\sigma_{yy}=C_{12}\varepsilon_{xx}+C_{11}\varepsilon_{yy}+C_{13}\varepsilon_{zz}\\
&\sigma_{zz}=C_{13}\varepsilon_{xx}+C_{13}\varepsilon_{yy}+C_{33}\varepsilon_{zz}\\
&\sigma_{yz}=\sigma_{zy}=2C_{44}\varepsilon_{yz}\\
&\sigma_{zx}=\sigma_{xz}=2C_{44}\varepsilon_{zx}\\
&\sigma_{xy}=\sigma_{yx}=2C_{66}\varepsilon_{xy}\\
&H_{xx}=K_1w_{xx}+K_2w_{yy}\\
&H_{yy}=K_2w_{xx}+K_1w_{yy}\\
&H_{yz}=K_4w_{yz}\\
&H_{xy}=\left(K_1+K_2+K_3\right)w_{xy}+K_3w_{yx}\\
&H_{xz}=K_4w_{xz}\\
&H_{yx}=K_3w_{xy}+\left(K_1+K_2+K_3\right)w_{yx}\\
&p_{xx}=-p+2\eta\dot{\xi}_{xx}-\frac{2}{3}\eta\dot{\xi}_{kk}\\
&p_{yy}=-p+2\eta\dot{\xi}_{yy}-\frac{2}{3}\eta\dot{\xi}_{kk}\\
&p_{zz}=-p+2\eta\dot{\xi}_{zz}-\frac{2}{3}\eta\dot{\xi}_{kk}\\
&p_{yz}=2\eta\dot{\xi}_{yz}\\
&p_{zx}=2\eta\dot{\xi}_{zx}\\
&p_{xy}=2\eta\dot{\xi}_{xy}\\
&\dot{\xi}_{kk}=\dot{\xi}_{xx}+\dot{\xi}_{yy}+\dot{\xi}_{zz}
\end{aligned}\right\}(22)$$

With the constitutive law the Hamiltonians defined by (19) can be written, and substitute them into (14)-(17) the equations of motion for soft-matter quasicrystals of 12-fold symmetry can be obtained, and combining the equation of state (18), so the governing equations of the dynamics of the material can be constructed, in which the fields variables are $u_x, u_y, u_z, w_x, w_y, V_x, V_y, V_z, \rho$ and $p$, the amount of the field variables is 10, and amount of field equations is 10 too, among them include the mass conservation equation, the momentum conservation equations or the generalized Navier-Stokes equations, the equations of motion of phonons due to the symmetry breaking, the phason dissipation equations, and the equation of state, respectively. The equations are consistent to be mathematical solvability, if there is lack of the equation of state, the equation system is not closed, and has no meaning mathematically and physically. This shows the equation of state is necessary.

Due to the lengthy of these equations, we do not list themhere, the detail can be referred to Appendix of this paper or monograph [55].

Similarly the governing equations of other systems of the first kind of soft-matter quasicrystals can also be derived. For detail, please refer to [55].

In the equations, the phonons and fluid phonon represent wave propagation with speeds $c_1 = \sqrt{\frac{A+L+2M-2B}{\rho}}, c_2 = c_3 = \sqrt{\frac{M}{\rho}}$ and $c_4 = \sqrt{\frac{\partial p}{\partial \rho}}$ andphasons represent diffusion with the diffusion coefficient $D = \frac{1}{\Gamma_w}$ , respectively, in which $L = C_{12}, \quad M = (C_{11} - C_{12})/2 = C_{66}$.

### 5.3 Dynamics equations of soft-matter quasicrystals with 18-fold symmetry

The 18-fold symmetry soft matter quasicrystals were observed from colloids, which is the typical structure of the second kind of soft-matter quasicrystals. Based on group representative theory, there is one coupling constant between two phason modes, labeled as G, and no coupling between phonon and either of phasons.The 2d constitutive law is

$$\left.\begin{aligned}
\sigma_{xx} &= \left(L+2M\right)\varepsilon_{xx} + L\varepsilon_{yy} \\
\sigma_{yy} &= L\varepsilon_{xx} + \left(L+2M\right)\varepsilon_{yy} \\
\sigma_{xy} &= \sigma_{yx} = 2M\varepsilon_{xy} \\
\tau_{xx} &= T_1 v_{xx} + T_2 v_{yy} + G\left(w_{xx} - w_{yy}\right) \\
\tau_{yy} &= T_2 v_{xx} + T_1 v_{yy} + G\left(w_{xx} - w_{yy}\right) \\
\tau_{xy} &= T_1 v_{xy} - T_2 v_{yx} - G\left(w_{yx} + w_{xy}\right) \\
\tau_{yx} &= -T_2 v_{xy} + T_1 v_{yx} + G\left(w_{yx} + w_{xy}\right) \\
H_{xx} &= K_1 w_{xx} + K_2 w_{yy} + G\left(v_{xx} + v_{yy}\right) \\
H_{yy} &= K_2 w_{xx} + K_1 w_{yy} - G\left(v_{xx} + v_{yy}\right) \\
H_{xy} &= K_1 w_{xy} - K_2 w_{yx} - G\left(v_{xy} + v_{yx}\right) \\
H_{yx} &= K_1 w_{yx} - K_2 w_{xy} + G\left(v_{xy} - v_{yx}\right) \\
p_{xx} &= -p + 2\eta(\dot{\xi}_{xx} - \frac{1}{3}\dot{\xi}_{kk}) \\
p_{yy} &= -p + 2\eta(\dot{\xi}_{yy} - \frac{1}{3}\dot{\xi}_{kk}) \\
p_{xy} &= p_{yx} = 2\eta\dot{\xi}_{xy} \\
\dot{\xi}_{kk} &= \dot{\xi}_{xx} + \dot{\xi}_{yy}
\end{aligned}\right\} \quad (23)$$

From this constitutive law we can calculate the corresponding Hamiltonians according to (19'), then substitute them into (14)-(18) (some of them should be extended), the final governing equations of the generalized dynamics of soft-matter quasicrystals can be presented for 18-fold symmetry soft-matter quasicrystals. Due to the lengthy of the equations, they are not listed here, please refer to Ref [55] for the detail.

In these equations there are 10 equations with 10 field variables. Ten field variables are $u_x, u_y, \mathrm{v}_x, \mathrm{v}_y\ w_x, w_x, V_x, V_y, \rho$ and $p$ . Ten equations consist of: mass conservation equation, momentum conservation equations or generalized Navier-Stokes equations, equations of motion of phonons due to symmetry breaking , first phason dissipation equations, second phason dissipation equations and the equation of state .

In the equations, the phonons and fluid phonon represent wave propagation with speeds $c_1 = \sqrt{\dfrac{A+L+2M-2B}{\rho}}, c_2 = c_3 = \sqrt{\dfrac{M}{\rho}}$ and $c_4 = \sqrt{\dfrac{\partial p}{\partial \rho}}$ andphasons represent diffusion with the diffusion coefficient $D_v = \dfrac{1}{\Gamma_v}, D_w = \dfrac{1}{\Gamma_w}$, respectively.

## 6 Application examples

The mathematical structure of the above mentioned governing equations looks quite complicated. To solve these equations, one needsthe initial- and boundary-value conditions. So far, there are very little experimental data to compare. We carried out some investigation on fundamental samples of soft-matter quasicrystals, which helps us to exam the governing equations, and to exam formulation of the solving system too. Our calculations show the consistence with the governing equations and formulation of the solving system in the rest of this session.

### 6.1 Transient response

A soft-matter quasicrystal with 10-fold symmetry under tension impact is simulated with the above governing equations. The specimen geometry is shown in Fig.4.

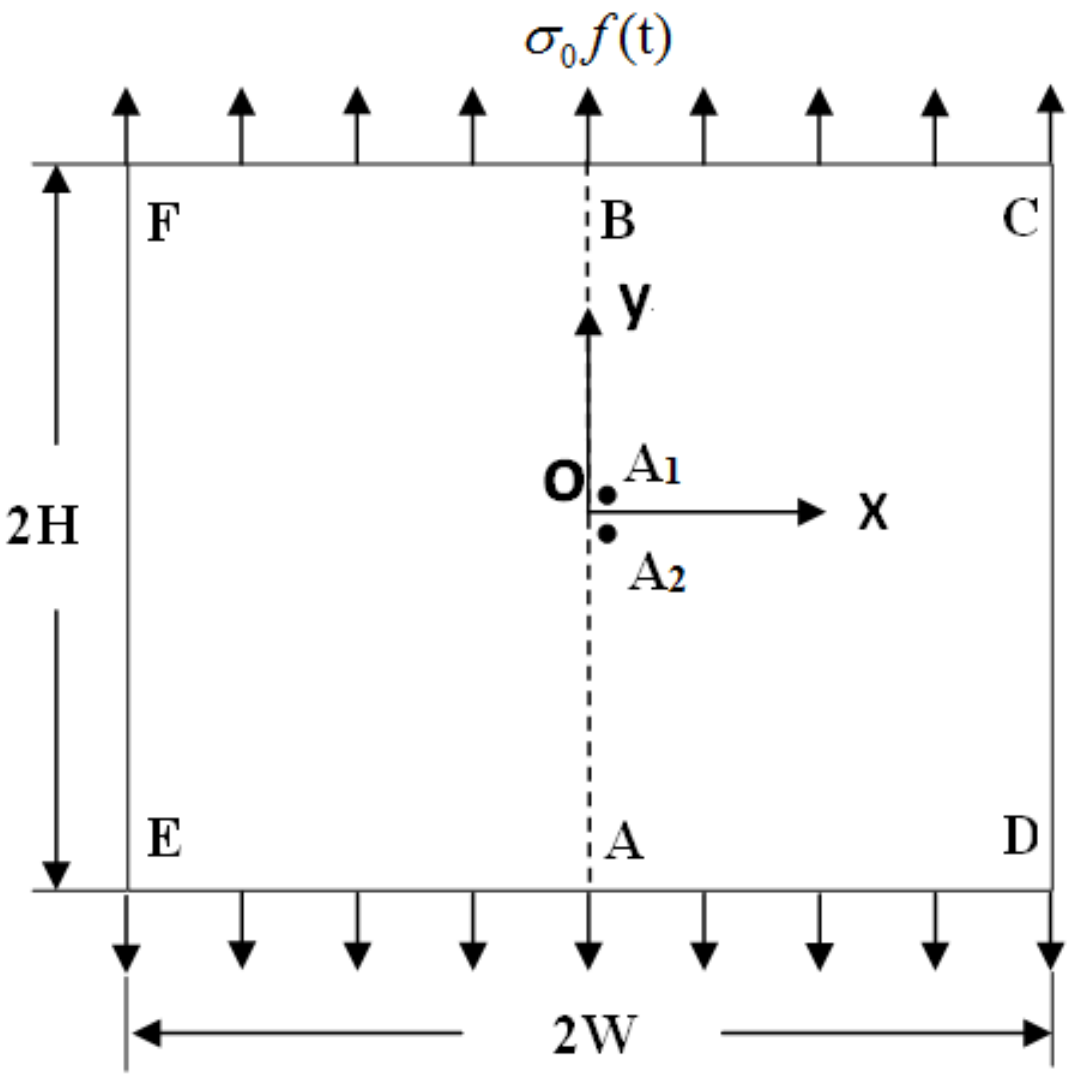

Fig.4Specimen geometry of a 10- fold symmetry soft-matter quasicrystalsunder dynamic loading.

Here we consider the specimen is a planar case, so the equations (14-17) (the detail can refer to Appendix of this paper )can be simplifiedwithout consideration of the z-components. The non-zero elastic constants are reduced to

$$C_{ijkl} = L\delta_{ij}\delta_{kl} + M(\delta_{ik}\delta_{jl}),$$
$$L = C_{12}, M = (C_{11} - C_{12})/2 = \mathrm{C}_{66},$$
$$K_{1111} = K_{2222} = K_{2121} = K_{1212} = K_1,$$
$$K_{1122} = K_{2211} = -K_{2112} = -K_{1221} = K_2,$$
$$\mathrm{R}_{1111} = R_{1122} = -R_{2211} = -R_{2222} = R_{1221} = R_{2121} = -R_{1212} = -R_{2112} = R$$

After simplification of thefield equations (the simplified equations have not been listed here for saving the space), we can solve the equations in $x-y$ plane under the initial and boundary conditions:

$$t = 0: V_x = V_y = 0, u_x = u_y = 0, w_x = w_y = 0, \ \mathrm{p=p_0};$$

$y = \pm H, |x| < W : V_x = V_y = 0, \sigma_{yy} = \sigma_0 f(t), \sigma_{yx} = 0, H_{yy} = H_{yx} = 0, \mathrm{p=p_0};$

$x = \pm W, |y| < H : V_x = V_y = 0, \sigma_{xx} = \sigma_{xy} = 0, H_{xx} = H_{xy} = 0, \mathrm{p=p_0}.$

We use finite difference method to construct numerical solution with the initial-boundary problem.

In the present computation we take $2H = 0.01m$ , $2W = 0.01m$ , $\sigma_0 = 0.01MPa$, $f(t) = H(t)$ the Heaviside function of time, $\rho_0 = 1.5\times10^3 kg/m^3$ , $\eta = 0.1Pa\cdot s$ , $L = 10MPa$ , $M = 4MPa$ , $K_1 = 0.5L$ , $K_2 = -0.1L$ , $R = 0.04M$, $\Gamma_u = 4.8\times10^{-17} m^3\cdot s/kg$ , $\Gamma_w = 4.8\times10^{-19} m^3\cdot s/kg$ , $\mathrm{A} \sim 0.2MPa$ , $\mathrm{B} \sim 0.2MPa$ , in which the fluid initial pressure is 1 atm.

The computation canshow the time-space evolution of any field variables in the specimen. Now we consider the field variables at point $A_1(10^{-4}m, 10^{-4}m)$ (or $A_2$ $(10^{-4}m, -10^{-4}m)$).

The numerical solution presents highly stability, this indicates the correctness of the equations and the formulation of corresponding initial-boundary value problem. The correctness of the solution can also be checked by some verification physically presented in Figs. 5-13.

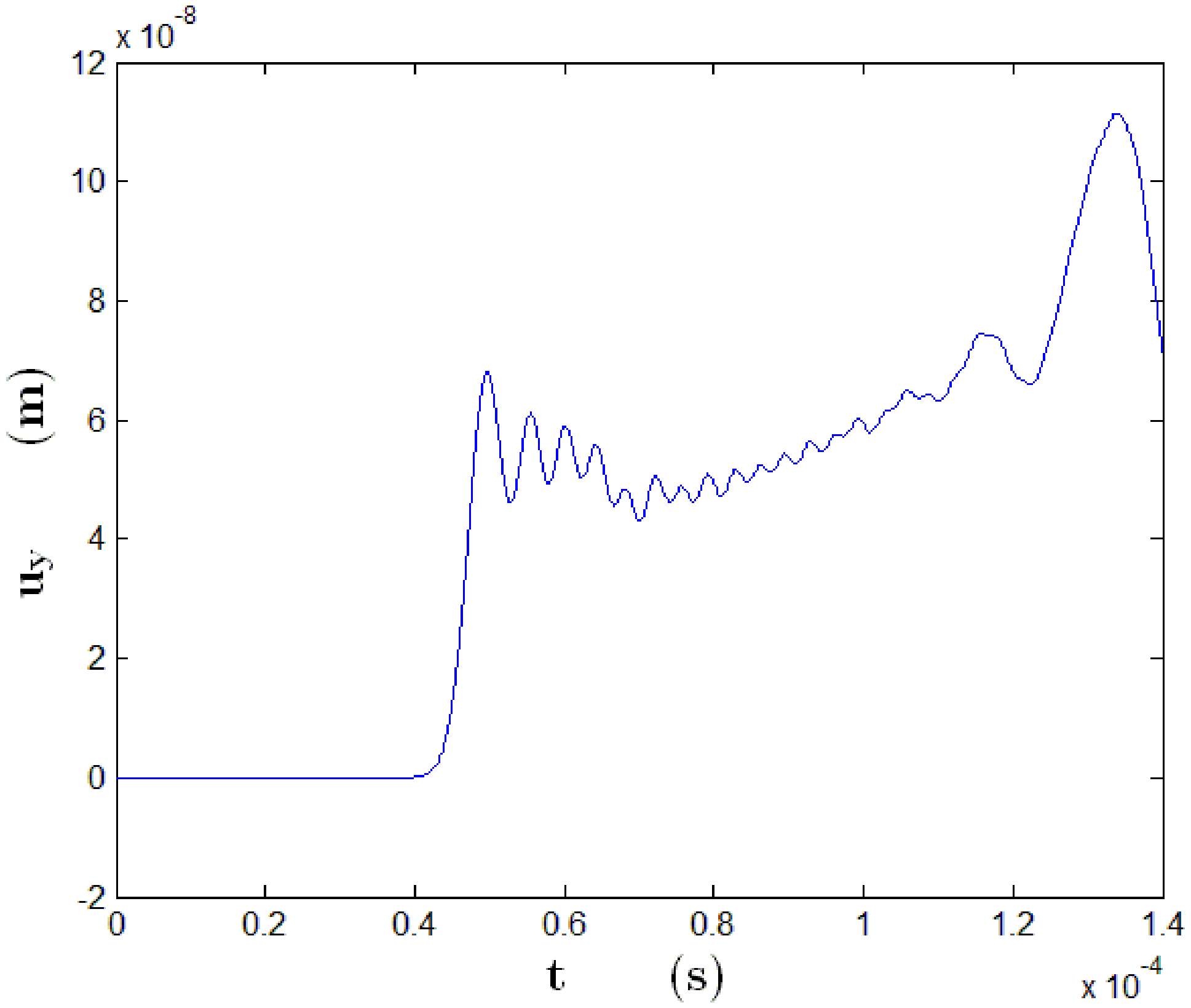

Fig.5 Response of normal displacement of phonon at point $A_1$

$(10^{-4}m, 10^{-4}m)$ (or A$_2$ $(10^{-4}m, -10^{-4}m)$) to the fluctuation due to applied stress

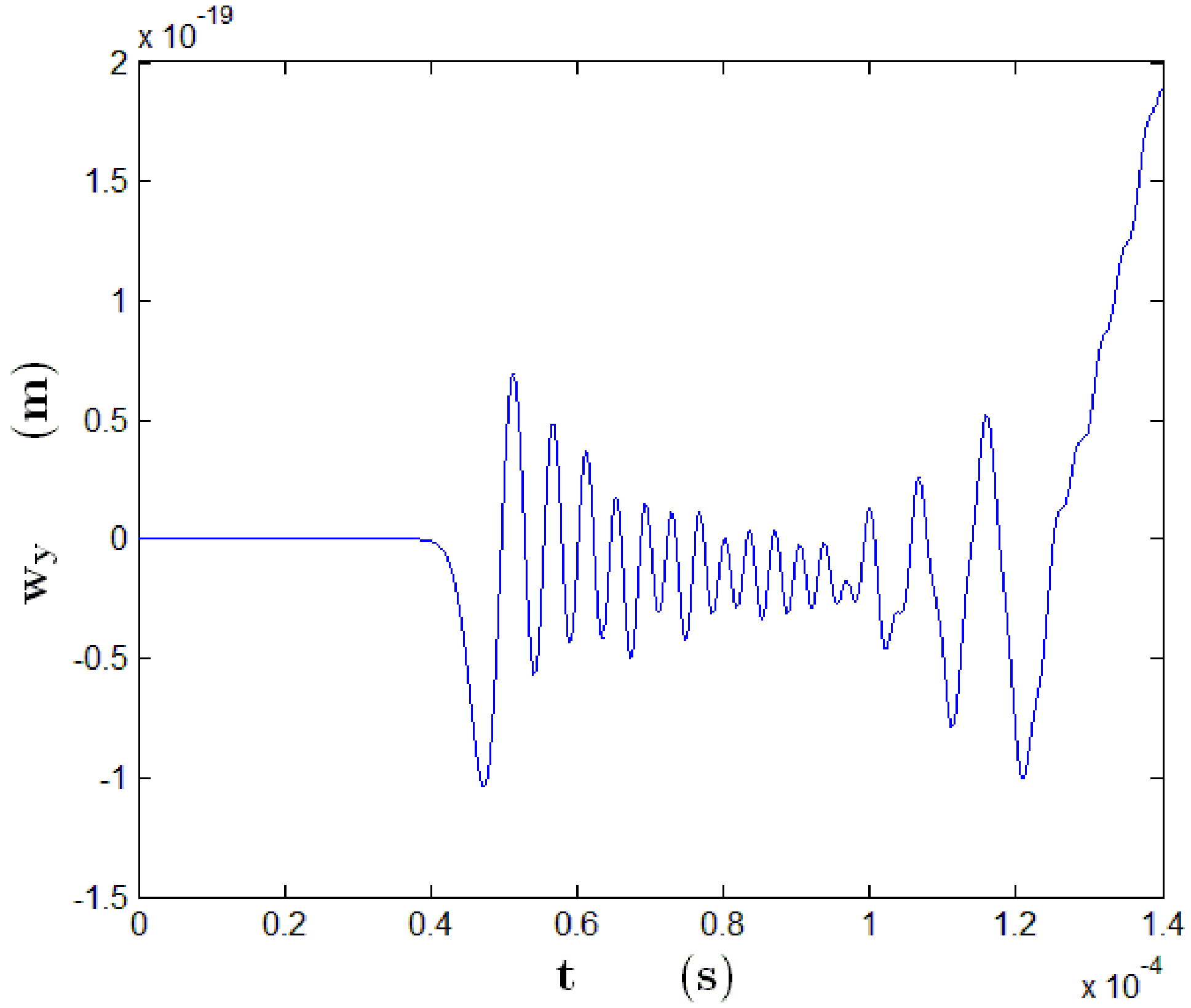

Fig.6 Response of normal phason displacement at A$_1$ $(10^{-4}m, 10^{-4}m)$ (or A$_2$ $(10^{-4}m, -10^{-4}m)$) to the applied impact loading

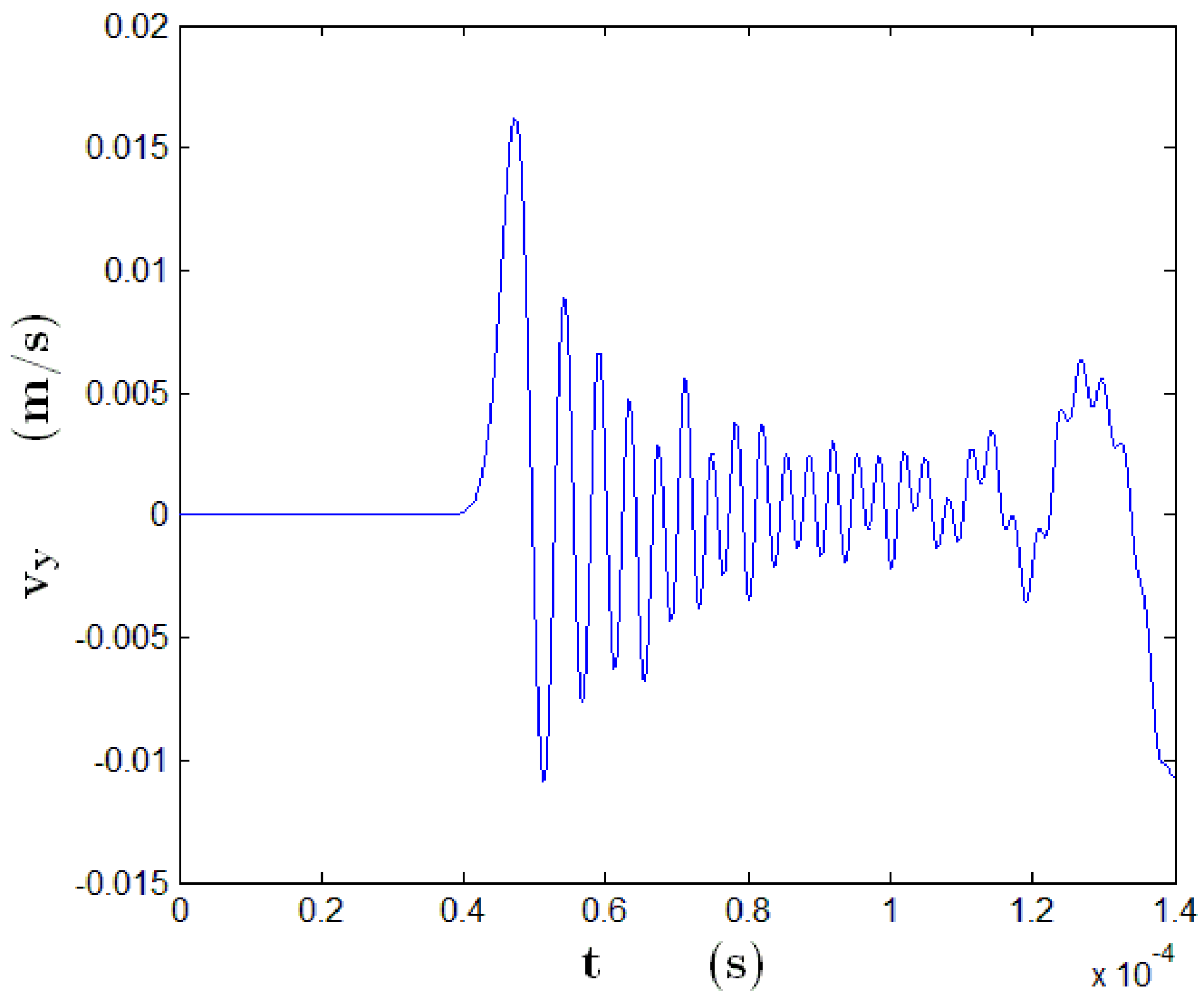

Fig. 7Time dependent fluid phonon speed $V_y$ at specimen point $A_1$.

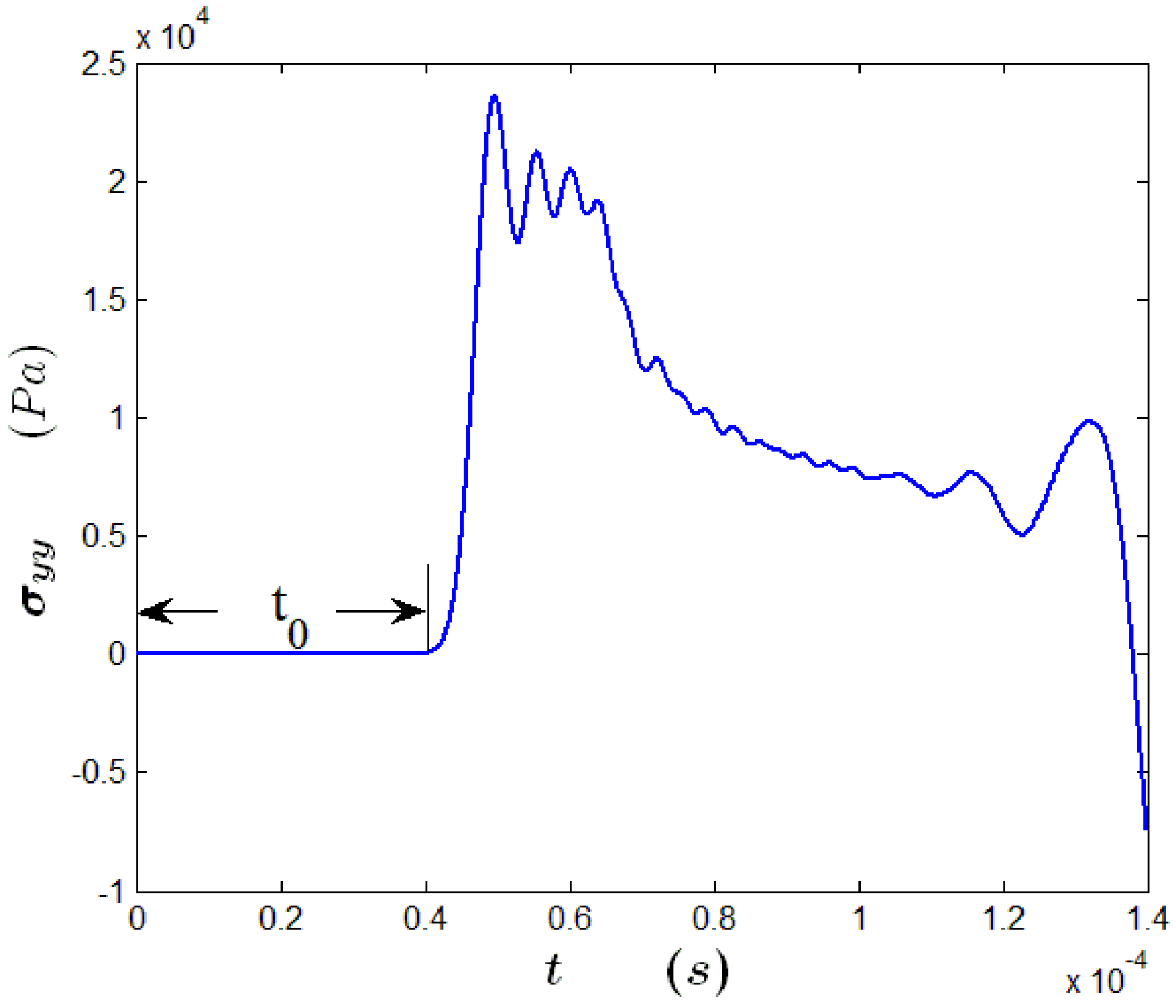

Fig.8Time dependent normal stress of phonon field at specimen point $A_1$ (or $A_2$).

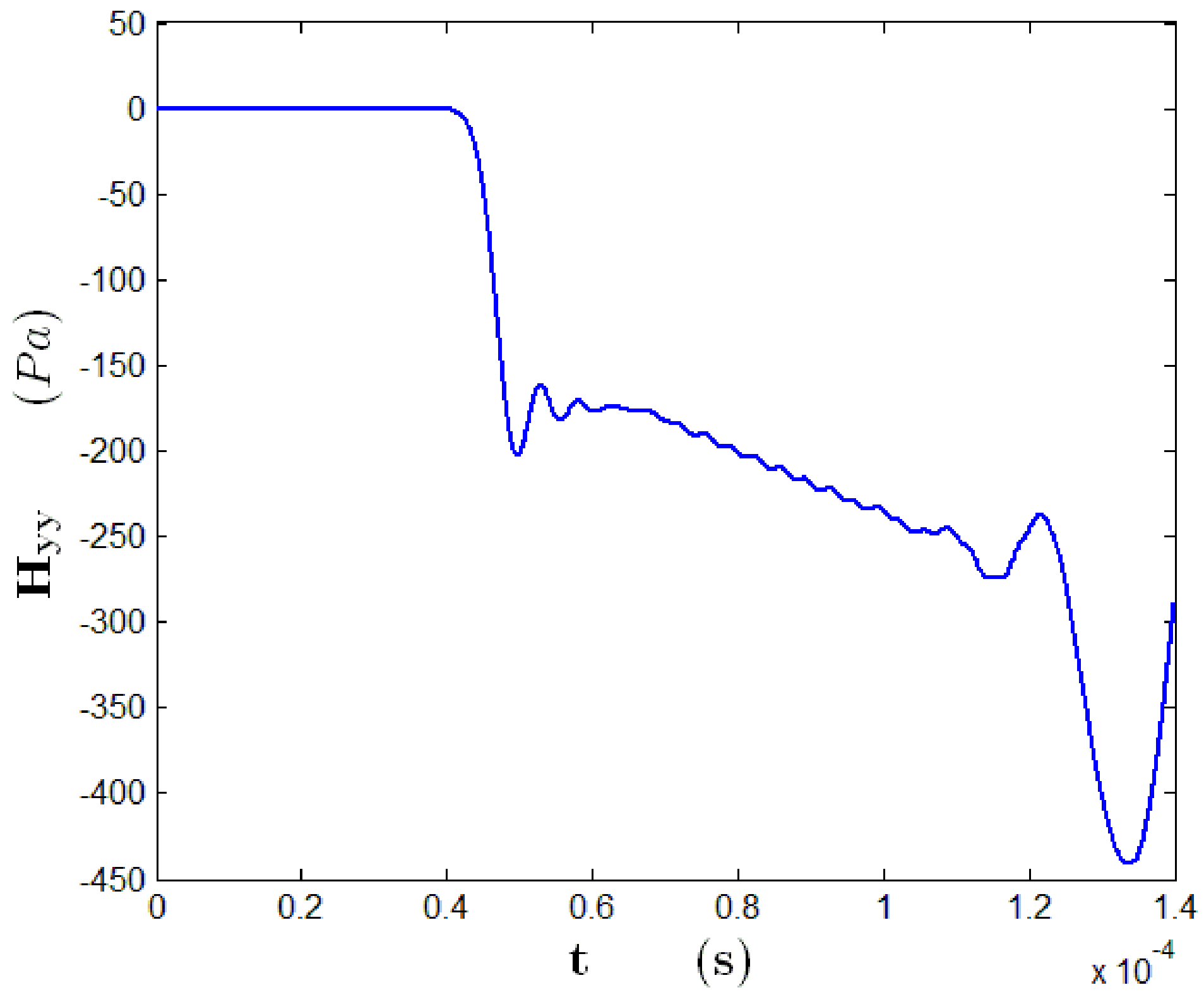

Fig.9Time dependent normal phason stress $H_{yy}$at specimen point $A_1$ (or $A_2$).

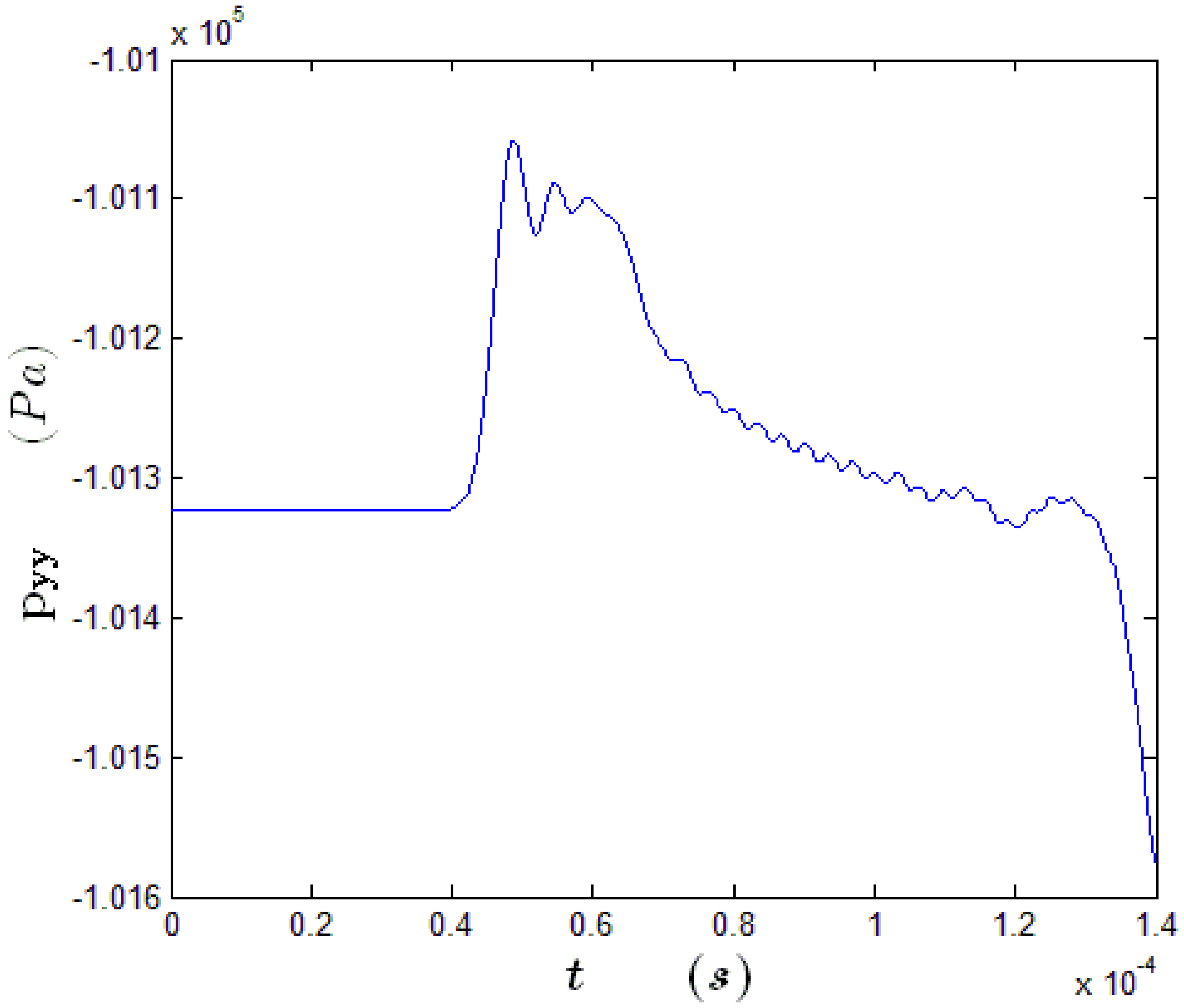

Fig.10Time dependent fluid normal stress$P_{yy}$at specimen point $A_1$ (or $A_2$).

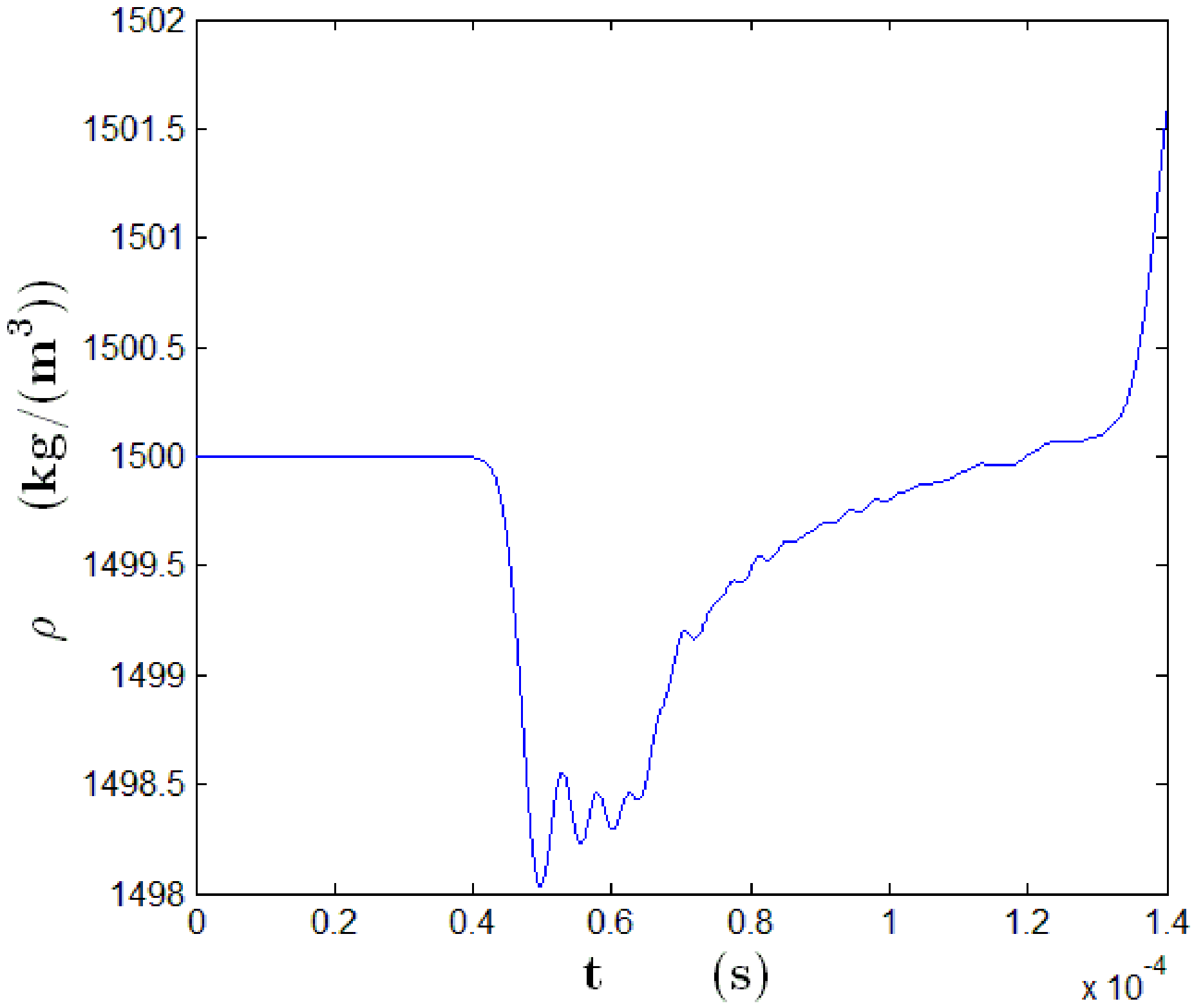

Fig.11（a）Time dependent mass density at specimen point $A_1$ (or $A_2$).

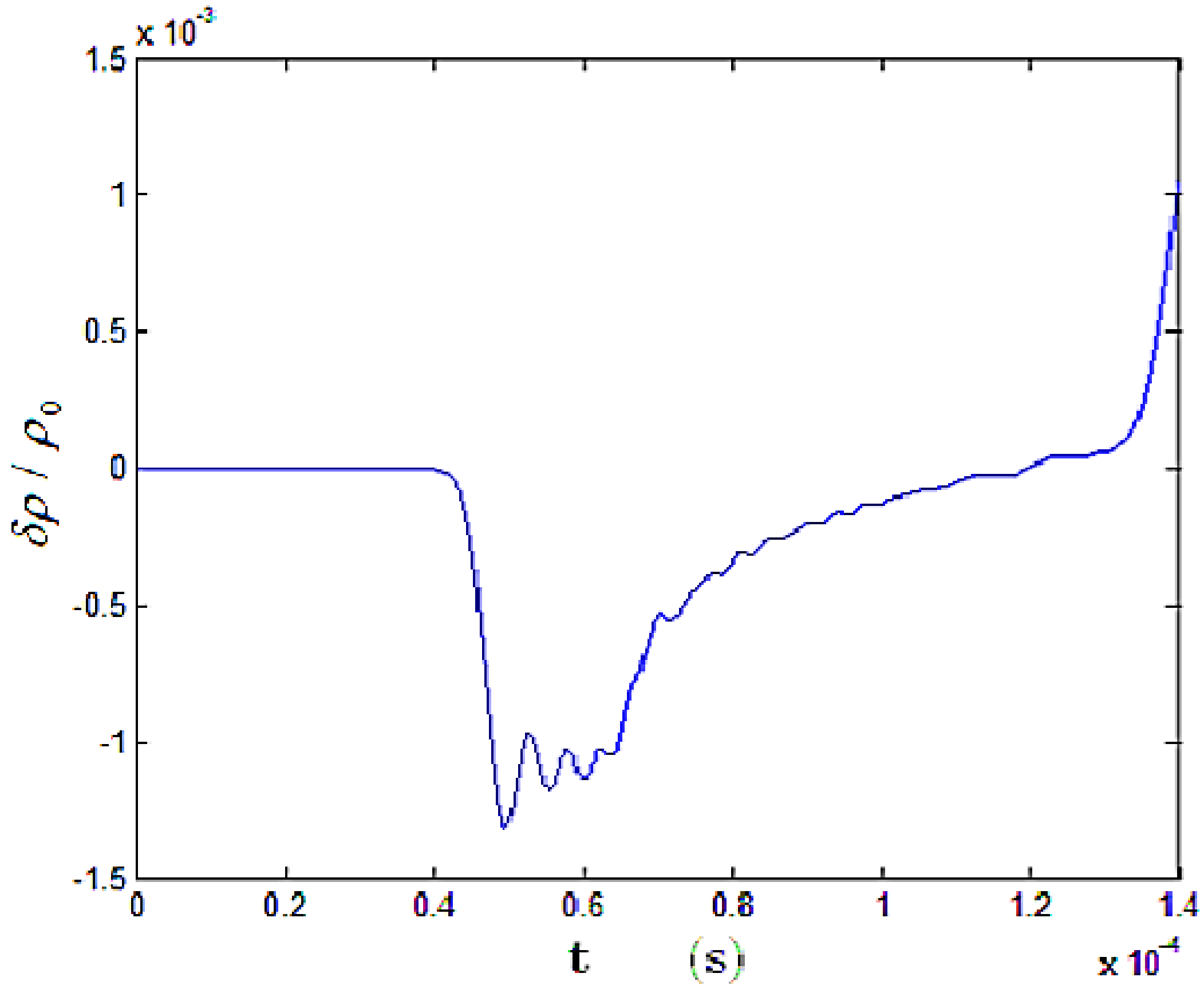

（b）Time depend relative mass density variation at specimen point $A_1$ (or $A_2$).

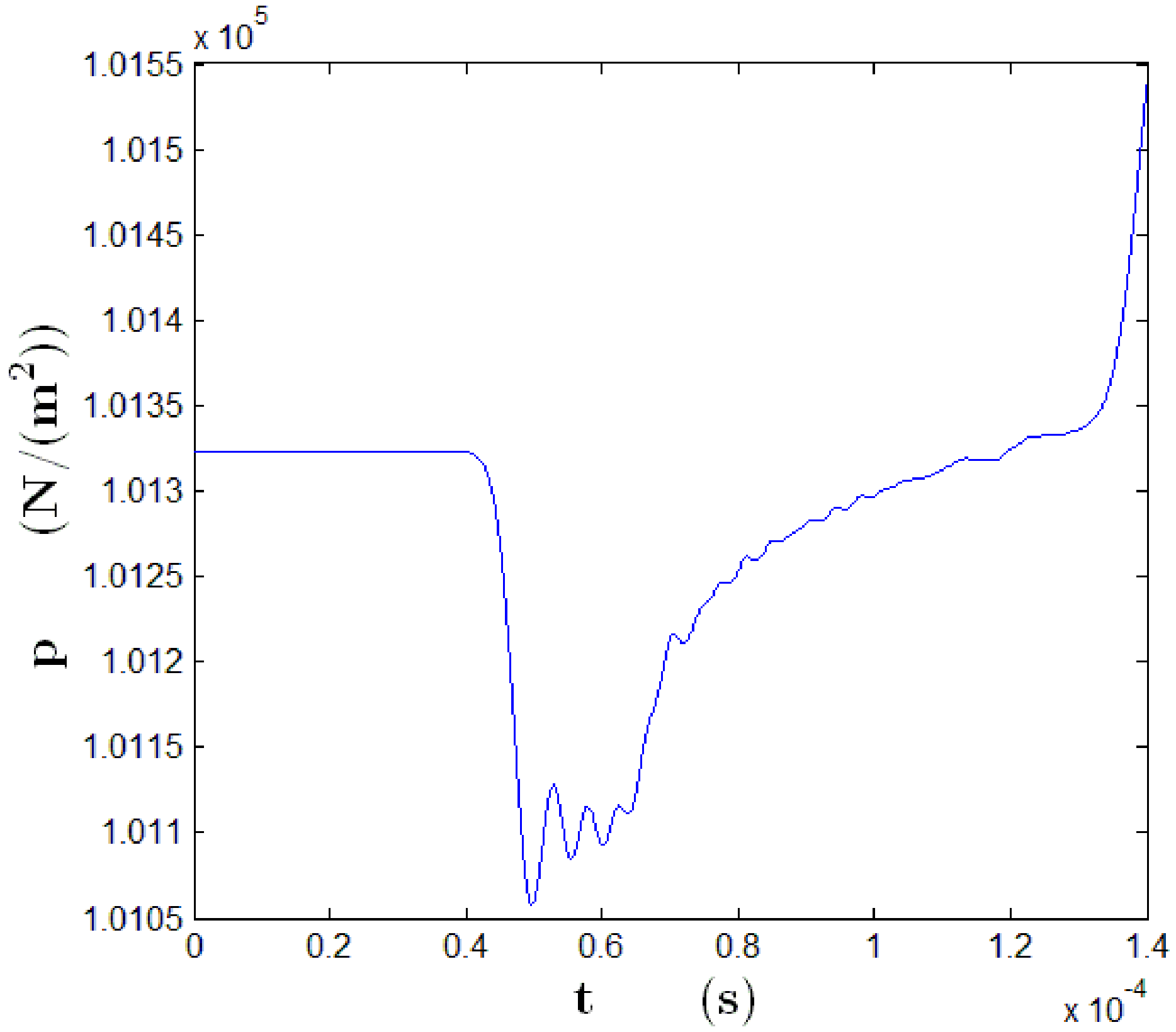

Fig.12Time dependent fluid pressure at specimen point $A_1$ (or $A_2$).

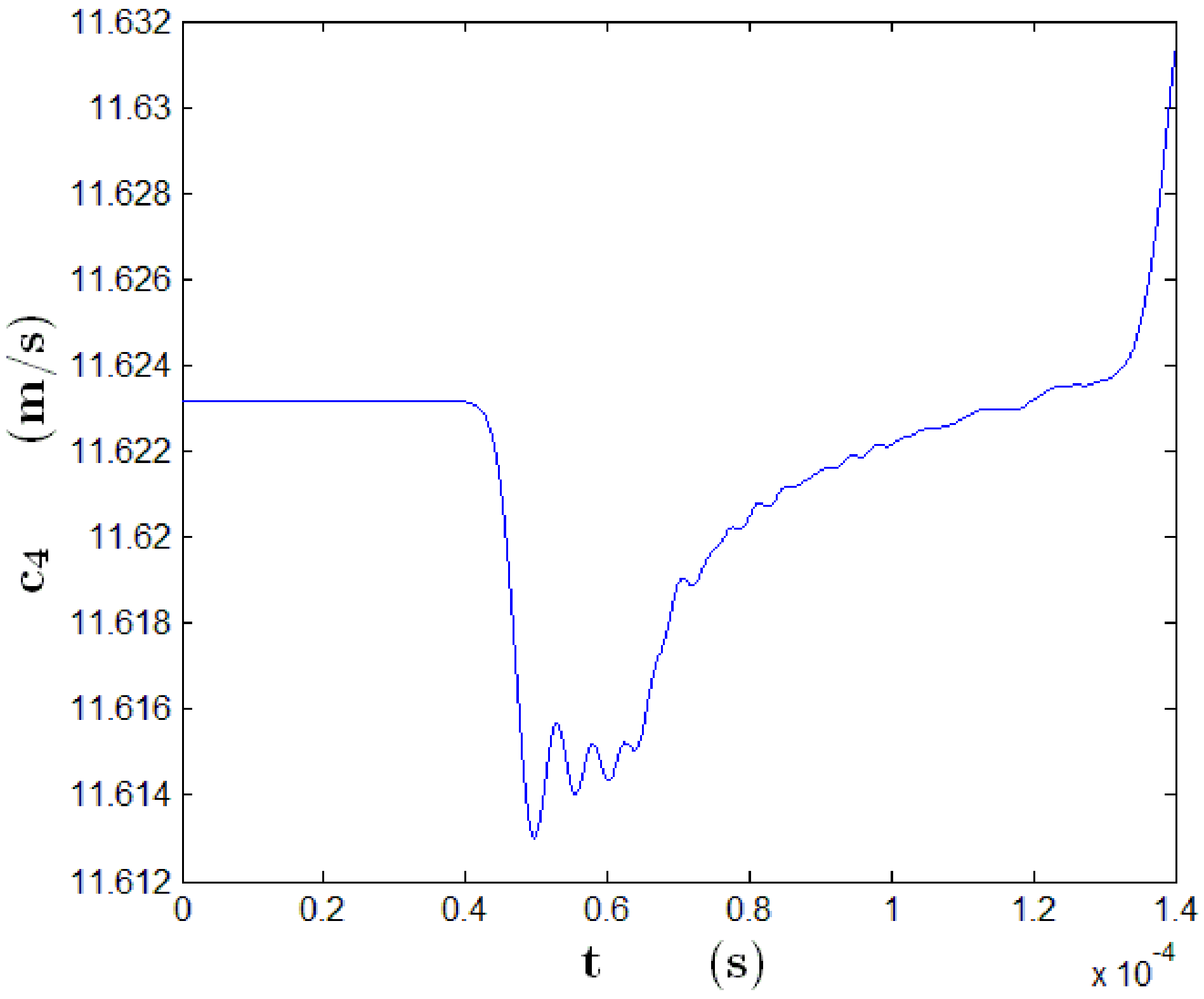

Fig.13Time dependent fluid acoustic wave velocity$C_4$at specimen point $A_1$ (or $A_2$).

In the Fig. 5 to Fig.13, the wave propagates from the upper or lower surface of the specimen.It takes $t_0 = 4.07\times10^{-5}s$ to reach $A_1$ $(10^{-4}m, 10^{-4}m)$ (or $A_2$ $(10^{-4}m, -10^{-4}m)$). We call the time $t_0$ as "response time" of the matter at the location to the dynamic loading. The wave propagating distance is $H_0 = H - 10^{-4} = 0.0049m$ , thus the wave propagation speed is $c = \frac{H_0}{t_0} = \frac{0.0049}{4.07\times10^{-5}} = 120.39m/s$ .The density of soft-matter quasicrystalsis $\rho_{\min} = 1498kg/m^3$, and the speed of elastic longitudinal wave can be estimated as $c_{1\max} = \sqrt{\frac{A+L+2M-2B}{\rho_{\min}}} = 109.6176m/s$ , which is very close to c within $(c - c_{1\max})/c = 6\times10^{-3}$. This indicates theoretical prediction is quite close. Before the wave arriving at the point, i.e., as $t < t_0$ the all field variables are equal to zero or their initial values (for displacements, velocities, stresses, mass density and fluid pressure). The responses of the field variables appear only as $t > t_0$. This simple fact implies a "response law", and shows the importance of phonon (elasticity) for the soft matter.

The system has the longitudinal wave speed $c_1 = \sqrt{\frac{A+L+2M-2B}{\rho}} = 109.5445m/s$ , two transverse speeds $c_2 = c_3 = \sqrt{\frac{M}{\rho}} = 51.6398m/s$ and the fluid longitudinal wave peed $(c_4)_0 = \sqrt{\left.\frac{\partial p}{\partial \rho}\right|_{\rho=\rho_0}} = 11.6232\text{m/s}$. Comparing the results, it is obvious that $c_1$ plays the dominated role although soft matter is an intermediate phase between solid and liquid, this also shows the importance of phonons.

In addition the tensile specimen discussed here, the shear loading specimen is also computed, the corresponding transient response is similar.This demonstration shows the dynamic response of soft matter presents same physical characters, in which the phonon and fluid phonon play important role, and the phason effect is relative weaker.

### 6.2Flow past obstacle, the Oseen flow in soft-matter quasicrystals

In the classical Stokes equations for viscous incompressible object, it fails to dealing with the problem of flow passing circular cylinder or other two-dimensional obstacles. This is the well-known Stokes paradox. Oseen [66] modified the Navier-Stokes equations and obtained the approximate analytic solution, the famous Oseen solution (refer to [67,68] in detail). The soft-matter quasicrystals are viscous and compressible, it would be interesting to apply the Oseen approximation to soft-matter quasicrystals. For this, Cheng and Fan constructed a generalized Oseen solution, i.e., the solution for a viscous and compressible conventional fluid flow passing a cylinder [55]. The generalized Oseen solution is compared with the classical Oseen solution. In generalthey are very similar, and with some noticeable differences due to the compressibility effect. Our results are also compared with those of the generalized Oseen solution.

Let's consider the 12-fold symmetry soft-matter quasicrystals, in two-dimensional case (i.e. the plane field case), where flows passes a circular cylinder as shown in Fig.14.

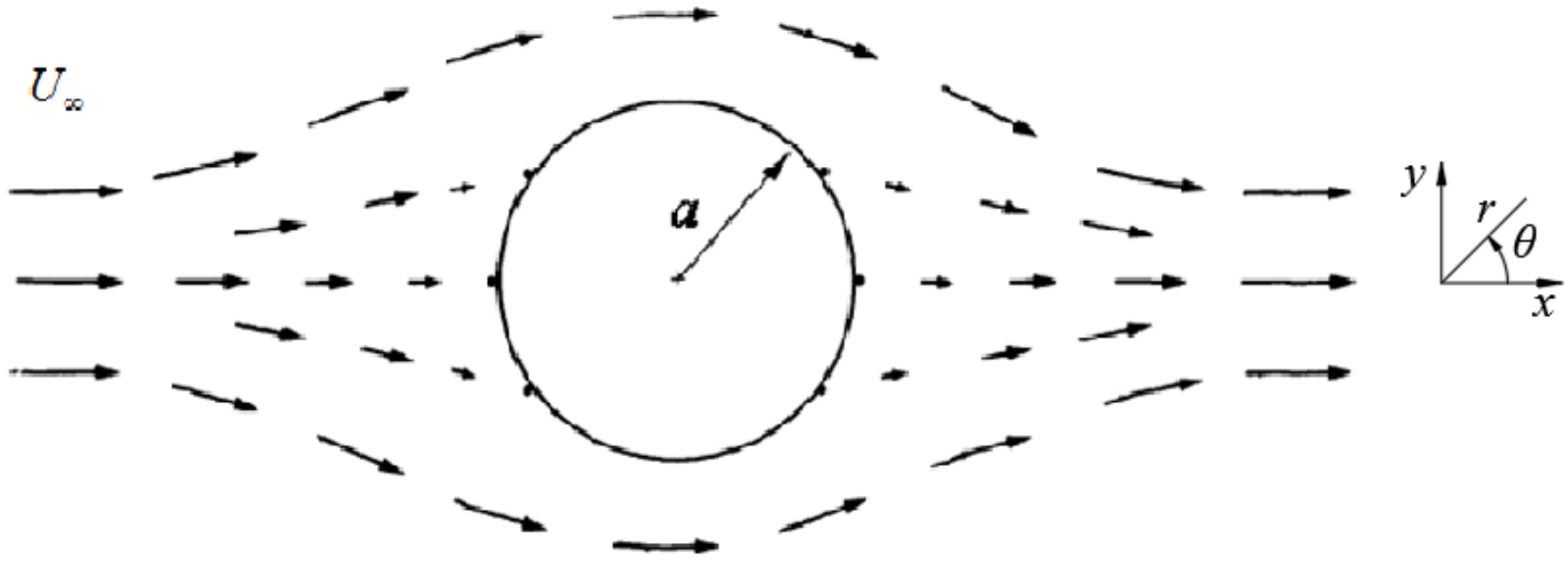

Fig. 14 Flow passes a circular cylinder with given velocity $U_\infty$ at infinity for 12-fold symmetry soft-matter quasicrystal.

First we apply Oseen modification to the dynamic equations of 12-fold symmetry soft-matter quasicrystals under the steady state

$$
\left.\begin{aligned}
&\frac{\partial \rho}{\partial t}+\nabla\bullet(\rho\mathbf{V})=0\\
&\frac{\partial(\rho V_x)}{\partial t}+\frac{\partial(U_x\rho V_x)}{\partial x}+\frac{\partial(U_y\rho V_x)}{\partial y}=-\frac{\partial p}{\partial x}+\eta\nabla^2V_x+\frac{1}{3}\eta\frac{\partial}{\partial x}\nabla\bullet\mathbf{V}+M\nabla^2u_x+(L+M-B)\frac{\partial}{\partial x}\nabla\bullet\mathbf{u}\\
&-(\mathrm{A}-\mathrm{B})\frac{1}{\rho_0}\frac{\partial\delta\rho}{\partial x}\\
&\frac{\partial(\rho V_y)}{\partial t}+\frac{\partial(U_x\rho V_y)}{\partial x}+\frac{\partial(U_y\rho V_y)}{\partial y}=-\frac{\partial p}{\partial y}+\eta\nabla^2V_y+\frac{1}{3}\eta\frac{\partial}{\partial y}\nabla\bullet\mathbf{V}+M\nabla^2u_y+(L+M-B)\frac{\partial}{\partial y}\nabla\bullet\mathbf{u}\\
&-(\mathrm{A}-\mathrm{B})\frac{1}{\rho_0}\frac{\partial\delta\rho}{\partial y}\\
&\frac{\partial u_x}{\partial t}+U_x\frac{\partial u_x}{\partial x}+U_y\frac{\partial u_x}{\partial y}=V_x+\Gamma_{\mathbf{u}}\left[M\nabla^2u_x+(L+M)\frac{\partial}{\partial x}\nabla\bullet\mathbf{u}\right]\\
&\frac{\partial u_y}{\partial t}+U_x\frac{\partial u_y}{\partial x}+U_y\frac{\partial u_y}{\partial y}=V_y+\Gamma_{\mathbf{u}}\left[M\nabla^2u_y+(L+M)\frac{\partial}{\partial y}\nabla\bullet\mathbf{u}\right]\\
&\frac{\partial w_x}{\partial t}+U_x\frac{\partial w_x}{\partial x}+U_y\frac{\partial w_x}{\partial y}=\Gamma_{\mathbf{w}}\left[K_1\nabla^2w_x+(K_2+K_3)\frac{\partial}{\partial y}\left(\frac{\partial w_x}{\partial y}+\frac{\partial w_y}{\partial x}\right)\right]\\
&\frac{\partial w_y}{\partial t}+U_x\frac{\partial w_y}{\partial x}+U_y\frac{\partial w_y}{\partial y}=\Gamma_{\mathbf{w}}\left[K_1\nabla^2w_y(K_2+K_3)\frac{\partial}{\partial x}\left(\frac{\partial w_x}{\partial y}+\frac{\partial w_y}{\partial x}\right)\right]\\
&p=f(\rho)=3\frac{k_BT}{l^3\rho_0^3}\left(\rho_0^2\rho+\rho_0\rho^2+\rho^3\right)
\end{aligned}\right\}\quad(24)
$$

in which $U_x$ and $U_y$ are given in the boundary conditions. The equation (24) is the Oseen modification of equation under two dimensional and steady state, the details refer to [55].

According to the modified dynamic equations of 12-fold symmetry soft-matter quasicrystals, we take a finite difference method, and the difference network is shown in Fig.15:

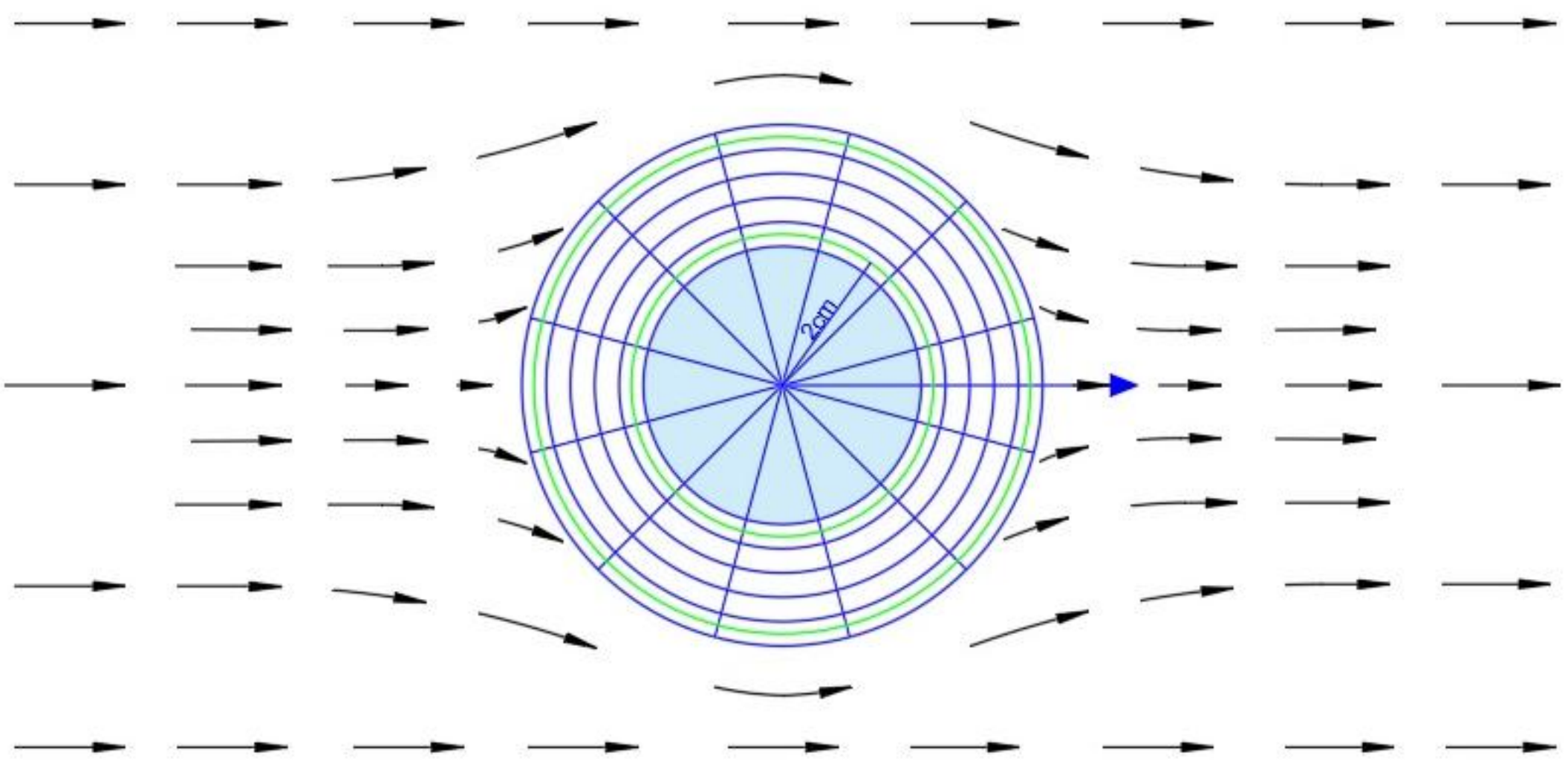

Fig.15 Finite difference network in polar coordinate

In the computation, we take the boundary conditions

$$\begin{cases} r=\sqrt{x^2+y^2+z^2}\to\infty: \\ V_r=U_\infty\cos\theta, V_\theta=-U_\infty\sin\theta, \sigma_{rr}=\sigma_{\theta\theta}=0, \mathrm{H}_{rr}=H_{\theta\theta}=0, \\ r=a: \\ V_r=V_\theta=0, \sigma_{rr}=\sigma_{r\theta}=0, \mathrm{H}_{rr}=H_{r\theta}=0 \end{cases} \quad (25)$$

and with the geometry and physical data

$U_x=U_\infty=0.01m/s, U_y=0$, $\rho_0=1.5g/cm^3$, $\eta=1\mathrm{Poise}$, $l=8nm$, $r/a=1.55$, $a=1\mathrm{cm}$, $k_B=1.38\times10^{-23}J/K$, $T=293\mathrm{K}$, $L=10MPa$, $M=4MPa$, $K_1=0.5L$, $K_2=-0.1L$, $K_3=0.05L$, $\Gamma_u=4.8\times10^{-17}m^3\cdot s/kg$, $\Gamma_w=4.8\times10^{-19}m^3\cdot s/kg$, $\mathrm{A}\sim0.2MPa$, $\mathrm{B}\sim0.2MPa$

The part computational results are shown in the Figs. 16-20.

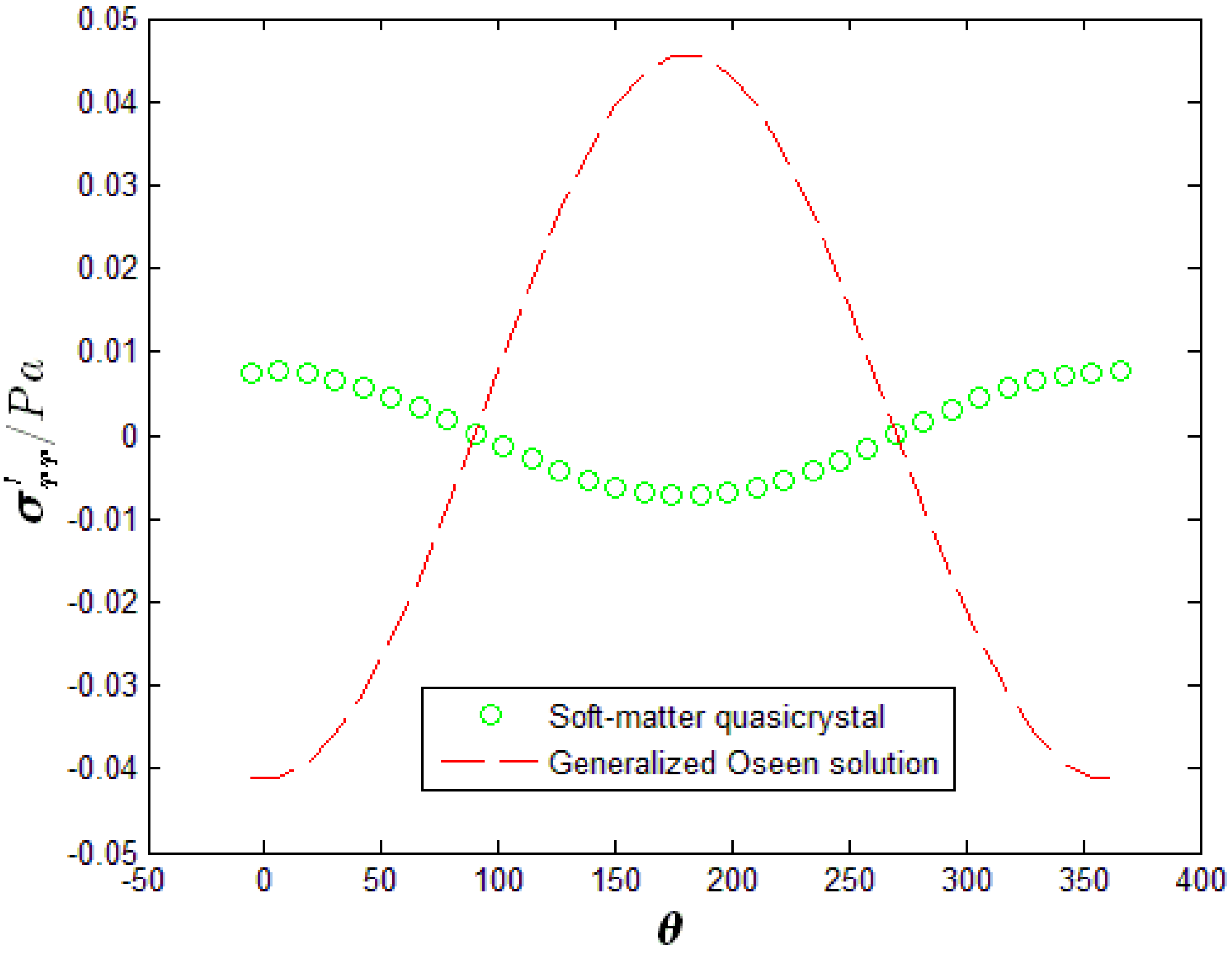

Fig.16 Angular distribution of radial normal stress due to fluid viscosity and comparison to that of solution of conventional liquid given by Ref [55]

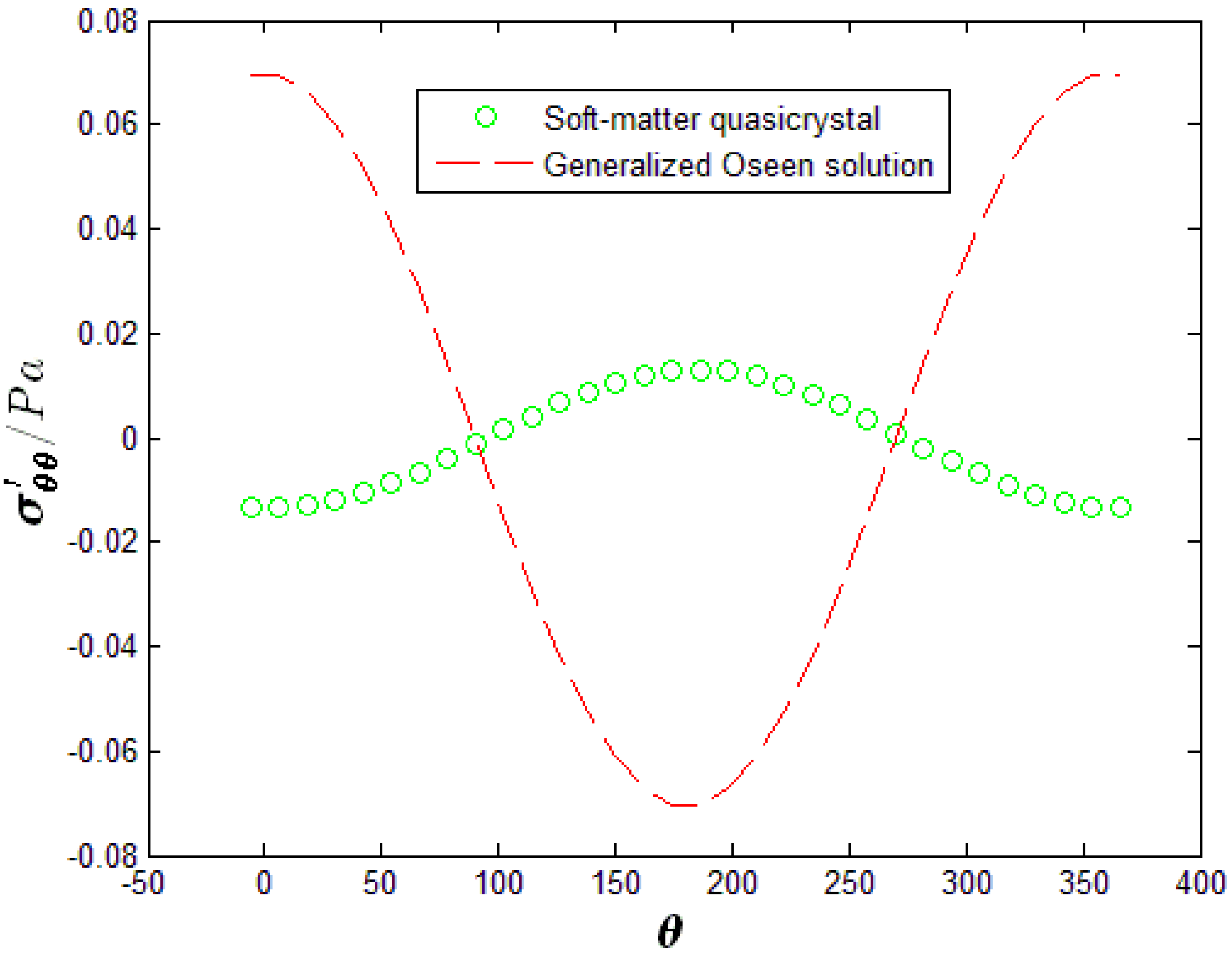

Fig. 17 Angular distribution of circumferential normal stress due to fluid viscosity and comparison to that of solution of conventional liquid given by Ref [55]

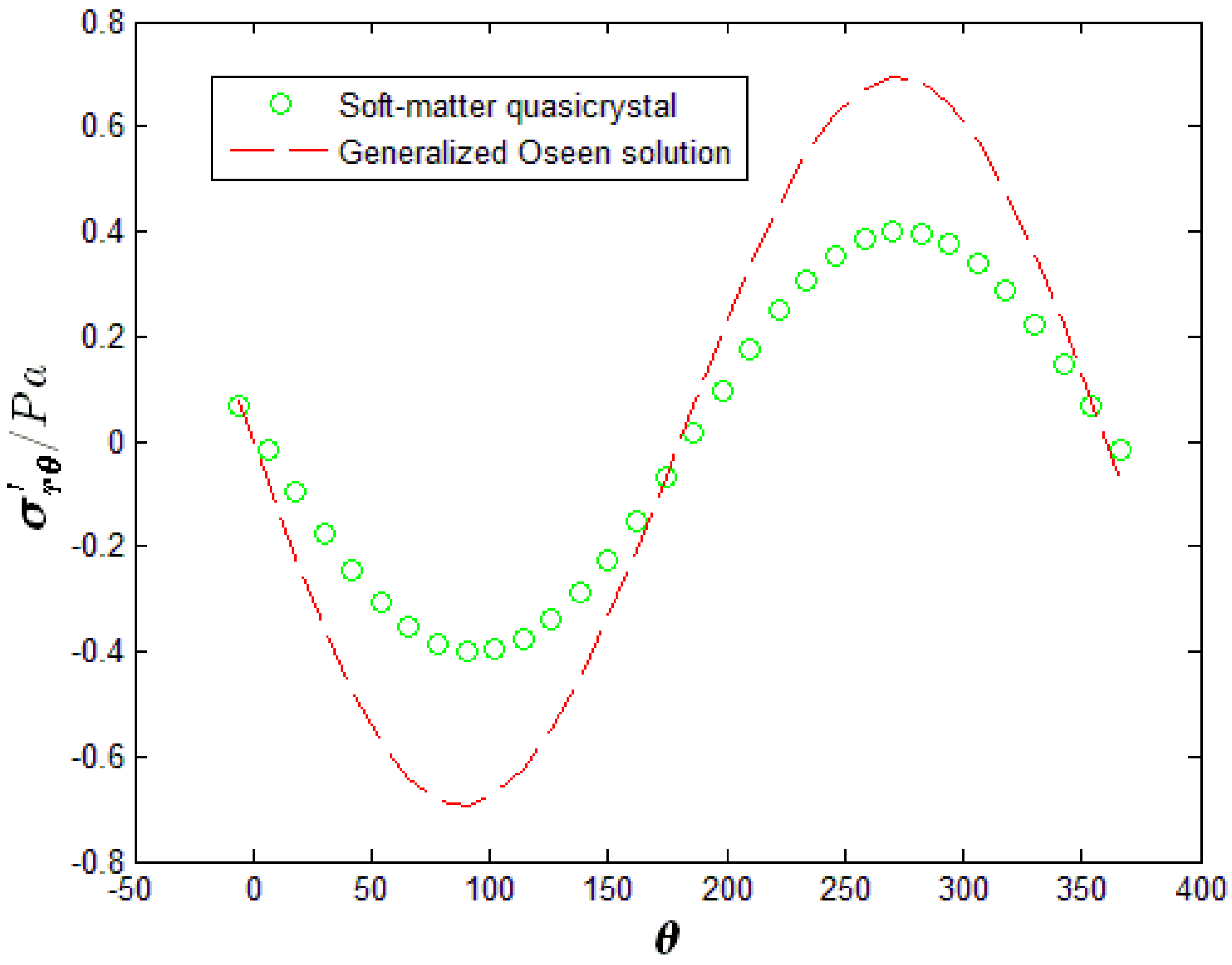

Fig.18 Angular distribution of shear stress due to fluid viscosity and comparison to that of solution of conventional liquid given by Ref [55]

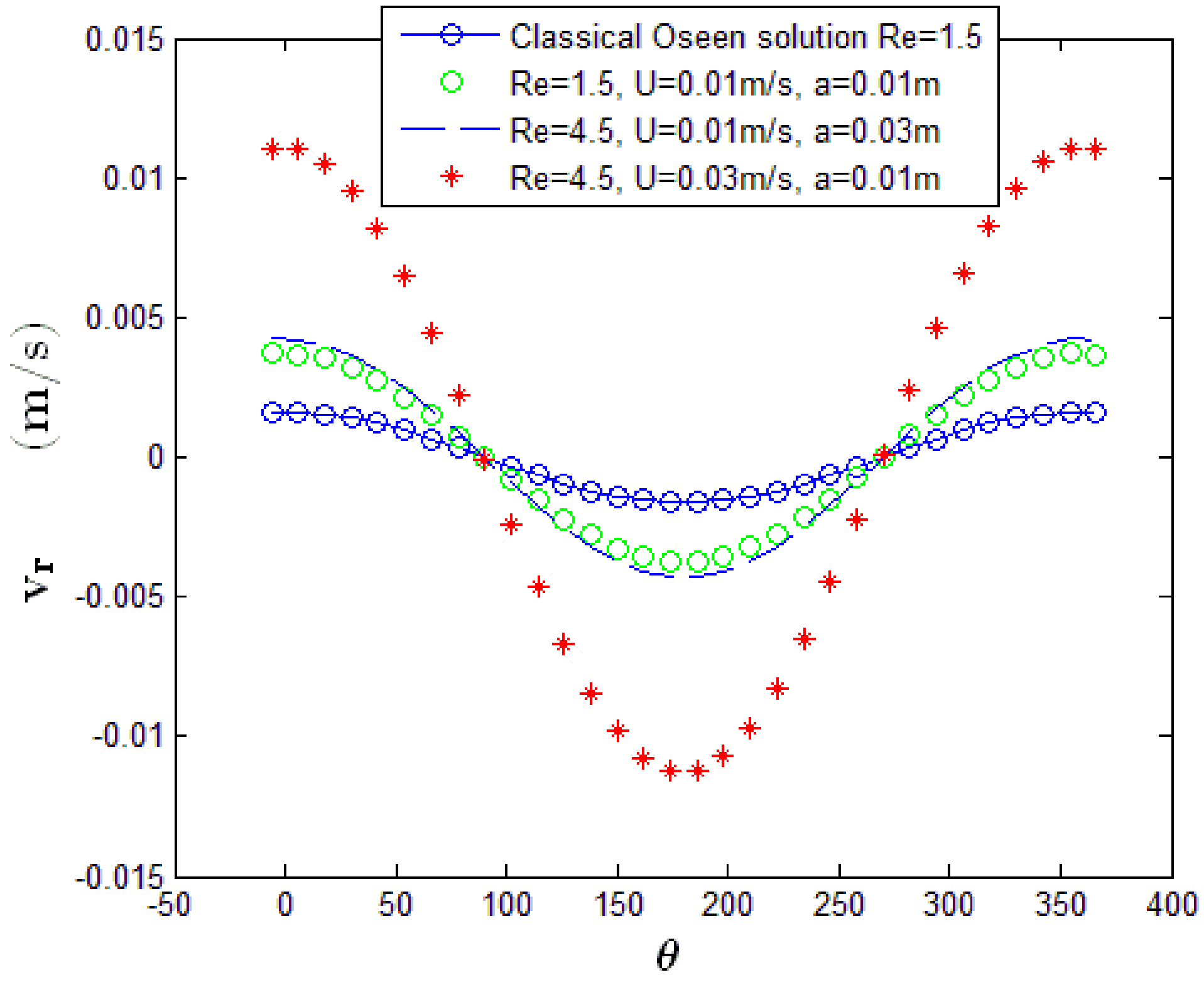

Fig.19 Angular distribution of radial velocity under different Reynolds number and comparison with the classical Oseen solution[55]

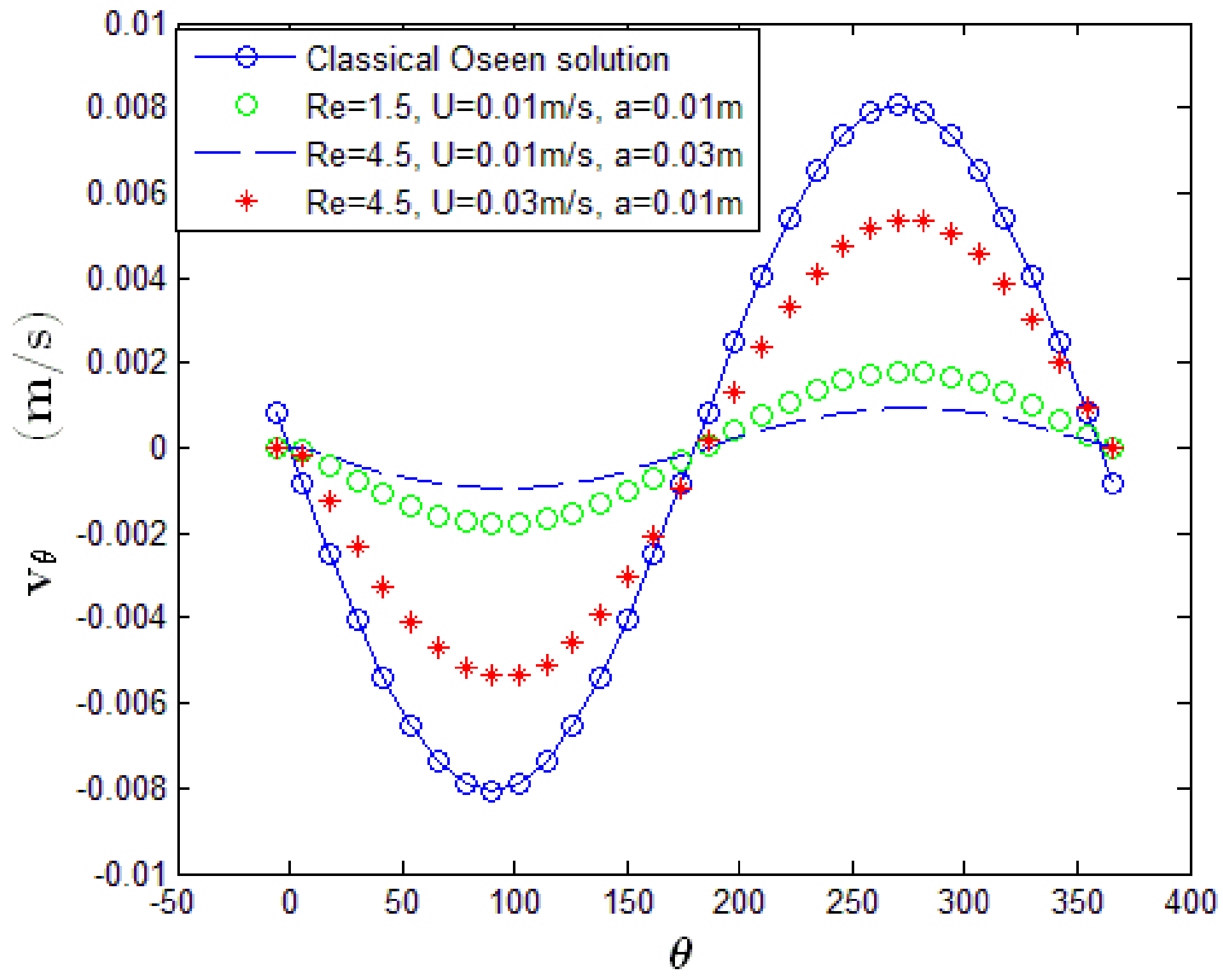

Fig.20 Angular distribution of circumferential velocity under different Reynolds number and comparison with the classical Oseen solution[55].

A part of the results can be compared with the classical and generalized Oseen solutions, the comparison shows the solution of complex fluid of quasiperiodic symmetry with the solution of the classical and generalized Oseen solutions presents in good agreement! Of course there are certain differences between results of soft-matter quasicrystals and conventional fluids.

The similarity between Oseen solution of conventional fluid and solution of soft-matter quasicrystals (a complex fluid with quasiperiodic symmetry), and the Oseen approach produces an approximate analytic solution,these facts may hint that one can carry out study on the approximate analytic solution of soft-matter quasicrystals past cylinder, and may estimate the effects of the quasiperiodic symmetry and interaction between phonons and fluid phonon.

The solution of flow around a circular cylinder in possible 8-fold symmetry soft-matter quasicrystals is also obtained, the features of the solution of phonons and fluid phonon are similar to those of 12-fold symmetry soft-matter quasicrystals, and found the solution of phasons, which has not been included in the solution of 12-fold symmetry soft-matter quasicrystals because of the decoupling between phonons and phasons. The solution of the problem for 7-fold symmetry soft-matter quasicrystals obtained by us presents some other features that the differences between solutions of quasicrystal and classical Oseen flow and generalized Oseen flow are obvious than those appearing in the solutions of the first kind of soft-matter quasicrystals.

The flow past a sphere is more interesting problem. It is well-known in the conventional fluid there was the famous Stokes solution, which plays an important role in modern physics [67]. Einstein made use of it in the theory of the Brownian motion which leads to a determination of Loschmidt's number. Later it has become important again in Millikan's determination of the electronic charge. The generalized Stokes problem of flow of soft-matter quasicrystals is significant too, and the work is been carrying out. For the purpose we developed a spherical coordinate form of the three-dimensional dynamic equations of the first kind of soft-matter quasicrystals [69], and the solution will be obtained.

**6.3 Defects, dislocations**

Dislocationas one kind of topological defectsis important for soft-matter quasicrystals [48]. The study on dislocations for this type material is much more complicated than the solid quasicrystals with additional fluidity anddynamic effect. At present, the static and elastic solutions of quasicrystals can be seen as an approximation of the dislocation solutions of soft-matter quasicrystals.

Here we introduce the dislocation solution of 9-fold symmetry soft-matter quasicrystals. Of course we can study the approximation of the dislocation solutions of soft-matter quasicrystals only at present.

Due to the decoupling for 9- and 18-fold symmetry soft-matter quasicrystals, the determination of phonon field and phason field can be separately carried out. So that from the extended equations of (14-18) (in detail refer to [55]) we need only to solve

$$M\nabla^2 u_x + (L+M)\frac{\partial}{\partial x}\nabla\cdot\mathbf{u} = 0, \qquad (26\text{-}1)$$

$$M\nabla^2 u_y + (L+M)\frac{\partial}{\partial y}\nabla\cdot\mathbf{u} = 0, \qquad (26\text{-}2)$$

and

$$T_1\nabla^2 v_x + G\left(\frac{\partial^2 w_x}{\partial x^2} - \frac{\partial^2 w_x}{\partial y^2}\right) + 2G\frac{\partial^2 w_y}{\partial x\partial y} = 0, \qquad (26\text{-}3)$$

$$T_1\nabla^2 v_y - 2G\frac{\partial^2 w_x}{\partial x\partial y} + G\left(\frac{\partial^2 w_y}{\partial x^2} - \frac{\partial^2 w_y}{\partial y^2}\right) = 0, \qquad (26\text{-}4)$$

$$K_1\nabla^2 w_x + G\left(\frac{\partial^2 v_x}{\partial x^2} - \frac{\partial^2 v_x}{\partial y^2}\right) - 2G\frac{\partial^2 v_y}{\partial x\partial y} = 0, \qquad (26\text{-}5)$$

$$K_1\nabla^2 w_y + 2G\frac{\partial^2 v_x}{\partial x\partial y} + G\left(\frac{\partial^2 v_y}{\partial x^2} - \frac{\partial^2 v_y}{\partial y^2}\right) = 0, \qquad (26\text{-}6)$$

under the dislocation conditions

$$\int_\Gamma du_j = b_j, \int_\Gamma dv_j = b_{1j}^{\perp}, \int_\Gamma dw_j = b_{2j}^{\perp}, j = 1,2. \quad (27)$$

The equations (31-1) and (31-2) under boundary conditions

$$\int_\Gamma du_j = b_j$$

have been solved in solid quasicrystals, refer to [20, 21] and [55], i.e., the solutions are known already, and here we focus on solving equations (26-3) to (26-6) under boundary conditions

$$\int_\Gamma dv_j = b_{1j}^{\perp}, \int_\Gamma dw_j = b_{2j}^{\perp}, j = 1,2 \quad (28)$$

i.e., we need only to determine the first and second phason solutions induced by the dislocation.

To obtain the dislocation-induced phason field (i.e., to solve equations (26-3) to (26-6) under boundary conditions (28)) , we introduce two biharmonic functions $\Phi$ and $\Psi$ and choose

$$v_x = -G\left(\frac{\partial^2\Phi}{\partial x^2} - \frac{\partial^2\Phi}{\partial y^2}\right) - 2G\frac{\partial^2\Psi}{\partial x\partial y}, \qquad (29\text{-}1)$$

$$v_y = 2G\frac{\partial^2\Phi}{\partial x\partial y} - G\left(\frac{\partial^2\Psi}{\partial x^2} - \frac{\partial^2\Psi}{\partial y^2}\right), \qquad (29\text{-}2)$$

$$w_x = T_1\nabla^2\Phi, w_y = T_1\nabla^2\Psi, \qquad (29\text{-}3)$$

where

$$\nabla^2\nabla^2\Phi = 0, \nabla^2\nabla^2\Psi = 0. \quad (30)$$

then the partial differential equations (26-3)-(26-6) are all automatically satisfied. Based on the above general representation, in order to fulfill the dislocation circuit condition, omitting concrete detail we readily obtain the dislocation phason displacement as follows [70]:

$$v_x = \frac{b_{11}^{\perp}}{2\pi}\arctan\frac{y}{x} + \frac{b_{21}^{\perp}G}{2\pi T_1}\frac{xy}{x^2+y^2} - \frac{b_{22}^{\perp}G}{2\pi T_1}\frac{x^2-y^2}{2\left(x^2+y^2\right)}, \quad (31\text{-}1)$$

$$v_y = \frac{b_{12}^{\perp}}{2\pi}\arctan\frac{y}{x} + \frac{b_{21}^{\perp}G}{2\pi T_1}\frac{x^2-y^2}{2\left(x^2+y^2\right)} + \frac{b_{22}^{\perp}G}{2\pi T_1}\frac{xy}{x^2+y^2}, \quad (31\text{-}2)$$

$$w_x = \frac{b_{11}^{\perp}G}{2\pi K_1}\frac{xy}{x^2+y^2} + \frac{b_{12}^{\perp}G}{2\pi K_1}\frac{x^2-y^2}{2\left(x^2+y^2\right)} + \frac{b_{21}^{\perp}}{2\pi}\arctan\frac{y}{x}, \quad (31\text{-}3)$$

$$w_y = -\frac{b_{11}^{\perp}G}{2\pi K_1}\frac{x^2-y^2}{2\left(x^2+y^2\right)} + \frac{b_{12}^{\perp}G}{2\pi K_1}\frac{xy}{x^2+y^2} + \frac{b_{22}^{\perp}}{2\pi}\arctan\frac{y}{x}. \quad (31\text{-}4)$$

The above displacement fields have no logarithmic singularity near the dislocation core. Moreover, using the following constitutive equations

$$\tau_{11} = T_1\frac{\partial v_x}{\partial x} + T_2\frac{\partial v_y}{\partial y} + G\left(\frac{\partial w_x}{\partial x} + \frac{\partial w_y}{\partial y}\right), \quad (31\text{-}5)$$

$$\tau_{22} = T_2\frac{\partial v_x}{\partial x} + T_1\frac{\partial v_y}{\partial y} - G\left(\frac{\partial w_x}{\partial x} + \frac{\partial w_y}{\partial y}\right), \quad (31\text{-}6)$$

$$\tau_{12} = T_1\frac{\partial v_y}{\partial x} - T_2\frac{\partial v_x}{\partial y} + G\left(\frac{\partial w_y}{\partial x} - \frac{\partial w_x}{\partial y}\right), \quad (31\text{-}7)$$

$$\tau_{21} = -T_2\frac{\partial v_x}{\partial y} + T_1\frac{\partial v_y}{\partial x} + G\left(\frac{\partial w_y}{\partial x} - \frac{\partial w_x}{\partial y}\right), \quad (31\text{-}8)$$

$$H_{11} = K_1\frac{\partial w_x}{\partial x} + K_2\frac{\partial w_y}{\partial y} + G\left(\frac{\partial v_x}{\partial x} - \frac{\partial v_y}{\partial y}\right), \quad (31\text{-}9)$$

$$H_{22} = K_2\frac{\partial w_x}{\partial x} + K_1\frac{\partial w_y}{\partial y} + G\left(\frac{\partial v_x}{\partial x} - \frac{\partial v_y}{\partial y}\right), \quad (31\text{-}10)$$

$$H_{12} = K_1\frac{\partial w_x}{\partial y} - K_2\frac{\partial w_y}{\partial x} - G\left(\frac{\partial v_x}{\partial y} + \frac{\partial v_y}{\partial x}\right), \quad (31\text{-}11)$$

$$H_{21} = -K_2 \frac{\partial w_x}{\partial y} + K_1 \frac{\partial w_y}{\partial x} + G\left(\frac{\partial v_x}{\partial y} + \frac{\partial v_y}{\partial x}\right). \text{(31-12)}$$

we get the corresponding stress field as follows

$$\tau_{11} = -\frac{b_{11}^{\perp}\left(K_1T_1 + G^2\right)}{2\pi K_1}\frac{y}{x^2+y^2} + \frac{b_{12}^{\perp}\left(K_1T_2 + G^2\right)}{2\pi K_1}\frac{x}{x^2+y^2}$$

$$-\frac{b_{21}^{\perp}\left(T_1+T_2\right)G}{\pi T_1}\frac{x^2y}{\left(x^2+y^2\right)^2} + \frac{b_{22}^{\perp}\left(T_1+T_2\right)G}{2\pi T_1}\frac{x\left(x^2-y^2\right)}{\left(x^2+y^2\right)^2}, \text{(31-13)}$$

$$\tau_{22} = -\frac{b_{11}^{\perp}\left(K_1T_2 + G^2\right)}{2\pi K_1}\frac{y}{x^2+y^2} + \frac{b_{12}^{\perp}\left(K_1T_1 - G^2\right)}{2\pi K_1}\frac{x}{x^2+y^2}$$

$$-\frac{b_{21}^{\perp}\left(T_1+T_2\right)G}{2\pi T_1}\frac{y\left(x^2-y^2\right)}{\left(x^2+y^2\right)^2} - \frac{b_{22}^{\perp}\left(T_1+T_2\right)G}{\pi T_1}\frac{xy^2}{\left(x^2+y^2\right)^2}, \text{(31-14)}$$

$$\tau_{12} = \frac{b_{11}^{\perp}\left(K_1T_1 - G^2\right)}{2\pi K_1}\frac{x}{x^2+y^2} + \frac{b_{12}^{\perp}\left(K_1T_2 + G^2\right)}{2\pi K_1}\frac{y}{x^2+y^2}$$

$$-\frac{b_{21}^{\perp}\left(T_1+T_2\right)G}{\pi T_1}\frac{xy^2}{\left(x^2+y^2\right)^2} + \frac{b_{22}^{\perp}\left(T_1+T_2\right)G}{2\pi T_1}\frac{y\left(x^2-y^2\right)}{\left(x^2+y^2\right)^2}, \text{(31-15)}$$

$$\tau_{21} = -\frac{b_{11}^{\perp}\left(K_1T_2 + G^2\right)}{2\pi K_1}\frac{x}{x^2+y^2} - \frac{b_{12}^{\perp}\left(K_1T_1 - G^2\right)}{2\pi K_1}\frac{y}{x^2+y^2}$$

$$-\frac{b_{21}^{\perp}\left(T_1+T_2\right)G}{2\pi T_1}\frac{x\left(x^2-y^2\right)}{\left(x^2+y^2\right)^2} - \frac{b_{22}^{\perp}\left(T_1+T_2\right)G}{\pi T_1}\frac{x^2y}{\left(x^2+y^2\right)^2}, \text{(31-16)}$$

$$H_{11} = -\frac{b_{11}^{\perp}\left(K_1-K_2\right)G}{\pi K_1}\frac{x^2y}{\left(x^2+y^2\right)^2} - \frac{b_{12}^{\perp}\left(K_1-K_2\right)G}{2\pi K_1}\frac{x\left(x^2-y^2\right)}{\left(x^2+y^2\right)^2}$$

$$-\frac{b_{21}^{\perp}\left(K_1T_1 - G^2\right)}{2\pi T_1}\frac{y}{x^2+y^2} + \frac{b_{22}^{\perp}\left(K_2T_1 - G^2\right)}{2\pi T_1}\frac{x}{x^2+y^2}, \text{(31-17)}$$

$$H_{22} = \frac{b_{11}^{\perp}\left(K_1-K_2\right)G}{2\pi K_1}\frac{y\left(x^2-y^2\right)}{\left(x^2+y^2\right)^2} - \frac{b_{12}^{\perp}\left(K_1-K_2\right)G}{\pi K_1}\frac{xy^2}{\left(x^2+y^2\right)^2}$$

$$-\frac{b_{21}^{\perp}\left(K_2T_1 - G^2\right)}{2\pi T_1}\frac{y}{x^2+y^2} + \frac{b_{22}^{\perp}\left(K_1T_1 - G^2\right)}{2\pi T_1}\frac{x}{x^2+y^2}, \text{(31-18)}$$

$$H_{12}=-\frac{b_{11}^{\perp}(K_1-K_2)G}{\pi K_1}\frac{xy^2}{\left(x^2+y^2\right)^2}-\frac{b_{12}^{\perp}(K_1-K_2)G}{2\pi K_1}\frac{y\left(x^2-y^2\right)}{\left(x^2+y^2\right)^2}$$

$$+\frac{b_{21}^{\perp}\left(K_1T_1-G^2\right)}{2\pi T_1}\frac{x}{x^2+y^2}+\frac{b_{22}^{\perp}\left(K_2T_1-G^2\right)}{2\pi T_1}\frac{y}{x^2+y^2}, \quad (31\text{-}19)$$

$$H_{21}=\frac{b_{11}^{\perp}(K_1-K_2)G}{2\pi K_1 r}\frac{x\left(x^2-y^2\right)}{\left(x^2+y^2\right)^2}-\frac{b_{12}^{\perp}(K_1-K_2)G}{\pi K_1}\frac{x^2y}{\left(x^2+y^2\right)^2}$$

$$-\frac{b_{21}^{\perp}\left(K_2T_1-G^2\right)}{2\pi T_1}\frac{x}{x^2+y^2}-\frac{b_{22}^{\perp}\left(K_1T_1-G^2\right)}{2\pi T_1}\frac{y}{x^2+y^2}. \quad (31\text{-}20)$$

It is clear that the above stress fields obey equilibrium equations. The solutions present the stress and strain singularity behaviour around the dislocation core.Fig.20shows the contour lines of the stress distribution induced by a component of the Burgers vector.

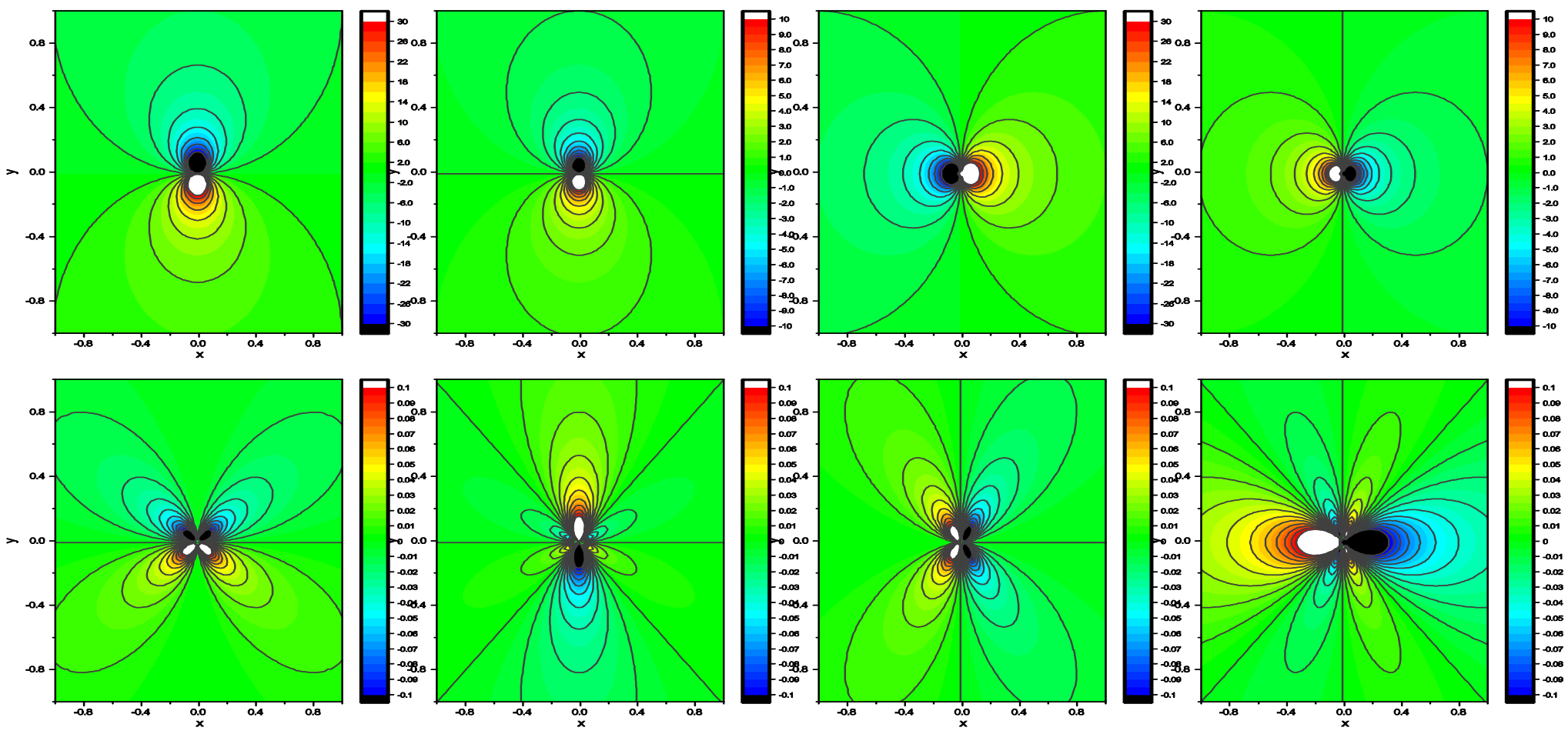

Fig.21 Contour lines of the stress distribution: X: $2\pi Y10^{-6}/b_{11}^{\perp}$, $Y:\tau_{11},\tau_{22},\tau_{12},\tau_{21},H_{11},H_{22},H_{12},H_{21}$, induced by a component $b_{11}^{\perp}$ of the Burgers vector, where the elastic constants were applied as $K_1=5$ MPa, $K_2=-1$ MPa, $T_1=4$ MPa, $T_2=1$ MPa, $G=0.04$ MPa [55].

The solution for possible 18-fold symmetry soft-matter quasicrystals is similar, and the results are not included here (for detail, please refer to reference[70]).

The approximate solutions of dislocations of the first kind of soft-matter quasicrystals have beendiscussed in detail previously [20,21,55], it is not needed to introduce any more. If we want to promote the research work of dislocation problems, some new techniques need to be developed, e.g. studying the fluid effect and dynamic effect on the dislocation field, of course this is a complex and difficult problem.

**6.4 Defects, cracks**

Although soft matter is not solid, the crack and rupture problems in this materials are very important and have been intensively studied [71-81]. It is different from the topological defects---dislocations, the crack is a kind of metric defects, which presents important application tool. So far the work reported in above references are qualitative only, as quantitative study only for a few of branches in soft matter science. Certainly it needs quantitative investigation to understand the mechanism. We have established generalized dynamics for soft-matter quasicrystals, this may provide a basis for quantitatively studying crack and rupture problems of the matter. The quantitative theory of crack began from the classical work of Griffith [82]. Following the Griffith point of view we consider the Fig.22 in which the cracked plate is subjected a tension.

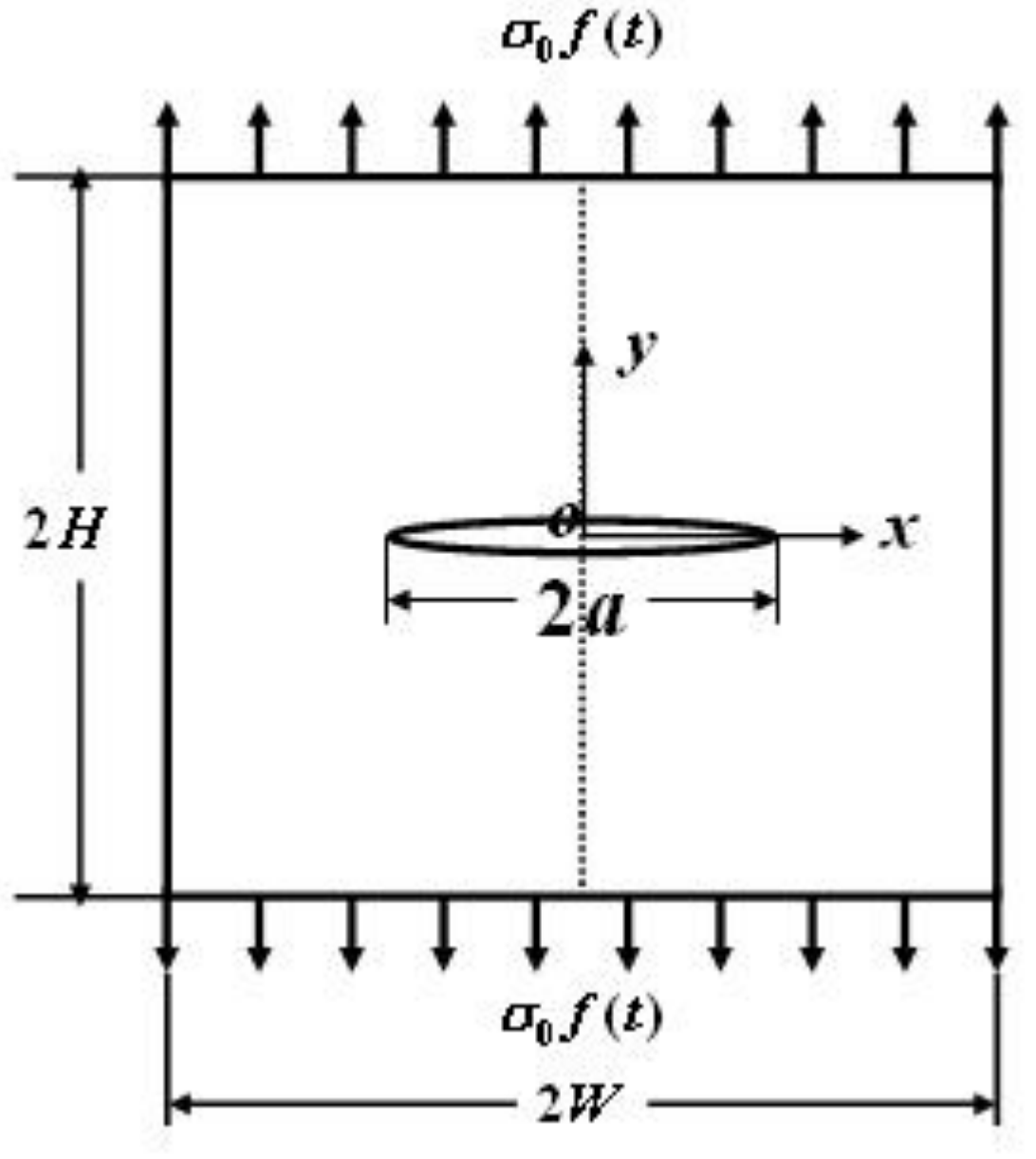

Fig.22 A Griffith crack in a plate of soft-matter quasicrystal

For the structural materials, if the crack surface is opened by a stress, then the crack will be propagated under a certain condition. In general, the stress is elastic stress and leads to induce the crack surface to be opening and leads to crack propagation when the applied stress beyond the limit threshold of the material which can be charged. There is called Griffith criterion

$$K(\sigma,a)=K_C(T,f(s)) \quad (32)$$

The left-side term in equation (32) $K(\sigma,a)$ is named working stress intensity factor as a function of applied stress $\sigma$ , and sample configuration related crack characteristic size $a$;the right-side term $K_C(T, f(s))$ is rupture toughness of the material as a function of temperature $T$ and the material structure factors $f(s)$.

Cheng et al [83] analyzed the sample of soft-matter quasicrystals, and found that there is a fluid stress apart from elastic stress applied at the crack surface at the same time. The elastic stress makes the crack surface to be opening, while the fluid stress makes the crack surface to be closing. Following the Griffith theory, the stress intensity consists of two parts:

$$K_e = \lim_{x \to a^+} \sqrt{\pi(x-a)}\sigma_{yy}(x,0,\mathrm{t})$$

isthe phonon stress intensity factor, and

$$K_f = \lim_{x \to a^+} \sqrt{\pi(x-a)}p_{yy}(x,0,\mathrm{t})$$

Isthe fluid phonon stress intensity factor. Because $K_f < 0$,

so that $K_e + K_f < K_e$

When $K_e + K_f < 0$ the crack surface will keep closed. Thus, the rupture process of soft matter may be a process of the competition between the phonon stress intensity factor $K_e$ and the fluid phonon stress intensity factor $K_f$. This rupture mechanism is new and has never been explored in structural materials. In the Ref [83] some further discussions about this are given.This might bring a new idea in rupture theory and develop the classical Griffith work. Based on detailed computation and the modified Griffith criterion the analysis may arrive in a quantitative level. The quantitative dynamics of soft-matter quasicrystals and the stress analysis tool will promote the development of crack and rupture study in soft matter including soft-matter quasicrystals, which can provide theoretical guideline for the practical applications.

The tensile cracked specimen is relative easier, while the shearing cracked specimen is much more complicated. The tensile crack propagates mainly along the direction of the main crack surface, which is called as a opening mode, and the effects of phonon and fluid phonon stresses are easy to characterize, as we conclude in the previous session. For the shearing crack cases, the crack propagation does not along the direction of the main crack surface, the material rupture mechanism is not so clear, and the classical Griffith theory cannot be simply applied, and the effects of phonon stress and fluid phonon stress and their interaction are not clear too. This suggests an in-depth experimental study is needed.

**6.5 Stability of soft-matter quasicrystals (modified and supplementary edition)**

Some results in Sections 1-5 can also be used to the study of stability of soft-matter quasicrystals. As we mentioned in Section 1, just after the discovery of soft-matter quasicrystals, Lifshitz and his group [48, 49] pointed out that because the significantdifferent formation mechanism of soft-matter quasicrystals with that of the solid ones, the originationof instability of soft-matter quasicrystal's formation remains unclear, and great debate exists today. Since then, up to now, the stability problem of soft-matter quasicrystals became a critical concern among a numberof researchers. They presenteddifferent approaches that show the problem is complicated. Lifshitz and Diamant suggested to study the problem from the point of view of effective free energy proposed by Lifshitz and Petrich [89].This energy is based on mass density $\rho$, and the governing equation describing the time-space evolution of $\rho$ was given by Lifshitz and Diamant. They performed many analyses on the stability of 12-fold symmetry of soft-matter quasicrystals. The other discussions on the stability in relevant references concerning the two natural length scales and two wave numbers, which show the importance of the combination between thermodynamics and dynamics in the study. Some researchers [90-93] extended the Lifshitz'spioneering work and promotedhis study.

Fan and his group [87, 88] recently presented studies in this topic, but they proposed a different way from that of Lifshitz et al.

### 6.5.1Thermodynamic stability of quasicrystals in soft matter and its variation

It is well-known that for the stability problem, in the system of classical elasticity, conventional crystallography and solid quasicrystallography, people have obtained remarkable achievements based on thermodynamics and variational point of view. The starting point of such work is also from free energy of the considered system. When the second order variation of the free energy density being greater than or equal to zero, i.e., $\delta^2 F \geq 0$, then the system will be stable thermodynamically. This condition is equivalent to the requirement of positive definite of quadratic form of the energy density for the system.

In the systems of classical elasticity, the conventional crystallography and solid quasicrystallography, the free energy density consists of the elastic energy density. For soft-matter quasicrystals, we must consider energy density part of fluid phonon and fluid phonon-phonon coupling.

### 6.5.2 Stability of 12-fold symmetry quasicrystals in soft matter

At first, we consider the case of 12-fold symmetry quasicrystals. By including the energy density of fluid phonon and fluid phonon-phonon coupling, we define an extended free energy density [87]

$$F_{ex} = U_{ex} - TS \qquad (33)$$

where $U_{ex}$ the extended inner energy density defined by

$$U_{ex} = \frac{1}{2}A\left(\frac{\delta\rho}{\rho_0}\right)^2 + B\left(\frac{\delta\rho}{\rho_0}\right)\nabla\bullet\mathrm{u} + C\left(\frac{\delta\rho}{\rho_0}\right)\nabla\bullet\mathrm{w} + U_{el} \quad (33a)$$

and $T$ the absolute temperature and $S$ the entropy, and $A, B$ are defined in previous sections of this paper , $C$ is the constant relating fluid-phason coupling but which is removed due to this coupling does not exist for quasicrystals, so that in the following we take $C = 0$. In addition the elastic energy density in (33a) is defined in the Subsection 5.1, that is

$$U_{el} = U_{\mathrm{u}} + U_{\mathrm{w}} + U_{\mathrm{uw}} \quad (33b)$$

i.e., the sum of densities of phonon, phason and phonon-phason coupling elastic energies, in which

$$\left.\begin{aligned} U_{\mathrm{u}} &= \frac{1}{2}C_{ijkl}\varepsilon_{ij}\varepsilon_{kl} \\ U_{\mathrm{w}} &= \frac{1}{2}K_{ijkl}w_{ij}w_{kl} \\ U_{\mathrm{uw}} &= R_{ijkl}\varepsilon_{ij}w_{kl} + R_{klij}w_{ij}\varepsilon_{kl} \end{aligned}\right\} \quad (33c)$$

whose concrete forms depend upon the constitutive law (22) (the elastic part of which), so that we obtain the concrete form of (33) (of course $C = 0$).

It is well-known, the energy density is a quadratic form of strain components (or stress components) for the system of classical elasticity, conventional crystallography and solid quasicrystallography. For the soft-matter quasicrystals in the quadratic form, the “strain components” are extended to cover the relative variation of the mass density, i.e., $\delta\rho / \rho_0$ because here the extended free energy density replaces the free energy density. The coefficients (including $A$ and $B$) of the energy density quadratic form constitutes so-called rigidity matrix, and for soft-matter quasicrystals the relevant matrix should be named as extended rigidity matrix denoted by $M$ , for the 12-fold symmetry soft-matter quasicrystals which is

$$M=\begin{pmatrix} A & B & B & B & 0 & 0 & 0 & 0 & 0 & 0 & 0 & 0 & 0 \\ B & C_{11} & C_{12} & C_{13} & 0 & 0 & 0 & 0 & 0 & 0 & 0 & 0 & 0 \\ B & C_{12} & C_{11} & C_{13} & 0 & 0 & 0 & 0 & 0 & 0 & 0 & 0 & 0 \\ B & C_{13} & C_{13} & C_{33} & 0 & 0 & 0 & 0 & 0 & 0 & 0 & 0 & 0 \\ 0 & 0 & 0 & 0 & 2C_{44} & 0 & 0 & 0 & 0 & 0 & 0 & 0 & 0 \\ 0 & 0 & 0 & 0 & 0 & 2C_{44} & 0 & 0 & 0 & 0 & 0 & 0 & 0 \\ 0 & 0 & 0 & 0 & 0 & 0 & C_{11}-C_{12} & 0 & 0 & 0 & 0 & 0 & 0 \\ 0 & 0 & 0 & 0 & 0 & 0 & 0 & K_1 & K_2 & 0 & 0 & 0 & 0 \\ 0 & 0 & 0 & 0 & 0 & 0 & 0 & K_2 & K_1 & 0 & 0 & 0 & 0 \\ 0 & 0 & 0 & 0 & 0 & 0 & 0 & 0 & 0 & K_4 & 0 & 0 & 0 \\ 0 & 0 & 0 & 0 & 0 & 0 & 0 & 0 & 0 & 0 & K_1+K_2+K_3 & 0 & K_3 \\ 0 & 0 & 0 & 0 & 0 & 0 & 0 & 0 & 0 & 0 & 0 & K_4 & 0 \\ 0 & 0 & 0 & 0 & 0 & 0 & 0 & 0 & 0 & 0 & K_3 & 0 & K_1+K_2+K_3 \end{pmatrix} \quad (34)$$

The positive definite requirement of the quadratic form of energy density is equivalent to that of the rigidity matrix.

The above discussion can be written as the lemma as follows

## Lemma

From (33), we have

$$S=-\frac{\partial F_{ex}}{\partial T},\sigma_{ij}=\frac{\partial F_{ex}}{\partial \varepsilon_{ij}},H_{ij}=\frac{\partial F_{ex}}{\partial w_{ij}},\delta^2 F_{ex}\geq 0 \quad (35)$$

and the theorem

## Theorem

If the rigidity matrix (34) is positive definite then $\delta^2 F_{ex}\geq 0$, and the system is stable thermodynamically.

For 12-fold symmetry soft-matter quasicrystals the stability condition is

$$\left.\begin{array}{l} A>0,\ A(C_{11}C_{33}+C_{12}C_{33}-2C_{13}^2)-B^2\left(C_{11}+C_{12}-4C_{13}+2C_{33}\right)>0,\ C_{11}-C_{12}>0, \\ C_{44}>0,\ K_1-K_2>0,\ K_1+K_2>0,\ K_1+K_2+2K_3>0,\ K_4>0 \end{array}\right\} \quad (36)$$

The proof of the theorem is straightforward and omitted here.

This result is given by Ref [87] and very simple and intuitive, that shows the stability depends only upon the material constants, which are known in the previous presentation and can be measured by experiments. The result (36) has been examined, which can be exactly reduced to those of crystals and solid quasicrystals.

### 6.5.3 Stability of 18-fold symmetry quasicrystals in soft matter

For18-fold symmetry quasicrystals in soft matter the discussion is similar, and we have the rigidity matrix of the system

$$M_1 = \begin{pmatrix}
A & B & B & B & & & & & & & & & & & & & & & \\
B & C_{11} & C_{12} & C_{13} & & & & & & & & & & & & & & & \\
B & C_{12} & C_{11} & C_{13} & & & & & & & & & & & & & & & \\
B & C_{13} & C_{13} & C_{33} & & & & & & & & & & & & & & & \\
 & & & & 2C_{44} & & & & & & & & & & & & & & \\
 & & & & & 2C_{44} & & & & & & & & & & & & & \\
 & & & & & & 2C_{66} & & & & & & & & & & & & \\
 & & & & & & & T_1 & T_2 & & & & & G & -G & & & & \\
 & & & & & & & T_2 & T_1 & & & & & G & -G & & & & \\
 & & & & & & & & & T_3 & & & & & & & & & \\
 & & & & & & & & & & T_1 & & -T_2 & & & & -G & & -G \\
 & & & & & & & & & & & T_3 & & & & & & & \\
 & & & & & & & & & & -T_2 & & T_1 & & & & G & & G \\
 & & & & & & & G & G & & & & & K_1 & K_2 & & & & \\
 & & & & & & & -G & -G & & & & & K_2 & K_2 & & & & \\
 & & & & & & & & & & & & & & & K_3 & & & \\
 & & & & & & & & & & -G & & G & & & & K_1 & & -K_2 \\
 & & & & & & & & & & & & & & & & & K_3 & \\
 & & & & & & & & & & -G & & G & & & & -K_2 & & K_1
\end{pmatrix} \tag{37}$$

The stability criterion for this system is

$$\left.\begin{aligned} &A > 0,\ C_{11} - C_{12} > 0,\ C_{44} > 0,\ A(C_{11}C_{33} + C_{12}C_{33} - 2C_{13}^2) - B^2\left(C_{11} + C_{12} - 4C_{13} + 2C_{33}\right) > 0, \\ &\quad K_1 + K_2 > 0,\ T_1 - T_2 > 0,\ \left(K_1 - K_2\right)\left(T_1 + T_2\right) - 4G^2 > 0,\ K_1 - K_2 > 0,\ T_3 > 0,\ K_3 > 0 \end{aligned}\right\} \tag{38}$$

The detail can be referred to Ref [88].

For other systems of soft-matter quasicrystals the criteria can be obtained in similar manner to those of the above presentation. Due to the limitation of space those results are not listed here.

#### 6.5.4 Summary

The results on stability of 12- and 18-fold symmetry quasicrystals demonstratedthat

1) The stability of 12-fold symmetry quasicrystals depends on the fluid effect, represented by constant $A$, fluid coupling to phonons represented by constant $B$, phonons represented by $C_{ijkl}$ and phasons represented by $K_{ijkl}$;

2) The stability of 18-fold symmetry quasicrystals depends on the fluid effect, fluid coupling to phonons, phonons and phasons, which is similar to those of 12-symmetry ones; but alsoupon the coupling between first and second phasons represented by $G_{ijkl}$;

3) These constants can be measured by experiments;

4) The differences between those of 12- and 18-fold symmetry quasicrystals lie in their different structures, explored by the constitutive laws, which are determined by the theory of group representation;

5) Due to the combination between thermodynamics and generalized dynamics of the matter, the approach determining the stability developed by our work presents systematic, direct and simple features, it does not need any simulation effort, and the results obtained present bright physical meaning, whose correctness, of course, has to be verified by experiments further.

6) The approach developed by us is different from Lifshitz and other researchers, which is originated from the traditional or classical approach, but we suggested some extensions, so that the present approach can be named extended traditional or extended classical approach. This approach might be able to serve in other branches of soft matter science in additional to soft-matter quasicrystals.

The study on stability is a thermodynamics problem, we here introduced some results of dynamics and that helped us to obtain the criteria of stability for various systems of soft-matter quasicrystals, in which the Hamiltonian and constitutive law of the systems are a basis. The constitutive law is applicable for stable state of thermodynamics, but quasicrystals are in sub-stable state in many cases. Although for the sub-stable state the variational principle is still held, the usage of the constitutive law may result in some questions, so that further study is needed. In practical applications, the constitutive law may be used in many cases of nonequilibrium states, and the validity range of the law may be disregarded. Of course, these developments are still approximate, and are not very vigorous as idealized in pure science.

## 6.6 Photon Band-Gap of Holographic Photonic Quasicrystals

The most attractive aspect of application of soft-matter quasicrystals may be in photon band-gap. The soft-matter quasicrystals observed so far are two-dimensional structure with quasiperiodic symmetry, and higher fold of orientational symmetry being greater than that of solid ones appeared, there is superiority than solid quasicrystals in this respect.

The two-dimensional photonic quasicrystals (PQCs) with different rotational symmetries are constructed based on holographic interference patterns. Their photonic band-gap (PBG) properties are obtained by using finite element method (FEM) to calculate the transmission and reflection spectra with different directions. The maximum gap-midgapratiosare given for these PQCs with different dielectric contrast and fill factorto compare their abilities to produce band-gaps. The results show that 10-fold (five beams interference) and 12-fold quasicrystals are easier to form band-gaps than others. On the other hand, 10-fold PQCs was prepared by using single-prism holographic lithography. This may provide a reference to select appropriate quasicrystal structures for photonic devices and optimize device performance.

### 6.6.1 On PBG

PBG materials have attracted significant interest for the ability to control the propagation of light. Compared to periodic photonic crystals (PCs), PQCs have the following advantages: the low dielectric constant contrast when a complete PBG appears [94-96], higher isotropy of PBG [96-98]and rich defect modes without introducing defect [99] etc. PQCs have been utilized to design waveguides [97,100], sensors [101], laser microcavities[102], filters [95,103], owing to the band-gap and localized state properties, showing great potential for small size, flexibility and effectiveness, thereby lays the foundation for the development of optical integratedcircuits.

Research of PQC band-gap is the basis of application, and has been discussed a lot. For the study of band-gap properties of different quasicrystals, Mikael C. Rechtsman et al. [94] used a novel method to find the optimized pattern with the widest TM polarized gap for two-component materials. The results show that 5-fold quasicrystal possesses lager gaps when the dielectric contrast is less than 5, while 8-fold quasicrystal possesses lager gaps with the contrast equal or greater than 5 among quasicrystals. Marian Florescu et al. [104] introduced an optimization method to design examples of PQCs with substantial complete PBGs. The results show the maximal complete band-gap in 5-fold symmetric structure is larger than that in 8-fold symmetric structure with dielectric contrast equals 11.56. These studies have concentrated two or three quasicrystals' band-gaps. It is necessary to expand the research of quasicrystal owing to the abundance of PQCs.

Since large-area PQC structures can be obtained by simple experimental setup[105,106]. PQCs based on interference method are chosen to study which quasicrystal is more likely to provide band-gaps. In this chapter, the multiple rotational symmetry quasicrystals with medium rods in air background are constructed by using multi-beam interfering method. The corresponding band-gap properties have been obtained by calculating their transmittance and reflectance spectra. The ability to form band-gaps was compared for different quasicrystals in the same conditions. In the same time, 10-fold PQCs was prepared by using single-prism holographic lithography. The work may helpful for the design and application of photonic device.

### 6.6.2 The design and formation of holographicPQCs[107]

The two-dimensional PQC structures are obtained by the multi-beam common-path interference method [105]. The interference intensity on the medium can be expressed as：

$$I(r)=\sum_{j,l}E_jE_l*\exp\left\{-i\left[\left(\overrightarrow{k_j}-\overrightarrow{k_l}\right)\cdot\vec{r}\right]\right\}, (39)$$

Where$E_j$ is the amplitude of beam j, $\overrightarrow{k_j}$ is the wave vector of beam j, $\vec{r}$ =(x,y,z) is the spatial position vector. The wave vectors of N beams are:

$$\overrightarrow{k_m} = \frac{2\pi n_w}{\lambda}\left(\cos\frac{2(m-1)\pi}{N}\sin\varphi, \sin\frac{2(m-1)\pi}{N}\sin\varphi, \cos\varphi\right) \quad (40)$$

wherem=1-N is corresponding to the ordinal number of beams,$n_w$ represents the refractive index of the medium in the writing laser wavelength, $\varphi$ is the crossing angle between the light beam and zaxis. The holographic interferogramcan be gotten by substituting $\varphi$=44.3°, $n_w$=1.5, $\lambda$=355nm into equation (39) and equation (40). The two-dimensional 10-,14-,18-,22- and 26-fold quasicrystal structures are obtained by 5,7,9,11 and 13 beams interference, while the 8-,10-*,12-fold quasicrystal structures are obtained by 8,10,12 beams interference. All of these structures possess even fold rotational symmetry. The two-dimensional multiple quasicrystal structures with the area of 10μm×10μm are shown in Fig. 23.Specifically, the quasicrystals formed by both five beams and ten beams interference are 10-fold, which are marked as 10-fold in Fig.12.1(b) and 10*-fold in Fig. 1(c), respectively. But ten-beams interference doesn't make the structure more abundant, because some light spots were so close to give rise to interference that the spots number are reduced and a more sparse structure is formed.

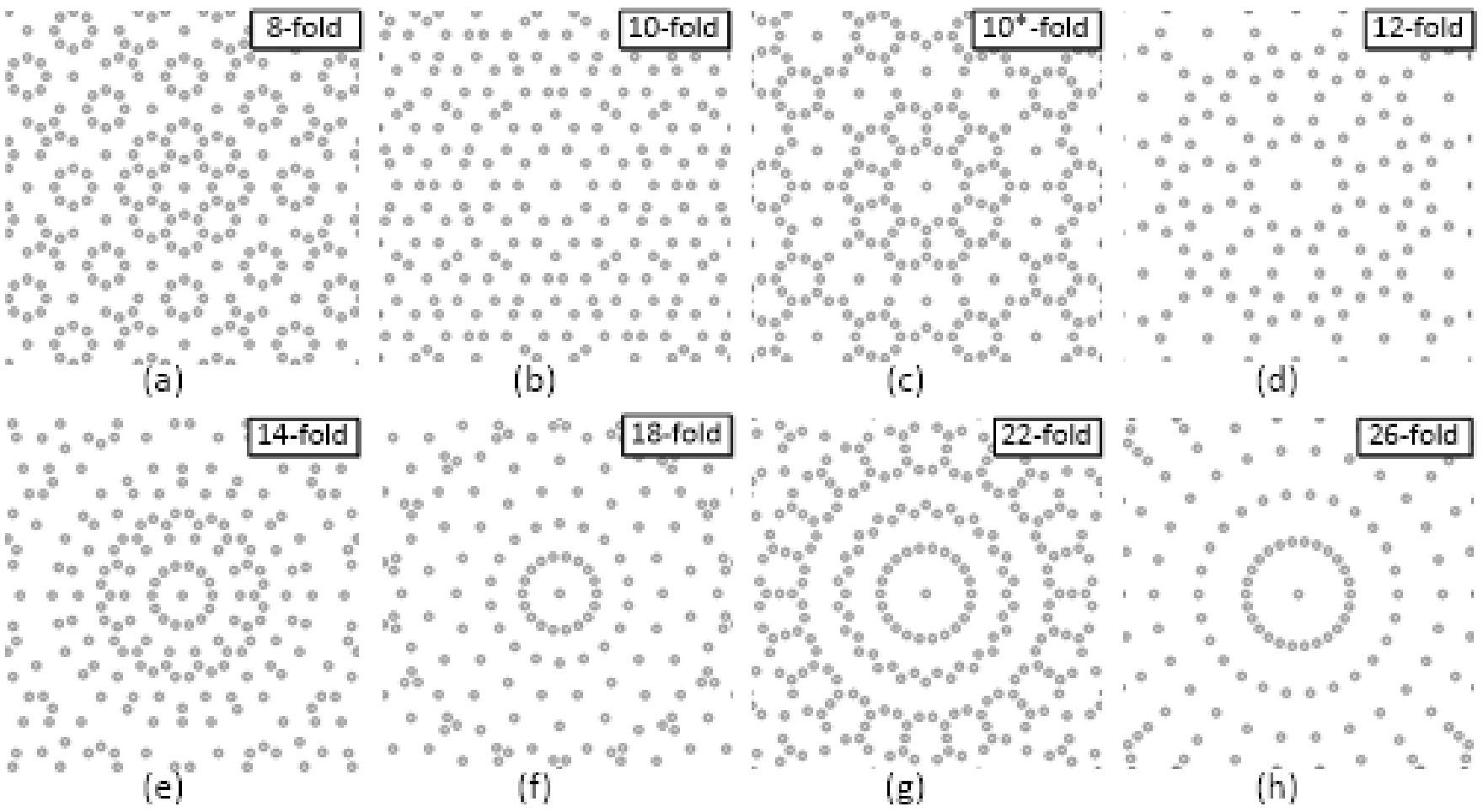

Fig. 23(a)-(h) are the two dimensional 8-, 10-, $10^*$-, 12-, 14-, 18-, 22-, and 26- fold quasicrystal structures based on the interferogram respectively. Among which, (b) is the 10-fold PQC by five beams interference, (c) is the $10^*$-fold PQC by ten beams interference.

### 6.6.3 Band-gap of 8-fold PQCs [107]

The 8-fold quasicrystal in Fig.(23)(a) is taken as an example to illustrate simulation settings in detail. As shown in Fig.24 (a), the central region in Fig.23(a) is selected for simulation. The more detailed setting is given in Fig.24 (b). The structure is composed of medium rods whose dielectric constant $\varepsilon_1$=15 with radius r=0.12 μm in air ($\varepsilon_0$=1) in the x-y plane. Light parallel to the x-axis is incident from the left side of the structure and the transmitted light is detected on the right side. Angle of the incident light relative to x-axis is represented by θ, which is depicted in Fig.24(b). Due to rotational symmetry in the range of 360° and mirror symmetry in the range of 45°, θ between 0° and 22.5° are sufficient to describe the propagation at all angles for 8-fold quasicrystal. The transmittance and reflection spectra were performed by Finite Element Method (FEM) to study PQCs band-gap properties.

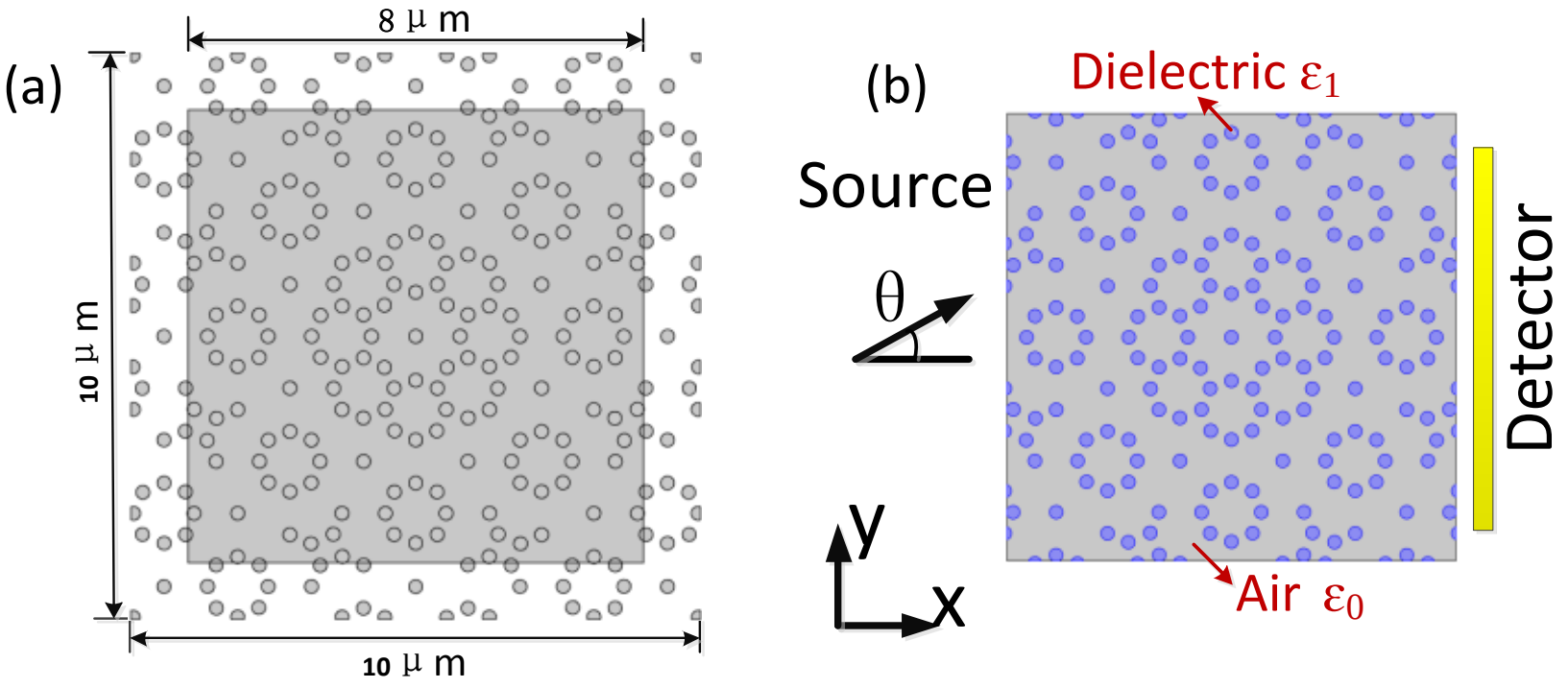

Fig. 24(a) The central modelling region of 8-fold quasicrystal in Fig.23(a).

(b) The simulation model and setup.

Fig.25 shows that the transmission and reflection spectrum versus source normalized frequency a/λ for TE polarized wave (E-field perpendicular to x-y plane) at different incident angels θ, where a=1μm is the lattice constant and λ is the wavelength of the source. Theoretically, the position where transmission equals zero and reflection equals one is considered to be photonic band-gap, and the band-gapfor different incident directions is thought as a real band-gap. Based on this point, the biggest PBGs are thought to be in the range of a/λ=0.476-0.523. Fig.26 shows the electric field distribution for TE polarization at incident frequency a/λ=0.507in the band-gap and incident frequency a/λ=0.730 not in the band-gap.

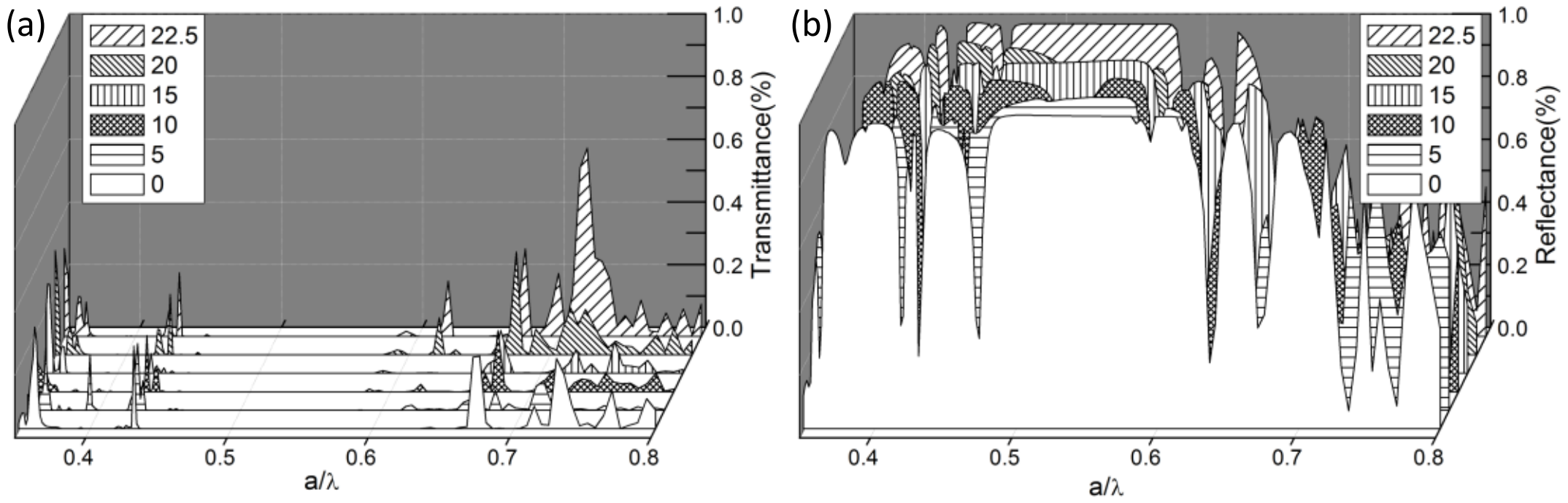

Fig.25 Transmission spectrum (a) and reflection spectrum (b) for TE polarizationat θ=0°, 5°,10°, 15°, 20°and 22.5° with r= 0.12μm under $\varepsilon_1$/ $\varepsilon_0$ =15 of 8-fold quasicrystal.

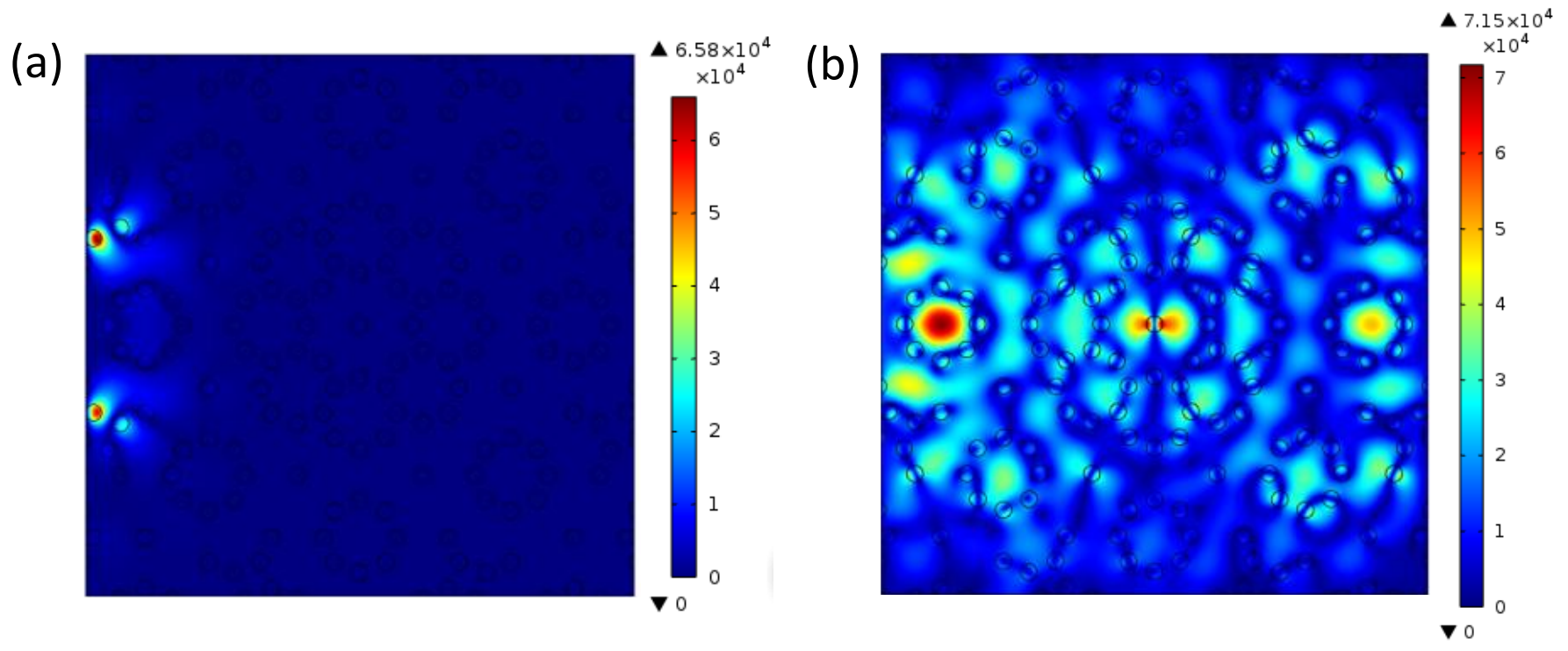

Fig. 26 The electric field distribution for TE polarized mode at θ=0° with r= 0.12μm under $\varepsilon_1/\ \varepsilon_0$ =15 of 8-fold quasicrystal. (a) when a/λ=0.507, (b) when a/λ=0.730.

Then the maximum band-gap of 8-fold quasicrystalbased on the change of dielectric contrast and fill factor was further investigated.Fig.27 shows the variation of maximal PBG size as a function of parameters $\varepsilon_1/\varepsilon_0$and r/a.According to Refs[94,95,98]$\Delta\omega/\omega_c$ of one band-gap increases firstly then tends to be invariant with the increase of dielectric contrast when keeping the fill factor constant, where, $\Delta\omega$ is the width of PBG, $\omega_c$ is the central frequency, respectively. Therefore in our calculation, the choice of $\varepsilon_1/\varepsilon_0$ between 6-18 is able to show the ability of quasicrystal to generate band-gaps. The upper limit of fill factor r/a =0.18 is determined by all the quasicrystal structures to keep rods from touching each other. There exists a maximum $\Delta\omega/\omega_c$ at r/a = 0.12 and $\varepsilon_1/\varepsilon_0$ =18 inFig.27.

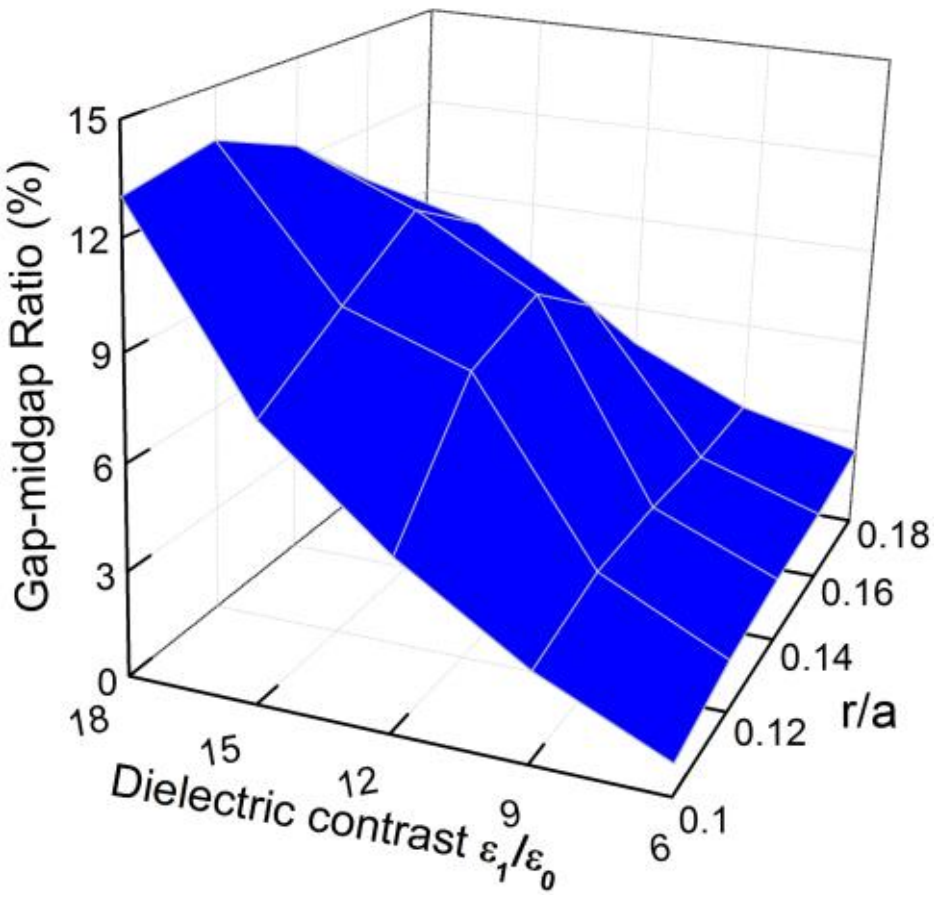

Fig. 27. Gap-midgap ratio as a function of parameters $\varepsilon_1/\varepsilon_0$ and r/a of 8-fold quasicrystal.

### 6.6.4 Band-gap of multi-fold complex PQCs[107]

In the same way, the maximum gap-midgapratio$\Delta\omega/\omega_c$among dielectric contrast 6-18 and fill factor0.1-0.18 is given to study the band-gap property of all the given PQCs. The results are compared in Fig.28and reveal the following information: (1) 10-fold PQC by five beams interference and 12-fold PQC show better ability to generate band-gaps than other structures; (2) 8-fold PQC, 10*-fold PQC by ten beams interference, 14-fold PQC and 18-fold PQC take the second place; (3) Band-gap in 22-fold and 26-fold PQCs is small because these structures are not locally self-similar, which were known to be critical conditions for PBG formation [108]. It can be said that higher rotational symmetry PQC formed by interference is hard to produce band-gap; (4) For the 10-fold and 10*-fold PQCs, the former possesses better self-similarity and wider band-gap than the latter. The reason is, as mentioned above, close light spots in the latter generate interference so as to destroy self-similarity to some extent. The results also explain to some extent why these PQCs are widely studied and even utilized for devices [94-105].

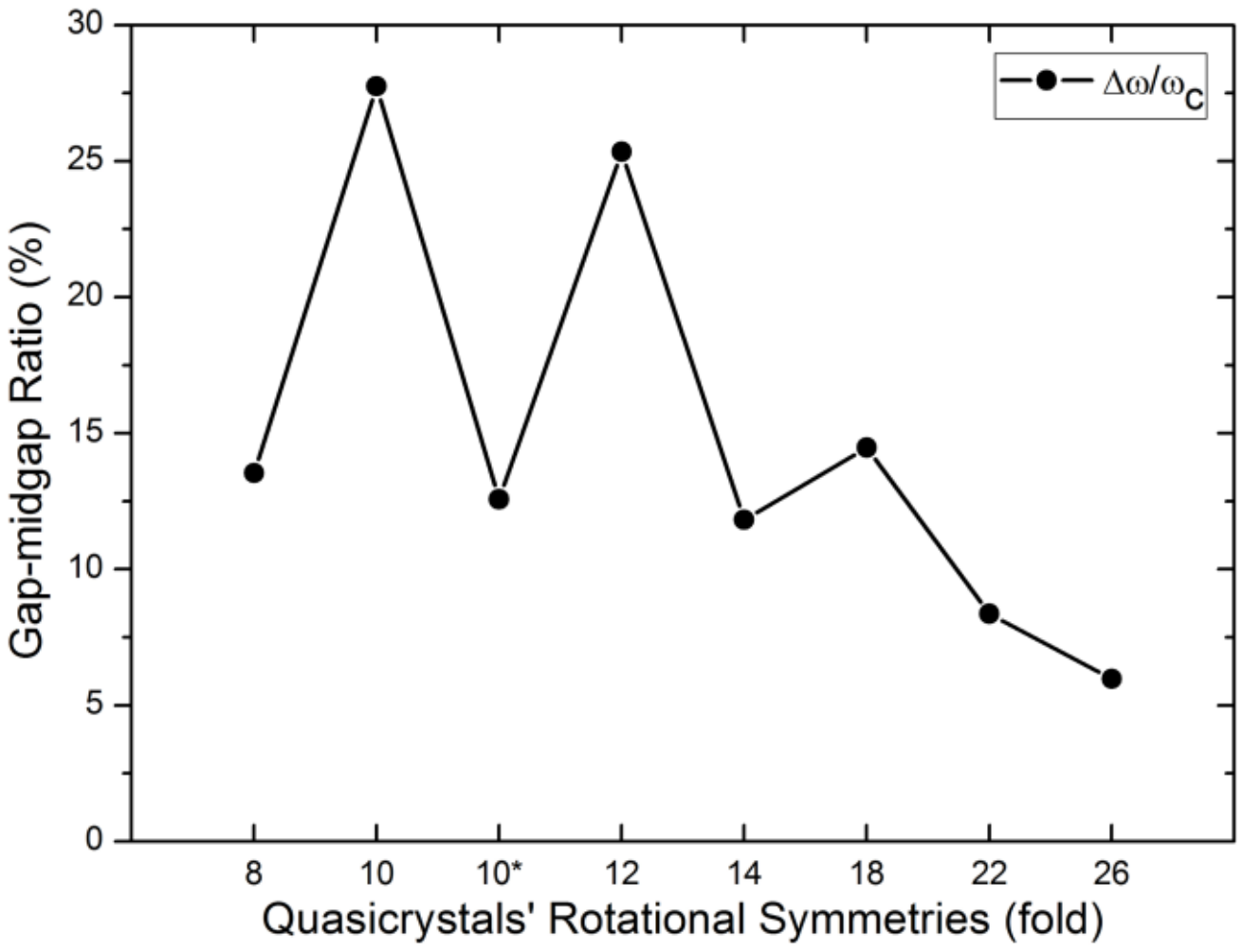

Fig.28 The maximum gap-midgap ratios $\Delta\omega/\omega_c$ of different quasicrystals.

### 6.6.5 Fabrication of 10-fold holographic PQCs [109]

#### 6.6.5.1Material and writing system

The material used in the fabrication is SU-8 photoresist, a high contrast, epoxy-based photo-resist resin from MICRO CHEM Cop. First, glass substrates were subjected to a piranha etch/clean (H2SO4 &H2O2) for 30 min at room temperature, followed by a thorough rinse of de-ionized water and dehydrated at 100 $^0$C for an hour in an oven. SU-8 photo-resist resin was first spin-coated on the pretreated substrate at 300 rpm for 5 s, then coated at 3000 rpm for 60 s, to ensure that the thickness of the SU8 was uniform. The total thickness of the SU-8 film was about 10 mm. Then the film was baked on an oven at 85$^0$Cfor 25 min.

The basic element of the writing system is the top-cut prism interferometer, comprising five identical side faces and two parallel different sized pentagonal bases, as shown in Fig.29. The top-cut prism is the main optical element and made of quartz with refractive index of 1.47. A 355 nm beam from a triple-frequency Nd:YAG laser (YG980, Quantel Co. France) was expanded and collimated to irradiate the sample film through the designed prism. The film was exposed with 20~40 mJ/cm$^2$ dose for about 4s. After exposure, it was post-baked at 85$^0$C for 30 min, developed in Propylene Glycol Monomethyl Ether Acetate (PGMEA, $C6H12O3$) for 50 min followed by rinsing in isopropanol and drying with a blower.

A collimated plane wave is normally incident from the top small base and is split into five umbrella-like beams. The angle between the side and central beam is $\phi = \alpha - arcsin(\sin\alpha / n_\omega)$, where $\alpha = 54.7°$ is the cutting angle of the top-cut prism and $n_\omega = 1.6$ is the refractive index of the SU-8 film at the writing wavelength. By blocking different surfaces of the prism, different interferograms can be obtained with various interference beams. For instance, if the central beam is blocked, as shown in Fig.29, 2D PQC structures can be obtained.

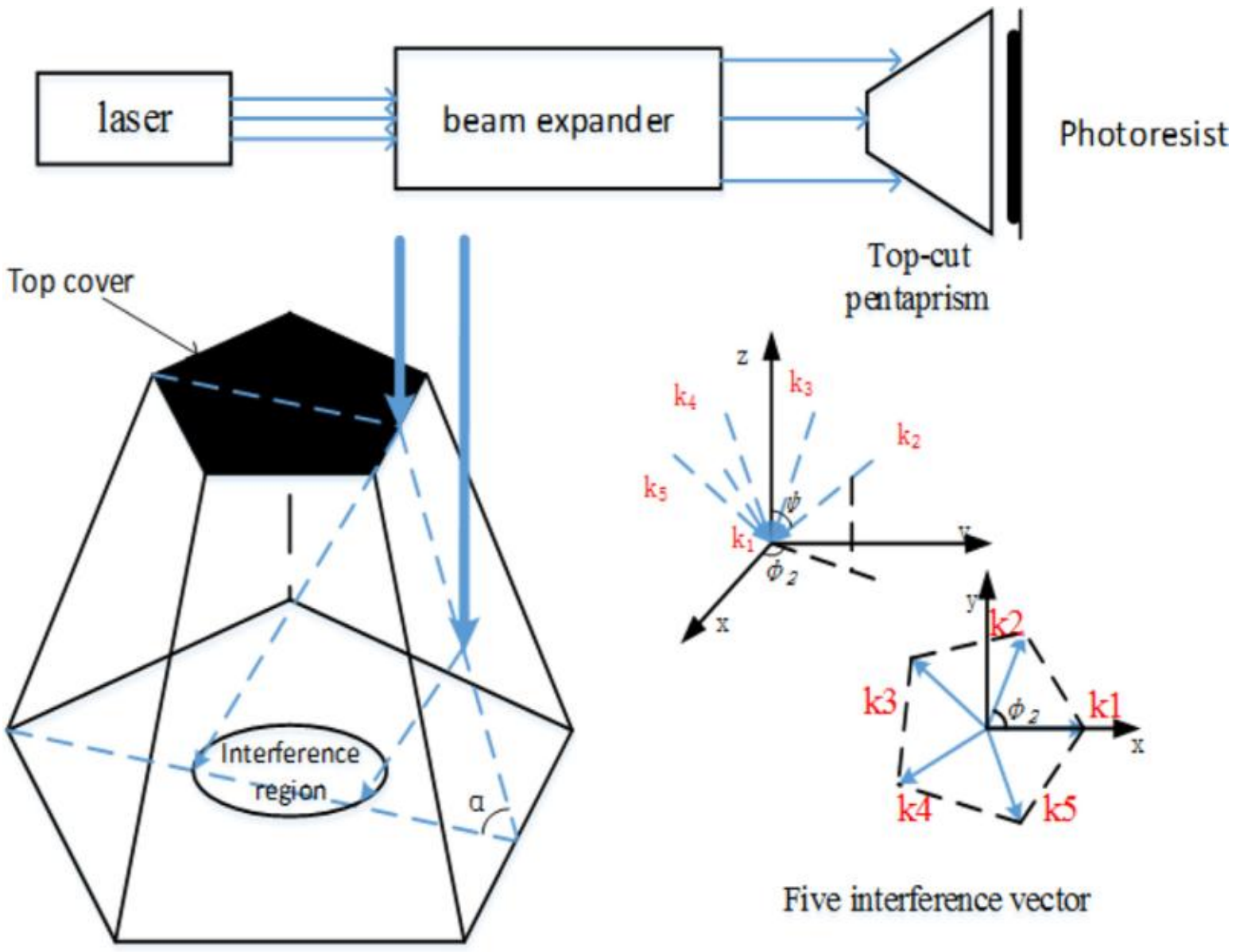

Fig.29 The schematic diagram of the top-cut prism interferometer.

### 6.6.5.2 Experimental Results

The exposed area is about 1.5 cm in diameter. The developed structure was cut into 1cm×1cm pieces for Scanning Electronic Microscopy (SEM S-4800, Hitachi Limited, Japan) and Scanning Probe Microscopy (Multimode 8 SPM, Bruker). For SEM, the sample is coated with a gold film and measured. For SPM, the film topography was obtained by using the tapping mode.

SEM micrographs at the different condition and area are given and shown in Fig.30and Fig.31. Fig.30 indicates the PQC structures with different magnifications on the condition that all beams have the same intensity (40 $mJ/cm^2$) and polarizations. The lattice unit is circled in red marks. This is a perfect 10-fold rotational PQC. When the intensity of two side beams is reduced to 10 $mJ/cm^2$ and the other beams remain unchanged, the PQC structure is shown in Fig.31. This is a poor 10-fold rotational PQC. But from these figures, it can be concluded that large area production can be realized by using this method.

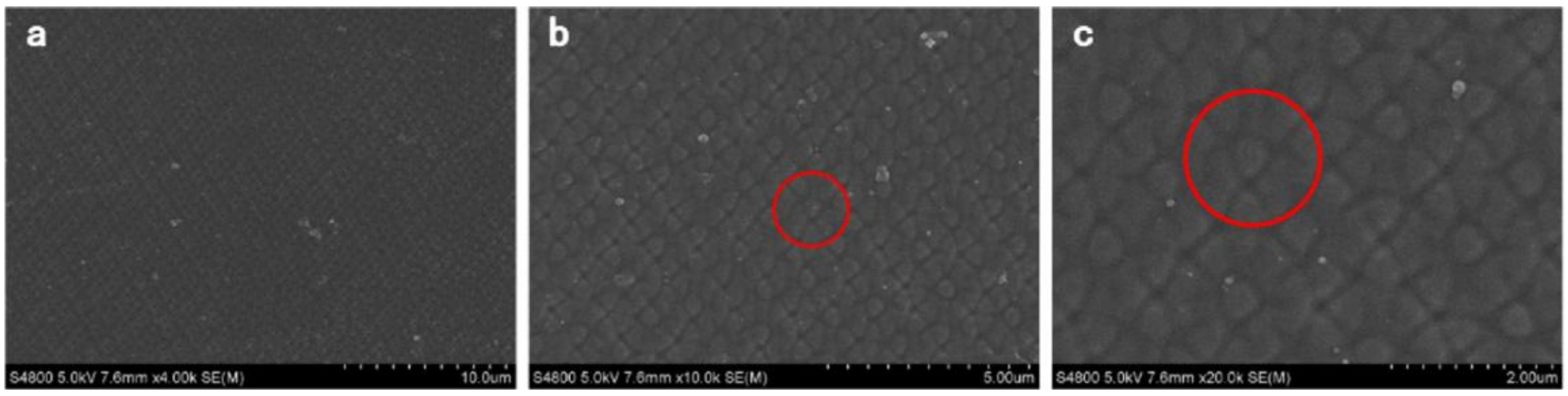

Fig.30 Different magnification of SEM micrograph of PQCs with the same beam intensity

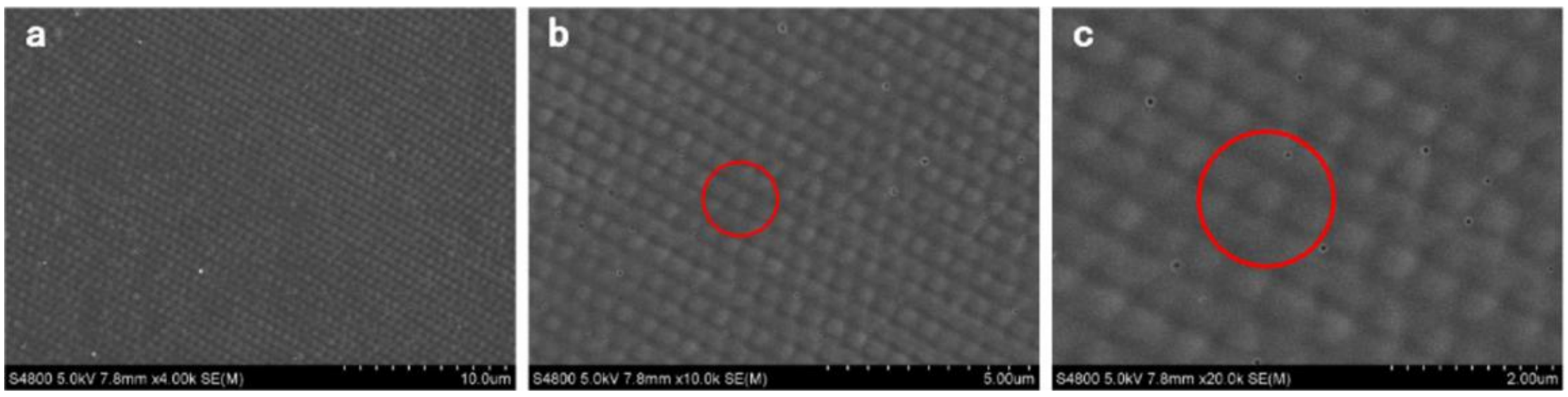

Fig.31Different magnification of SEM micrograph of PQC structure with two beams weaker

At the same time, SPM photographs are also provided to give a clear observation on the PQC structures. Fig.32a and b show the SPM micrographs with different magnification of the same sample. The sample is obtained on the condition that all beams have the same intensity and polarizations and the exposure dose is 20 mJ/cm$^2$ for 4s. Fig. 32b is the magnification of the marked area in Fig.32a. This is a 10-fold rotational symmetry structure with polymer rods in the air background. The bright and dark areas represent polymer and air, respectively. In Fig.32b, the diameter of the circled unit is about 1.6 mm. The distance between two adjacent rods in the circled area is about 490nm. The diameter of the rods is about 400 nm. To signify the 10-fold rotational symmetry more clearly, the diffraction patterns are measured and shown in Fig.32c. From this figure, it also can be seen that ten diffraction spots have equal brightness. This proves that five interfering beams have the same intensity. Fig.32d provides the calculated intensity distributions. White and black areas stand for small and large intensity distributions, respectively. The origin of the calculated intensity distributions is corresponding to the intersection of the optical axis (Z axis) and the sample plane (XY plane). The calculated structure is rotationally symmetrical along the origin. Only one part of the structure is shown in Fig.32d. The corresponding size of the 10-fold unit and adjacent rods is 1.62 mm and 500 nm, respectively. The values are a little larger than the experimental results. This is caused by the shrinkage of the photoresist film in the developing and drying process.

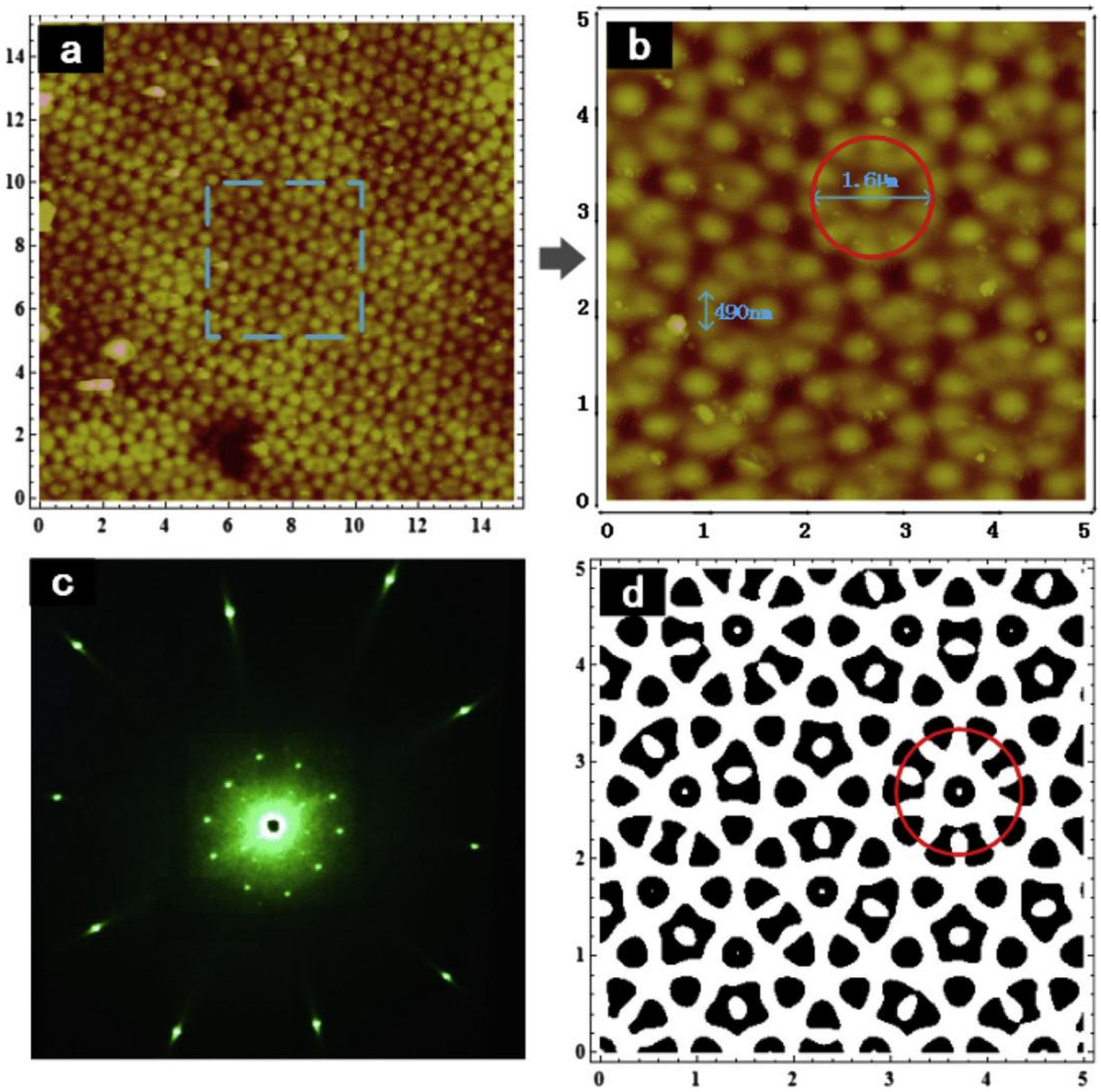

Fig.32 Interference for five beams with the equal intensity. (a) 15 mm×15 mm (b) 5 mm×5 mm SPM micrographs for different magnifications (c) Diffraction patterns(d) the calculated intensity distribution.

The 2D multiple quasicrystal structures based on the holographic interferogram have been constructed and investigated for the band-gap properties. In the course of simulation, the parameter settings are same except for the structure itself. The maximum gap-midgap ratio $\Delta\omega/\omega_c$ within a certain range of dielectric contrast and fill factor is given for different quasicrystals. The results show that 10-fold (five beams interference) and 12-fold quasicrystals are easier to produce band-gaps than others. On the other hand, 10-fold PQCs was prepared by using single-prism holographic lithography. This possibly provides some guidance to the design and fabrication of quasicrystal based photonic device.

**An additional explanation**: in Subsection 6.6.2 we mentioned the so-called "22- and 26-fold symmetry quasicrystals", which have not been observed experimentally so far, and belong to "man-made quasicrystals"due tothe need of engineering application rather than chemistry-physically quasicrystals according tothe group theory.

## 7 Conclusion and discussion

The soft-matter quasicrystal study is a new branch across soft matter and quasicrystal, which presents very interesting and attractive features and extremely enlarge the scope of quasicrystal research, especially on the dynamic study. The meaning of the dynamics is not only for the topic itself but also for the soft matter science. The generalized dynamics becomes a quantitative branch in soft matter study apart from the liquid crystals. The establishment of the dynamics lies in the high symmetry of the matter, for which the symmetry breaking and elementary excitation principles provide the basis of the dynamics, and group theory and group representation theory play important role too.

After establishment of the governing equations, with initial and boundary conditions we have solved certain fundamental problems with the partial differential equations in terms of analytic and numerical methods for some transient, steady dynamic and static cases. These solutions show a consistent examination on both the equations and the solving formulation including the computer program that we created. The computation produces the first group of basic data, for example, for soft-matter quasicrystals $\delta\rho/\rho_0 \sim 10^{-4} - 10^{-3}$ , which may be seen as the compressibility of the matter, while for solid quasicrystals $\delta\rho/\rho_0 \sim 10^{-13}$ according to our calculation, and other basic data, like the order of magnitude of phonon displacements, phason displacements, fluid velocities and so on. As there are orders of magnitude difference of the results, the physical properties of soft-matter quasicrystals are close to those of fluid rather than solid, this is understandable because in the computation we take the material constants from liquid crystals due to lack of the measured material constants of soft-matter quasicrystals. These examples mainly aimed to check the validity of the theory and methods,it does not mean that they belong to the most important subjects in the study. Even so, the sample shown by Fig. 4 can be realized and measured experimentally, one can obtain the measured values, and carry out the comparison between theoretical and experimental results. The sample given by Fig. 14 enhances the correlation between soft-matter and conventional fluid. The numerical analysis of the sample reveals the nature of complex fluid of soft-matter quasicrystals, and explores that the solutions of conventional fluid may provide some rough approximation of soft-matter quasicrystals or a starting point for probing the approximate solutions of the latter.The systematical results obtained here demonstrate a continuous development on the methodology of the new disciplinary of condensed matter physics and chemistry.

The study on the defect of soft-matter quasicrystals is an important subject, which is related with singularity of some field variables and changes the space homogeneity at local places, even leads to the initiation of material breaking in the locations, so greatly influences on material properties. In Section 6 the discussion on application is in a beginning stage. Due to lack of experimental observation, the dislocation study is not in-depth. The practical needs on crack and rupture study in soft matter will push the development further, and promote the quantitative dynamics study in soft-matter quasicrystals.

The governing equations including the equation of state need further examining. Due to the limitation of experimental input, the theoretical examination and comparison is not enough. Our recent work [85, 86] provide a good comparison with those of conventional fluids and solid quasicrystals as the first step and show computation results are significant, and more advanced development can be expected.

More recently, we also pay an effort for studying connection and possible connection between thermodynamics and dynamics of soft-matter quasicrystals, which leads to a development of study on stability of soft-matter quasicrystals [87,88] starting from the discussion of Refs [48.49], in Refs [87,88] the authors proposed two theorems to describe the stability of the first and second kind of soft-matter quasicrystals respectively, which may be an interesting and important topic of quasicrystal science.

Because quasicrystals belong to a material with photon band-gap effect, present important applications in the devices and quantum electronics, in the Subsection 6.6 some applications in this respect are introduced, which are attractive and interesting very much.

**Acknowledgements**

The funding by National Natural Science Foundation of China over the years and Alexander von Humboldt Foundation to the collaboration with German colleagues are gratefully acknowledged.

Many thanks to the discussions with Profs. T. Lubensky, S. Z. D. Cheng and H. H. Wensink. During the work Profs.U. Messerschmidt, H.-R. Trebin, R.Lifshitz, C. Z. Hu, X. F. Li, W. Q. Chen and Dr Z.L. Wang provided many helpfuladvices.

## Appendix Some mathematical supplements

The equations of motion of soft-matter quasicrystals are (14)-(17), which are derived by Poisson bracket method of condensed matter physics.

The concept of Poisson brackets comes from the classical analytic mechanics**,** i.e., for two mechanical quantities $f,g$ there is the following relation

$$\{f,g\}=\sum_i\left(\frac{\partial f}{\partial q_i}\frac{\partial g}{\partial p_i}-\frac{\partial f}{\partial p_i}\frac{\partial g}{\partial q_i}\right)\text{(A1)}$$

which is thePoissonbracket**,** where $p_i, q_i$ denote the canonic momentumand canonic coordinate.

According to the terminology of physics, (A1) is named classical Poisson bracket hereafter.

Relative to the classical Poisson bracket (A1), there is a quantum Poisson bracket, which is related to the commutator in quantum mechanics

$$\left[\hat{A},\hat{B}\right]=\hat{A}\hat{B}-\hat{B}\hat{A}\text{(A2)}$$

in which $\hat{A},\hat{B}$ represent two operators, e.g. $\hat{A}$ represents coordinate operator $x_\alpha$, $\hat{B}$ the momentum operator $p_\beta$, then

$$\left[x_\alpha,p_\beta\right]=i\hbar\delta_{\alpha\beta},\left[x_\alpha,x_\beta\right]=0,\left[p_\alpha,p_\beta\right]=0\text{(A3)}$$

where $i=\sqrt{-1},\hbar=h/2\pi,h$ the Planck constant， $\delta_{\alpha\beta}$ theunit tensor. Equation (A3) is named quantum Poisson bracket. In quantum mechanics, mechanical quantities representoperators. Equation (A3) holds for any operators, in general.

There is inherent connection between the quantum Poisson bracket and classical Poisson bracket，i.e.,

$$\lim_{\hbar\to 0}\frac{i\left[\hat{A}\hat{B}-\hat{B}\hat{A}\right]}{\hbar}=\{A,B\} \quad \text{(A4)}$$

This is well-known result in the quantum mechanics.

Landau and his school developed the limit passing over（A4）from quantum Poisson bracket to the classical Poisson bracket in deriving the hydrodynamic equations of some condensed matter. He took the expansions of mass density and momentum such as：

$$\hat{\rho}(\mathrm{r})=\sum_{\alpha}m_{\alpha}\delta(\mathrm{r}_{\alpha}-\mathrm{r}) \quad \text{(A5)}$$

$$\hat{g}_k(\mathrm{r})=\sum_{\alpha}\hat{p}_k^{\alpha}\delta(\mathrm{r}_{\alpha}-\mathrm{r}) \quad \text{(A6)}$$

whose quantum Poisson brackets are

$$\begin{aligned}&[\hat{\rho}(\mathrm{r}_1),\hat{\rho}(\mathrm{r}_2)]=0\\&[\hat{p}_k(\mathrm{r}_1),\hat{\rho}(\mathrm{r}_2)]=i\hbar\hat{\rho}(\mathrm{r}_1)\nabla_k(\mathrm{r}_1)\delta(\mathrm{r}_1-\mathrm{r}_2)\\&[\hat{p}_k(\mathrm{r}_1),\hat{p}_l(\mathrm{r}_2)]=i\hbar\left(\hat{p}_l(\mathrm{r}_1)\nabla_k(\mathrm{r}_1)-\hat{p}_k(\mathrm{r}_2)\nabla_k(\mathrm{r}_2)\right)\delta(\mathrm{r}_1-\mathrm{r}_2)\end{aligned} \quad \text{(A7)}$$

where $\nabla_k(\mathrm{r}_1)$ represents derivative carrying out on coordinate $\mathrm{r}_1$ , and $\nabla_l(\mathrm{r}_2)$ oncoordinate $\mathrm{r}_2$ .

By using the limit passing over （A4）from the quantum Poisson to the classical Poisson bracket， from（A7）one can obtain the corresponding classical Poisson brackets：

$$\begin{aligned}&\{p_k(\mathrm{r}_1),\rho(\mathrm{r}_2)\}=\rho(\mathrm{r}_1)\nabla_k(\mathrm{r}_1)\delta(\mathrm{r}_1-\mathrm{r}_2)\\&\{p_k(\mathrm{r}_1),p_l(\mathrm{r}_2)\}=\left(p_l(\mathrm{r}_1)\nabla_k(\mathrm{r}_1)-p_k(\mathrm{r}_2)\nabla_k(\mathrm{r}_2)\right)\delta(\mathrm{r}_1-\mathrm{r}_2)\end{aligned} \quad \text{(A8)}$$

Lubensky et al [22] extended the discussion to solid quasicrystals. In solid quasicrystals apart from phonon elementary excitation, there is another excitation---phason excitation, in which the phonon type displacements are simply called phonon field $u_i$, and phason type displacements are simply called phason field $w_i$, respectively.

Similarly expanse the displacement vectors $u_i$ and $w_i$

$$u_k(\mathrm{r})=\sum_{\alpha}u_k^{\alpha}\delta(\mathrm{r}_{\alpha}-\mathrm{r}) \quad \text{（5.5-1）(A9)}$$

$w_k(\mathrm{r}) = \sum_{\alpha} w_k^{\alpha} \delta(\mathrm{r}_{\alpha} - \mathrm{r})$ （5.5-2）(A10)

By using the limit passing over （A4）from the quantum Poisson bracket to the classical Poisson bracket，from (A9)and(A10)one can find whose corresponding classical Poisson brackets as follows

$$\{u_k(\mathrm{r}_1), g_l(\mathrm{r}_2)\} = (-\delta_{kl} + \nabla_l(\mathrm{r}_1)u_k)\delta(\mathrm{r}_1 - \mathrm{r}_2) \text{ (A11)}$$

$$\{w_k(\mathrm{r}_1), g_l(\mathrm{r}_2)\} = (\nabla_l(\mathrm{r}_1)w_k)\delta(\mathrm{r}_1 - \mathrm{r}_2) \text{ (A12)}$$

It is evident that (A12)is quite different from (A11), this leads tothe dissipation equations of phasons given in the subsequent discussion are quite different from those of equations of motion of phonons due to symmetry breaking. The relevant derivations were carried out by Lubensky et al [22].

Here we give an outline in deriving the equations of motion of soft-matter quasicrystals (14) to (17) through the Poisson brackets, this method presents generality physically. The details will be given individually in Chapters 7-11 for different quasicrystal systems of [55]. The application of the Poisson bracket method is combined with generalized Langevin equation which will be introduced as below.

Apart from Poisson brackets, it needs another formula in the derivation of hydrodynamic equations of quasicrystals, which is related to the Langevin equation or generalized Langevin equation, refer to [55].

We know the Langevin equation

$$\frac{\partial \Psi(\mathrm{r},t)}{\partial t} = -\Gamma \Psi(\mathrm{r},t) + F_s \text{ (A13)}$$

in which $\Psi(\mathrm{r},t)$ is a mechanical quantity, $\Gamma$ represents a resistant force, $F_s$ a stochastic force. We also know the equation with multivariables

$$\frac{\partial \Psi_{\alpha}(\mathrm{r},t)}{\partial t} = -\Gamma_{\alpha\beta} \frac{\delta H}{\delta \Psi_{\beta}(\mathrm{r},t)} + (F_s)_{\alpha} \text{ (A14)}$$

where $H = H[\Psi(\mathrm{r},t)]$ denotes a energy functional, which can also be named Hamiltonian, $\frac{\delta H}{\delta \Psi_{\beta}(\mathrm{r},t)}$ represents a variation of $H = H[\Psi(\mathrm{r},t)]$ on $\Psi_{\beta}(\mathrm{r},t)$ , $\Gamma_{\alpha\beta}$ the elements of resistant matrix （or dissipation kinetic coefficient matrix）, the meanings of definitions of other quantities are the same as before.

In $d$ dimensional space, the partial derivative of macroquantity $\Psi_{\alpha}(\mathrm{r},t)$ on time

$$\frac{\partial \Psi_{\alpha}(\mathrm{r},t)}{\partial t}$$

stands for

$$\frac{\partial \Psi_{\alpha}(\mathrm{r},t)}{\partial t} = -\int \left( \left\{ \Psi_{\beta}(\mathrm{r}'), \Psi_{\alpha}(\mathrm{r}) \right\} \frac{\delta H}{\delta \Psi_{\beta}(\mathrm{r}',t)} \right) d^d\mathrm{r}' + \int \left( \frac{\delta \left\{ \Psi_{\beta}(\mathrm{r}'), \Psi_{\alpha}(\mathrm{r}) \right\}}{\delta \Psi_{\beta}(\mathrm{r}',t)} \right) d^d\mathrm{r}' - \Gamma_{\alpha\beta} \frac{\delta H}{\delta \Psi_{\beta}(\mathrm{r},t)} + (F_s)_{\alpha}$$

(A15)

where $d^d\mathrm{r}' = dV$ represents volume element of the integral. Based on the formulas ,

（A8）,（A11） and（A12）, utilized equation(A15)to derive the hydrodynamic equations of quasicrystals. This will be given later. In the derivation the last term in (A15) is omitted.

The derivation of equation of mass conservation is the same as that of simple fluid, so which is omitted.

At first we give the derivation on the equations of motion of phonons due to symmetry breaking.

Put $\Psi_{\alpha}(\mathrm{r},t) = u_i(\mathrm{r},t), \Psi_{\beta}(\mathrm{r}',t) = g_j(\mathrm{r}',t)$ in (A15)and omitting the second and fourth terms in the right-hand side of the equation, then

$$\frac{\partial u_i(\mathrm{r},t)}{\partial t} = -\int \left( \left\{ u_i(\mathrm{r}'), g_j(\mathrm{r}) \right\} \frac{\delta H}{\delta g_j(\mathrm{r}',t)} \right) d^d\mathrm{r}' - \Gamma_{\mathrm{u}} \frac{\delta H}{\delta u_i(\mathrm{r},t)}$$

Substituting bracket (A11) into the integral of right-hand side yields

$$\begin{aligned} \frac{\partial u_i(\mathrm{r},t)}{\partial t} &= \int \left( -\delta_{ij} + \nabla_j(\mathrm{r}) u_i \right) \delta(\mathrm{r}-\mathrm{r}') \frac{g_j(\mathrm{r}')}{\rho(\mathrm{r}')} d^d\mathrm{r}' + \Gamma_{\mathrm{u}} \frac{\delta H}{\delta u_i(\mathrm{r},t)} \\ &= -V_j \nabla_j(\mathrm{r}) u_i - \Gamma_{\mathrm{u}} \frac{\delta H}{\delta u_i(\mathrm{r},t)} + V_i \end{aligned} \quad \text{(A16)}$$

where $\Gamma_{\mathrm{u}}$ denotes the phonon dissipation kinematic coefficient, and the Hamiltonian is defined by

$$\begin{aligned} H &= H[\Psi(\mathrm{r},t)] = \int \frac{\mathrm{g}^2}{2\rho} d^d\mathrm{r} + \int \left[ \frac{1}{2} A \left( \frac{\delta\rho}{\rho_0} \right)^2 + B \left( \frac{\delta\rho}{\rho_0} \right) \nabla \bullet \mathrm{u} \right] d^d\mathrm{r} + F_{el} \\ &= H_{kin} + H_{density} + F_{el} \end{aligned} \quad \text{(A17)}$$

$$F_{el} = F_{\mathrm{u}} + F_{\mathrm{w}} + F_{\mathrm{uw}}, \mathrm{g} = \rho \mathrm{V}$$

and $\mathrm{V}$ represents the fluid velocity field, $A, B$ the constants describe mass density variation, the last term of (A17) represents elastic energies, which consists of phonons, phasons and phonon-phason coupling parts ,

$$\begin{aligned} F_{\mathrm{u}} &= \int \frac{1}{2} C_{ijkl} \varepsilon_{ij} \varepsilon_{kl} d^d\mathrm{r} \\ F_{\mathrm{w}} &= \int \frac{1}{2} K_{ijkl} w_{ij} w_{kl} d^d\mathrm{r} \\ F_{\mathrm{uw}} &= \int \left( R_{ijkl} \varepsilon_{ij} w_{kl} + R_{klij} \varepsilon_{ij} w_{ij} \varepsilon_{kl} \right) d^d\mathrm{r} \end{aligned} \quad \text{(A18)}$$

respectively, $C_{ijkl}$ the phonon elastic constants, $K_{ijkl}$ phason elastic constants, and $R_{ijkl}, R_{klij}$ the phonon-phason coupling elastic constants, and the strain tensors $\varepsilon_{ij}, w_{ij}$ are defined by

$$\varepsilon_{ij} = \frac{1}{2}\left(\frac{\partial u_i}{\partial x_j} + \frac{\partial u_j}{\partial x_i}\right), w_{ij} = \frac{\partial w_i}{\partial x_j} \quad \text{(A19)}$$

the associated stress tensors are related through the constitutive law for soft-matter quasicrystals including 5-, 8-, 10- and 12-fold symmetries (and the 7-, 9-, 14- and 18-fold symmetry quasicrystals are not included but are similar, the difference is resulted in the Hamiltonians between these two kinds of quasicrystals)

$$\left.\begin{aligned} &\sigma_{ij} = C_{ijkl}\varepsilon_{ik} + \mathrm{R}_{ijkl}\, w_{kl}, \\ &H_{ij} = K_{ijkl} w_{ij} + \mathrm{R}_{klij}\, \varepsilon_{kl}, \\ &p_{ij} = -p\delta_{ij} + \sigma_{ij}^{'} = -p\delta_{ij} + \eta_{ijkl}\dot{\xi}_{kl}, \\ &\varepsilon_{ij} = \frac{1}{2}\left(\frac{\partial u_i}{\partial x_j} + \frac{\partial u_j}{\partial x_i}\right), w_{ij} = \frac{\partial w_i}{\partial x_j}, \dot{\xi}_{ij} = \frac{1}{2}\left(\frac{\partial V_i}{\partial x_j} + \frac{\partial V_j}{\partial x_i}\right) \end{aligned}\right\} \quad \text{(A20)}$$

where $\eta_{ijkl}$ denotes the viscosity coefficient tensor of fluid phonon. Now consider the derivation of phason dissipation equations.

In (A15) put $\Psi_\alpha(\mathrm{r},t) = w_i(\mathrm{r},t), \Psi_\beta(\mathrm{r}',t) = g_j(\mathrm{r}',t)$, neglecting the second and fourth terms in the right-hand side, then substituting the Poisson bracket (A12) into it leads to

$$\frac{\partial w_i(\mathrm{r},t)}{\partial t} = -\int\left(\left\{w_i(\mathrm{r}'), g_j(\mathrm{r})\right\}\frac{\delta H}{\delta g_j(\mathrm{r}',t)}\right)d^d\mathrm{r}' - \Gamma_{\mathrm{w}}\frac{\delta H}{\delta w_i(\mathrm{r},t)}$$

then

$$\begin{aligned} \frac{\partial w_i(\mathrm{r},t)}{\partial t} &= \int\left(\nabla_j(\mathrm{r})w_i\right)\delta(\mathrm{r}-\mathrm{r}')\frac{g_j(\mathrm{r}')}{\rho(\mathrm{r}')}d^d\mathrm{r}' - \Gamma_{\mathrm{w}}\frac{\delta H}{\delta w_i(\mathrm{r},t)} \\ &= -V_j\nabla_j(\mathrm{r})w_i - \Gamma_{\mathrm{w}}\frac{\delta H}{\delta w_i(\mathrm{r},t)} \end{aligned} \quad \text{(A21)}$$

in which $\Gamma_{\mathrm{w}}$ denotes the phason dissipation coefficient, and Hamiltonian is defined by (A17).

By comparing (A16) and (A21), it is found that the physical meanings of phonons and phasons in hydrodynamic sense are quite different to each other. According to the explanation of Lubensky et al [22], the phonon represents wave propagation, while phason represents diffusion.

The momentum conservation equations are

$$\frac{\partial g_i(\mathrm{r},t)}{\partial t}=-\nabla_k(\mathrm{r})(V_k g_i)+\nabla_j(\mathrm{r})\left(-p\delta_{ij}+\eta_{ijkl}\nabla_k(\mathrm{r})g_l\right)-\left(\delta_{ij}-\nabla_i u_j\right)\frac{\delta H}{\delta u_i(\mathrm{r},t)}-$$
$$\left(\nabla_i w_j\right)\frac{\delta H}{\delta w_i(\mathrm{r},t)}-\rho\nabla_i(\mathrm{r})\frac{\delta H}{\delta\rho(\mathrm{r},t)},g_j=\rho V_j \quad \text{(A22)}$$

where $\eta_{ijkl}$ denotes the viscosity coefficient tensor of fluid, and the fluid phonon stress tensor is

$$p_{ij}=-p\delta_{ij}+\sigma_{ij}^{'}=-p\delta_{ij}+\eta_{ijkl}\dot{\xi}_{kl} \quad \text{(A23)}$$

with the deformation velocity tensor

$$\dot{\xi}_{kl}=\frac{1}{2}\left(\frac{\partial V_k}{\partial x_l}+\frac{\partial V_l}{\partial x_k}\right) \quad \text{(A24)}$$

Equations (A22) is different from the momentum conservation equation of solid quasicrystals due to the introducing of the fluid phonon. The equation (A22) can be understood as generalized Navier-Stokes equations.

Equations (A17), (A21), (A22) and mass density conservation equation

$$\frac{\partial\rho}{\partial t}+\nabla_k\left(\rho V_k\right)=0 \quad \text{(A 25)}$$

are the equations of motion of soft-matter quasicrystals, i.e., the equations (14) to (17) in the text.

This derivation can also be completed by the Lie group method, this shows there is more general meaning of the method.

As an example, we here list the governing equations of generalized dynamics of 12-fold symmetry quasicrystals in the three-dimensional case (which can be obtained by substituting the constitutive law (22) into the Hamiltonian then into equations (14)-(17) and including (18)) as follows

$$
\left.
\begin{aligned}
&\frac{\partial\rho}{\partial t}+\nabla\bullet(\rho\mathbf{V})=0\\
&\frac{\partial(\rho V_x)}{\partial t}+\frac{\partial(V_x\rho V_x)}{\partial x}+\frac{\partial(V_y\rho V_x)}{\partial y}+\frac{\partial(V_z\rho V_x)}{\partial z}=-\frac{\partial p}{\partial x}+\eta\nabla^2V_x+\frac{1}{3}\eta\frac{\partial}{\partial x}\nabla\bullet\mathbf{V}\\
&+\left(C_{66}\frac{\partial^2}{\partial y^2}+C_{44}\frac{\partial^2}{\partial z^2}\right)u_x+\left(C_{12}+C_{66}\right)\frac{\partial^2u_y}{\partial x\partial y}+(C_{13}+C_{44}-C_{11})\frac{\partial^2u_z}{\partial x\partial z}+(C_{11}-B)\frac{\partial}{\partial x}\nabla\bullet\mathbf{u}-(A-B)\frac{1}{\rho_0}\frac{\partial\delta\rho}{\partial x}\\
&\frac{\partial(\rho V_y)}{\partial t}+\frac{\partial(V_x\rho V_y)}{\partial x}+\frac{\partial(V_y\rho V_y)}{\partial y}+\frac{\partial(V_z\rho V_y)}{\partial z}=-\frac{\partial p}{\partial y}+\eta\nabla^2V_y+\frac{1}{3}\eta\frac{\partial}{\partial y}\nabla\bullet\mathbf{V}\\
&+\left(C_{12}+C_{66}\right)\frac{\partial^2u_x}{\partial x\partial y}+\left(C_{66}\frac{\partial^2}{\partial x^2}+C_{11}\frac{\partial^2}{\partial y^2}+C_{44}\frac{\partial^2}{\partial z^2}\right)u_y+(C_{13}+C_{44})\frac{\partial^2u_z}{\partial y\partial z}+(C_{11}-B)\frac{\partial}{\partial y}\nabla\bullet\mathbf{u}-(A-B)\frac{1}{\rho_0}\frac{\partial\delta\rho}{\partial y}\\
&\frac{\partial(\rho V_z)}{\partial t}+\frac{\partial(V_x\rho V_z)}{\partial x}+\frac{\partial(V_y\rho V_z)}{\partial y}+\frac{\partial(V_z\rho V_z)}{\partial z}=-\frac{\partial p}{\partial z}+\eta\nabla^2V_z+\frac{1}{3}\eta\frac{\partial}{\partial z}\nabla\bullet\mathbf{V}\\
&+\left(C_{44}\frac{\partial^2}{\partial x^2}+C_{44}\frac{\partial^2}{\partial y^2}+(C_{33}-C_{13}-C_{44})\frac{\partial^2}{\partial z^2}\right)u_z+(C_{13}+C_{44}-\mathrm{B})\frac{\partial}{\partial z}\nabla\bullet\mathbf{u}-(A-B)\frac{1}{\rho_0}\frac{\partial\delta\rho}{\partial z}\\
&\frac{\partial u_x}{\partial t}+V_x\frac{\partial u_x}{\partial x}+V_y\frac{\partial u_x}{\partial y}+V_z\frac{\partial u_x}{\partial z}=V_x+\Gamma_{\mathbf{u}}\left[\left(C_{11}\frac{\partial^2}{\partial x^2}+C_{66}\frac{\partial^2}{\partial y^2}+C_{44}\frac{\partial^2}{\partial z^2}\right)u_x+(C_{11}-C_{66})\frac{\partial^2u_y}{\partial x\partial y}+(C_{13}+C_{44})\frac{\partial^2u_z}{\partial x\partial z}\right]\\
&\frac{\partial u_y}{\partial t}+V_x\frac{\partial u_y}{\partial x}+V_y\frac{\partial u_y}{\partial y}+V_z\frac{\partial u_y}{\partial z}=V_y+\Gamma_{\mathbf{u}}\left[(C_{11}-C_{66})\frac{\partial^2u_x}{\partial x\partial y}+\left(C_{66}\frac{\partial^2}{\partial x^2}+C_{11}\frac{\partial^2}{\partial y^2}+C_{44}\frac{\partial^2}{\partial z^2}\right)u_y+(C_{13}+C_{44})\frac{\partial^2u_z}{\partial y\partial z}\right]\\
&\frac{\partial u_z}{\partial t}+V_x\frac{\partial u_z}{\partial x}+V_y\frac{\partial u_z}{\partial y}+V_z\frac{\partial u_z}{\partial z}=V_z+\Gamma_{\mathbf{u}}\left[(C_{13}+C_{44})\left(\frac{\partial^2u_x}{\partial x\partial z}+\frac{\partial^2u_y}{\partial y\partial z}\right)+\left(C_{44}\frac{\partial^2}{\partial x^2}+C_{44}\frac{\partial^2}{\partial y^2}+C_{33}\frac{\partial^2}{\partial z^2}\right)u_z\right]\\
&\frac{\partial w_x}{\partial t}+V_x\frac{\partial w_x}{\partial x}+V_y\frac{\partial w_x}{\partial y}+V_z\frac{\partial w_x}{\partial z}=\Gamma_{\mathbf{w}}\left[K_1\nabla_1^2w_x+(K_2+K_3)\frac{\partial^2w_x}{\partial y^2}+K_4\frac{\partial^2w_x}{\partial z^2}+2K_3\frac{\partial^2w_y}{\partial x\partial y}\right]\\
&\frac{\partial w_y}{\partial t}+V_x\frac{\partial w_y}{\partial x}+V_y\frac{\partial w_y}{\partial y}V_z\frac{\partial w_y}{\partial z}=\Gamma_{\mathbf{w}}\left[(K_2+K_3)\frac{\partial^2w_x}{\partial x\partial y}+K_3\frac{\partial^2w_x}{\partial y\partial z}+K_1\nabla_1^2w_y+(K_2+K_3)\frac{\partial^2w_y}{\partial x^2}+(K_1+K_2+K_3))\frac{\partial^2w_y}{\partial x\partial z}\right]\\
&p=f(\rho)=3\frac{k_BT}{l^3\rho_0^3}\left(\rho_0^2\rho+\rho_0\rho^2+\rho^3\right)
\end{aligned}
\right\}\text{(A 26)}
$$

in which $\nabla^2=\frac{\partial^2}{\partial x^2}+\frac{\partial^2}{\partial y^2}+\frac{\partial^2}{\partial z^2},\nabla_1^2=\frac{\partial^2}{\partial x^2}+\frac{\partial^2}{\partial y^2}$, $\nabla=\mathbf{i}\frac{\partial}{\partial x}+\mathbf{j}\frac{\partial}{\partial y}+\mathbf{k}\frac{\partial}{\partial z}$ , $\mathbf{V}=\mathbf{i}V_x+\mathbf{j}V_y+\mathbf{k}V_z$ , $\mathbf{u}=\mathbf{i}u_x+\mathbf{j}u_y+\mathbf{k}u_z$ , and $C_{11},C_{12},C_{13},C_{33},C_{44},C_{66}=\left(C_{11}-C_{12}\right)/2$ the phonon elastic constants, $K_1,K_2,K_3,K_4$ the phason elastic constants, $\eta$ the fluid dynamic viscosity, and $\Gamma_u$ and $\Gamma_w$ the phonon and phason dissipation coefficients, $A$ and $B$ the material constants due to variation of mass density, respectively, and the coupling constant between phonons and phasons is equal to zero due to the decouple for this quasicrystal system..

The equations (A26) are the final governing equations of dynamics of soft-matter quasicrystals of 12-fold symmetry in three-dimensional case with fields variables $u_x,u_y,u_z,w_x,w_y,V_x,V_y,V_z,\rho$ and $p$ , the amount of the field variables is 10, and amount of field equations is 10 too, among them: (A26a) is the mass conservation equation, (A26b)- (A26d) the momentum conservation equations or the generalized Navier-Stokes equations, (A26e) - (A26g) the equations of motion of phonons due to the symmetry breaking, (A26h) and (A26i) the phason dissipation equations, and (A26j) the equation of state, respectively. The equations are consistent to be mathematical solvable, if there is lack of the equation of state, the equation system is not closed, and has no meaning mathematically and physically. This shows the equation of state is necessary.

In equations (A26) we can see that there is lack of coupling between phonons and phasons, which belong to the simplest dynamics equations in all of systems of soft-matter quasicrystals. Of course these equations are quite complicated mathematically, this leads to a great difficulty in solving. The difficulty is not only coming from the mathematical structure of the equations but also from the coupling of initial and boundary conditions. When we solve equations (A26) the appropriate initial and boundary conditions must be considered, the meaningful solutions must satisfy both governing equations as well as the initial- and boundary-value conditions. In Section 6 in the text some solved examples were introduced and show the well-conditional initial- and boundary-value problems of the governing equations are solvable and some meaningful solutions physically are obtained by analytic and numerical methods respectively.